\documentclass[%
 aip,
 amsmath,amssymb,
 reprint,%
 onecolumn
]{revtex4-2}

\usepackage{setspace}

\usepackage{graphicx}
\usepackage{caption}
\usepackage{subcaption}
\usepackage{dcolumn}
\usepackage{bm}
\usepackage{subcaption}
\usepackage{changepage}
\usepackage[utf8]{inputenc}
\usepackage[T1]{fontenc}
\usepackage{mathptmx}
\begin{document}

\spacing{1.5}

\preprint{AIP/123-QED}

\title[]{Effect of Vortex and Entropy Sources in Sound Generation for Compressible Cavity Flow}

\author{Nitish Arya}
 \affiliation{Department of Aerospace Engineering, Indian Institute of Technology Kanpur, 208016, Uttar Pradesh,  India}
\author{Ashoke De}%
 \email{ashoke@iitk.ac.in}
\affiliation{Department of Aerospace Engineering, Indian Institute of Technology, Kanpur, 208016, Uttar Pradesh,
India
}%


\begin{abstract}
The present study reports the effect of different source terms on the near and far-field acoustic characteristics of compressible flow over a rectangular cavity using hybrid Computational Aero Acoustics (CAA) methodology. We use a low dispersive and dissipative compressible fluid flow solver in conjunction with an Acoustic Perturbation Equation (APE) solver based on the spectral/hp element method. The hybrid approach involves calculating the base fields and the acoustic sources from a fluid simulation in the first step. In the next step, the acoustic solver utilizes the variables to predict the acoustic propagation due to the given sources. The validation of the methodology against benchmark cases provides quite accurate results while compared against existing literature. The study is then extended to assess the importance of the entropy source term for the flow over a rectangular cavity. The predictions of hybrid simulations with vortex and entropy source terms reproduce the perturbation pressure values very close to the existing Direct Numerical Simulation (DNS) results. Moreover, the results suggest that the use of just the vortex source terms over-predicts the perturbation pressure near the source region. Finally, we have carried out detailed simulations with all the source terms to investigate the noise sources for compressible flow over the cavity for different Mach number range ($M = 0.4,0.5,0.6,0.7,1.5$). The obtained acoustic spectra and the sound directivity are in close agreement with the reference experiment.
\end{abstract}

\maketitle

\begin{quotation}
	
\end{quotation}

\section{\label{sec:level1}Introduction}

The aircraft industry has been growing at an unprecedented rate since the last decade. The increase in the number of take-off and landings at major airports has contributed vastly to the sound pollution in an urban environment. Therefore, sound mitigation is one of the most critical concerns of the aircraft industry and is the subject of active research \cite{li2020noise, laurendeau2008subsonic, narayanan2015airfoil,panigrahi2019effects,baskaran2019effects,baskaran2019aeroacoustic,li2013mechanism}. The aerodynamic sound generated by an aircraft consists of jet noise, fan noise, and airframe noise. While the categorization of sound based on the different components of an aircraft might seem relatively easy and intuitive, the process of aerodynamic sound generation itself is rather complicated. In fact, it is so elusive that a complete theory of aerodynamic sound generation is still absent. One of the earliest attempts at understanding the flow-generated noise was that of Sir James Lighthill  \cite{lighthill1952sound,lighthill1954sound} to study jet noise. He rearranged the equations of mass and momentum conservation to arrive at an inhomogeneous wave equation. In this equation, the sound propagation was separated from its generation; thus, it was analogous to classical acoustics. Many acoustic analogies have been developed since then, incorporating realistic effects such as solid boundaries, moving walls, refraction due to variable density and convection \cite{curle1955influence,ffowcs1969sound,lilley1974nouse,phillips1960generation,goldstein2003generalized}.Powell, \cite{powell1964theory} in his analogy, suggested that the fluctuations in the vorticity (in the form of Coriolis acceleration or Lamb vector) are the dominant sources of sound generation. Even though the assumptions of this analogy were similar to that of Lighthill's, it was compelling because to describe the sound field of a source region adequately; just the vorticity distribution is required, which is highly concentrated in a small region of the flow domain as opposed to the velocity or pressure distribution which is extensively spread across the whole domain.    

The source terms in any acoustic analogy can be obtained via fluid simulation in the first step. The calculation of the convolution of the Green's Function with the source terms in the second step would provide the predicted sound at any observer point. Such an approach is called a hybrid method contrary to a Direct Noise Calculation in which the source and the near propagation region are entirely resolved using a scale resolving simulation. Since a fraction of the total energy of a flow is radiated as sound energy, a linearized form of equations can be solved in the acoustic propagation region. This forms the basis of a hybrid approach in which source region and near acoustic field are simulated using a fluid flow solver, and equations such as Linearized Euler Equations (LEE) and Acoustic Perturbation Equations (APE)\cite{ewert2003acoustic} are solved in the far-field. The advancements in the non-reflective boundary conditions \cite{giles1990nonreflecting, richards2004evaluation} and damping due to sponge layers \cite{colombo2016assessment} have made this approach quite popular for aeroacoustics simulations.

More often than not, LEE's solution is contaminated by the excited hydrodynamic instabilities due to the source terms, which result from non-linear unsteady fluid flow equations. With the source terms mentioned above, the LEE describes the propagation of acoustic, vorticity, and entropy modes, which render them unstable for arbitrary mean flows. APE circumvents this problem with filtered sources such that neither the vorticity nor the entropy modes describe convection. The far-field domain is, thus, associated with only the propagation of the acoustic modes.

Hybrid CAA methods have primarily focussed on low Mach number flow regimes in which the convection effects are not that pronounced with the accompanying acoustic sources being compact. In such conditions, the vorticity source of Powell's analogy is the most dominating noise source. However, as the compressibility effects in the flow become more dominant, the entropy fluctuations become significant \cite{bayly1992density}. The source term composed of spatio-temporal changes in the entropy becomes large compared to the vortex source terms dictated by the APE system. Morfey \cite{morfey1973amplification} has proposed that the aerodynamic sound is more pronounced when entropy inhomogeneities are convected. A few researchers have attempted to study the effect of entropy fluctuations in compressible jet flows as a noise source \cite{lew2007investigation,uzun2004coupling}.   

Therefore, the present study investigates the noise sources in a compressible flow over a rectangular cavity using a hybrid methodology. The flow over a cavity is a simplistic yet fascinating flow configuration with applications in many high-speed engineering systems such as weapons bay in military aircraft, landing gear housing, scramjet combustors, etc. The mechanism of sound production due to the non-uniformities in the entropy is not well understood. The studies involving the effect of such a term in systems without any combustion process are rare. A correlation between the source term evolution and the sound generation is attempted, aiming to identify the regions and flow structures responsible for the noise generation inside a cavity. A hybrid approach involving fluid flow solver and APE system with vorticity and entropy fluctuations as the source terms are employed. While the vorticity source terms are calculated using the cross product of vorticity and velocity, the entropy source term is calculated using the NASA polynomials \cite{allison1972polynomial}. The methodology is first validated against several benchmark cases and then employed for the rectangular cavity of aspect ratio $(L/D)$ 2 for Mach numbers ranging from 0.4 to 1.5.

\section{\label{sec:level1}Numerical Method}
\subsection{\label{sec:level2}Flow Solver}
\quad Unsteady compressible Navier-Stokes Equations, solved in the OpenFoam framework employing Finite Volume Method over unstructured grids, are represented as  \begin{equation}
\frac{d}{dt}\int_V\vec{U}dV + \sum_{i = 1}^{3} \int_\Omega (\vec{F}_{i}^{c}-\vec{F}_{i}^{v})nd\Omega
\end{equation}
with
\begin{equation}
\begin{aligned}
\vec{U}=\begin{Bmatrix} \rho \\ \rho u_{i} \\ \rho E\end{Bmatrix},\vec{F}_{i}^{c}=\begin{Bmatrix} \rho u_{i} \\ \rho u_{i}u_{j} + p\delta_{ij} \\ \rho u_{i}H \end{Bmatrix},\\ 
\vec{F}_{i}^{v}=\begin{Bmatrix} 0 \\ \sigma_{ij} \\ \sigma_{ik}u_{k} - q_{i} \end{Bmatrix}
\end{aligned}
\end{equation} representing the vectors of conservative flux, inviscid flux, and viscous flux, respectively. Here, $u_{i}$ is the velocity component along the $i-th$ cartesian coordinate, $\rho$ is the density, E is the sum of kinetic and internal energy, and H is the total enthalpy.

The solver \cite{modesti2017low} uses Advection Upstream Splitting Method (AUSM) \cite{liou1993new} for the splitting of the convective fluxes. Pressure and convective artificial diffusion is added while solving the equations depending on the flow Mach number. A detailed description of the artificial diffusion and the application of shock sensor for flows involving discontinuities can be found out in the work of Modesti and Pirozzoli \cite{modesti2017low}. The solver employs low storage, third order, four-stage Runge-Kutta method for temporal discretization and has been tested against several cases over a wide range of Mach numbers \cite{modesti2017low}.

\subsection{\label{sec:level2}Acoustic Solver}

\quad The acoustic solver employed in the current study solves Acoustic Perturbation Equations \cite{ewert2003acoustic}. Out of the different forms of the APE, a variant is chosen in which the source terms and the base fields are obtained from a compressible simulation. The APE solver is available in the nektar++ \cite{ cantwell2015nektar++}, an open-source higher-order spectral/hp elements framework. Most traditional-higher order discretization schemes have difficulty in handling wave propagation for complex geometries. In nektar++, this problem is mitigated with the help of the Discontinuous Galerkin Scheme \cite{cockburn2000development}. It allows discontinuity at the element boundary while conserving the flux through the element surface. This scheme is, therefore, accompanied by relatively more straightforward implementation and higher stability. The APE system is given by 
\begin{equation}
\frac{\partial p'}{\partial t} + \overline c^{2}\nabla \cdot \left(\overline \rho \mathbf{u}' + \overline{\mathbf{u}} \frac{p'}{\overline c^{2}} \right) = \overline{c}^{2}q_{c}
\end{equation}

\begin{equation}
\frac{\partial \mathbf{u'}}{\partial t} + \nabla \left(\overline{\mathbf{u}} \cdot \mathbf{u}' \right) + \nabla \left( \frac{p'}{\overline \rho} \right)  = \mathbf {q_{m}}
\end{equation}
where the sources are given by-
\begin{equation}
q_{c} = -\nabla \cdot \left(\rho' \mathbf{u}' \right)' + \frac{\overline \rho}{C_{p}} \frac{Ds'}{Dt}
\end{equation}

\begin{equation}
\mathbf{q_{m}} = -\left(\mathbf{\omega} \times \mathbf{u} \right)' + T'\nabla \overline s - s'\nabla \overline T - \left(\nabla \frac{u'^{2}}{2} \right)' + \left(\frac{\nabla \cdot \overline \tau}{\rho} \right)'
\end{equation}

The primed variables are the fluctuating variables, whereas an overbar represents the time-averaged quantities. In the sources, the terms containing two primed quantities are usually much lesser than the other terms. They, therefore, do not contribute significantly to the overall sources. Finally, the viscous contributions to the sources are also neglected. Thus, the fluctuating Lamb Vector $(\omega \times u)'$ and the spatio-temporal changes in fluctuating entropy are used as source terms for the present case.

\subsection{\label{sec:level2}Coupling Method}

Traditionally, in the hybrid approach, the source terms are saved at different time steps to calculate the acoustic fields. This procedure is commonly referred to as file-based coupling. The major drawback of such a coupling is the considerable space requirement, sometimes of the order of a few Terabytes. The reading and writing of this data set usually require large memory. This requirement can be by-passed by sending asynchronous data from the fluid solver to the acoustic solver with a coupling library's help. An essential feature of such a library is the ability to couple codes written in different languages and platforms. Many code agnostic coupling libraries are available, out of which preCICE \cite{bungartz2016precice,uekermann2017official,chourdakis2017general} and CWIPI \cite{duchaine2011first}are the most popular choices. For specific reasons mentioned subsequently, the present study employs the CWIPI library for the coupling. The key features \cite{lackhove2018hybrid} of the coupling process are summarised below. 

\paragraph{Asynchronous field transfer:}
Vortical structures with some characteristic length scale generate acoustic waves with specific wavelengths depending on the flow Mach number \cite{ewert2003acoustic}. Thus, the time scales and the grid requirements for both the simulation might differ, which gives rise to different time steps. The fluid solver data can be sent to the acoustic solver depending on both the solvers' time steps. For instance, if the time step of the flow solver is ten times that of the acoustic solver, then the fields are transferred at every 10th acoustic solver's time step. This procedure also ensures the solver's higher stability by sending a linear combination of the fields during different time steps.

\paragraph{Filtering:}The time-averaged base fields and the unsteady source terms sent from the fluid to the acoustic solver contain high-frequency oscillations detrimental to the stability. Therefore, a spatial low pass filter is applied to the fields before being used in the acoustic solver. This filtering process is applied judiciously so that the essential features are not smeared off. The base fields are just sent once from the fluid solver and saved in a file from where they can be read subsequently. Then, in each iteration, source terms are read from the fluid solver with a larger filter size corresponding to the resolution of the wavelength of the acoustic waves and the order of the interpolating polynomial used in the spectral/hp element method \cite{moura2015linear}.  

\paragraph{Oversampling:} The data transfer from a finite volume code to a spectral code can be tricky. The variables are solved at the cell centers in a finite volume code, which must be interpolated to the cell vertices before they can be sent to the spectral code. The CFD and CAA's grid resolution may vary, which means that the representation of a field with an insufficient number of points might introduce an aliasing error. This problem is circumvented by receiving the fields at a higher number of points, reducing aliasing error.

\section{\label{sec:level1}Computational Details}
\subsection{\label{sec:level2}Data Handling}

Ideally, in a hybrid approach, the source terms obtained from the CFD simulation in the small source region only are transferred to the Acoustic Solver. It makes sense, too, as discussed earlier, that the vorticity source terms are concentrated in a smaller region, thereby allowing us to limit the region from which the source terms are obtained. However, the data exchange between the solvers handled by the CWIPI library first reads the mesh for both the solvers and establishes a communication channel. Data exchange from a particular part of the mesh of one solver to a specific part of the mesh belonging to a different solver involves specific computational challenges that have not been addressed in this particular study. Thus, even though not required, the data exchange is done between the total number of grid points for both the solvers if the number of grid points is equal for both the meshes. If the number of grid points is unequal, then the data is either interpolated from the nearest neighbor, or a uniform value is provided for the non-overlapping points. In the latter case, the computational overhead for treating the non-overlapping points increases, thereby making the simulation much slower for larger meshes. 

Since the CWIPI library provides fully parallel, asynchronous data transfer between different codes, both the solvers operate simultaneously in parallel using an MPI environment. For all the cases included in the present study, the fluid solver is maintained at a constant time step of $10^{-8}$ seconds, and the acoustic solver runs at a constant time step of $10^{-9}$ seconds. Thus, the acoustic solver receives data at every $10^{th}$ time step from the fluid solver. The intermediate expansion on which the data is received consists of $10$ points in each direction to avoid aliasing error.

\subsection{\label{sec:level2}Boundary Treatment}

Boundary treatment is of paramount importance in hybrid CAA. The region where the acoustic data is recorded must be free from reflections from the boundary. It not only corrupts the data but also affects the stability of the numerical procedure. Since a compressible system of equations is being used to make the fluid simulation free from the wave reflections from the boundaries, a wave-transmissive boundary condition is used at the exit boundaries \cite{soni2017characterization,soni2018investigation,soni2018role,seshadri2020investigation,ashoke2016numerical}. This ensures that the source region is not corrupted by the boundary reflections while limiting the distance of the exit boundaries from the region of acoustic source generation. 

For the cases considered in this study, the source generation occurs far from any inlet or outlet boundaries. This means that at all the boundaries of the acoustic solver, non-reflecting boundary conditions are imposed. The application of non-reflecting boundary conditions in acoustic solvers is generally preceded by a damping zone or a sponge layer to dampen the acoustic waves reaching the boundary. Such a methodology is preferred to have stable simulation and a reduction in the computational domain size.

However, the data exchange is done from the whole fluid domain to its equivalent acoustic domain for the present study. The standard procedure of using a smaller fluid domain and a larger acoustic domain is problematic for two reasons.

\paragraph{Computational overhead:}The smaller fluid domain would mean that the values at the acoustic mesh points not located on the fluid mesh are interpolated or assigned a value as mentioned by the user every time the data is exchanged. This increases the data exchange time between the two solvers making the coupled simulation very slow for larger meshes.

\paragraph{Boundary values of the fluid domain:}The values at the boundaries of the fluid domain can generate small disturbances in the computational domain, affecting the stability of the acoustic solver. These disturbances are more pronounced if turbulence generating conditions are used for the fluid domain and the acoustic domain extends beyond the inlet region of the fluid domain.

\begin{figure}
	\includegraphics[scale=0.6]{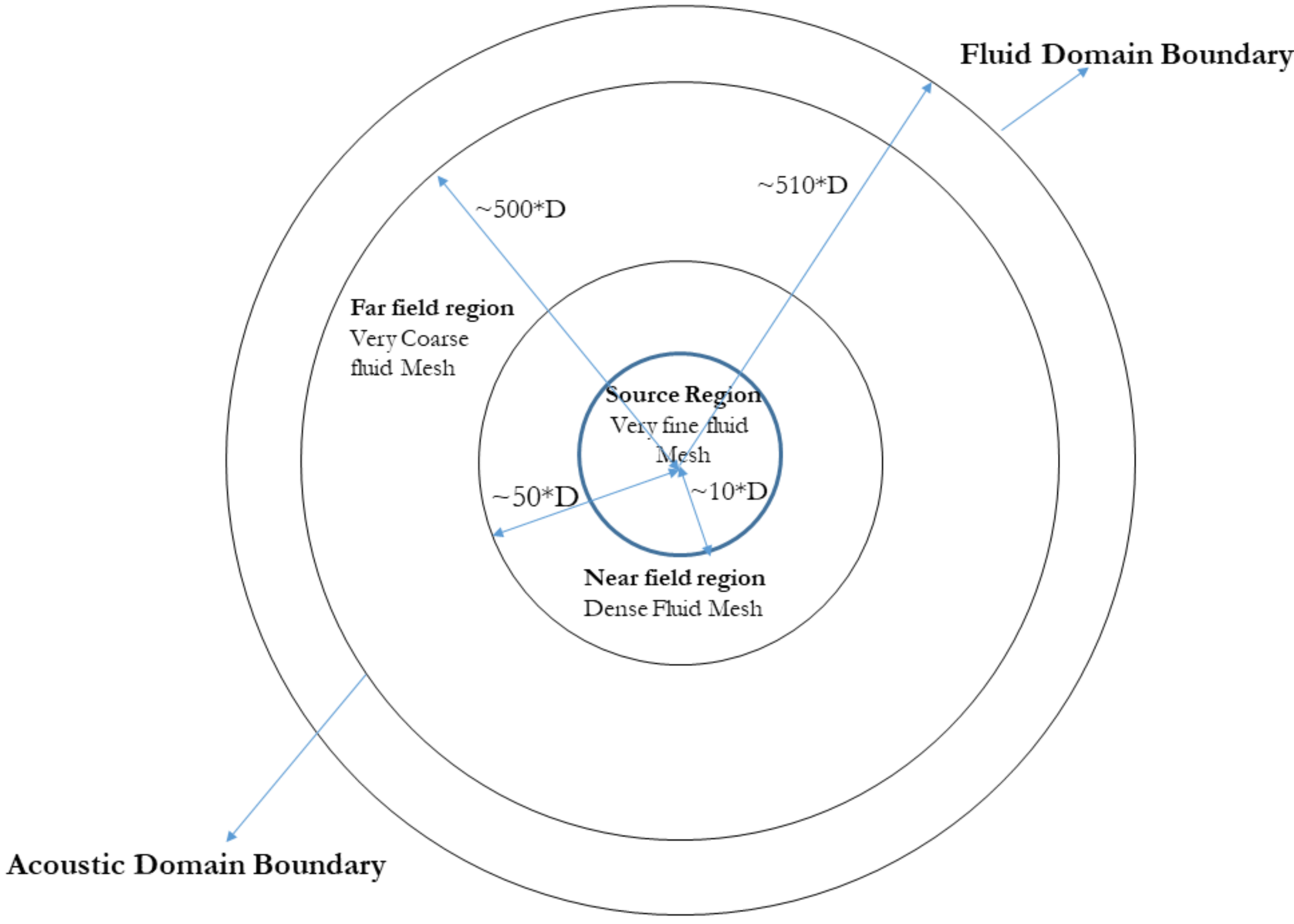}
	\caption{\label{fig:1} Domain comparison for the simulations}
\end{figure}

To circumvent the problems mentioned above, an innovative technique is designed for the present study. The fluid domain is made larger than the acoustic domain so that the data transfer from fluid mesh to acoustic mesh will not involve the transfer of any boundary data (since the points at the boundary do not exist in the acoustic domain), making the acoustic domain free from any disturbances occurring in the fluid domain. And since there are no points in the acoustic domain outside the fluid domain, there is no computational overhead while data transfer. Fig. \ref{fig:1} illustrates the representative boundaries for all the simulations. Since no common boundaries are shared for both the solvers, there is no need for a sponge layer or a damping function at the common boundary. In the later sections, it will be observed that even without applying a damping region, the acoustic pressure values are very close to the DNS results, proving that the hybrid CAA is free from any reflections from the boundaries. The larger domain makes the number of grid points in the fluid domain to increase. Still, again if we remember that the vorticity source terms are concentrated in a small region, we can use a larger number of grid points near the source region and progressively stretch the grid such that the number of grid points is not exorbitantly high. The relatively coarser grid far from the source region also ensures that the source terms, even though small at the far-field regions, are dissipated. The whole methodology is assumes that the source terms are essential only near the region of interest. Therefore, even with the present arrangement, the number of grid points for the fluid domain is relatively less with a smooth, reflection-free hybrid CAA simulation.

\subsection{\label{sec:level2}Geometry and Boundary Conditions}

The present study first validates the coupling for which three benchmark cases are employed. The cases considered are laminar flow around a 2D cylinder (circular and square), turbulent flow past a forward-backward facing step, and turbulent flow over a cavity with a lip. After the coupling strategy validation, we have studied laminar compressible (subsonic and supersonic) flow over a cavity. Fig. \ref{fig:2} shows the geometry for all the cases considered herein. While  Tab. \ref{table:1} tabulates the computational domain for fluid simulations, and Tab. \ref{table:2} presents the computational domain for acoustic simulations for all the cases. The tables highlight the distance of any boundary from the center of the geometry considered. The streamwise direction for all the cases is along the x-axis with wall-normal direction along the y-axis and spanwise direction along the z-axis. For the cavity cases, the origin is at the leading edge of the cavity, while for the cylinder cases, the origin coincides with the center of the cylinder. For fluid simulation, fixed inlet conditions are used for the laminar cases and turbulent inlet with a power-law profile for turbulent cases. In contrast,  wave-transmissive boundary conditions are used at the outlet and far-field with no-slip conditions at the walls and periodic condition in the spanwise plane. For the acoustic simulations, a fully reflecting boundary condition is used at the walls, while a non-reflecting boundary condition is used at the outlet and the far-field. The number of grid points for all the simulations is summarized in Tab. \ref{table:3}. The polynomial order is the order of the interpolating polynomial for the acoustic solver. To maintain the error in resolving all the necessary wavelengths for any simulation below 1 percent, an appropriate filter width is used before the source terms obtained from the fluid solution can be used for the acoustic simulation \cite{moura2015linear}.

\section{\label{sec:level1}Results and Discussion}

\subsection{\label{sec:level2}Validation of the Hybrid Method}

For the validation of the hybrid method, few benchmark cases are chosen corresponding to different flow regimes. These are summarized in the subsequent sections.

\subsubsection{\label{sec:level3}Cylinder in crossflow}

\begin{figure}
	\includegraphics[scale=0.5]{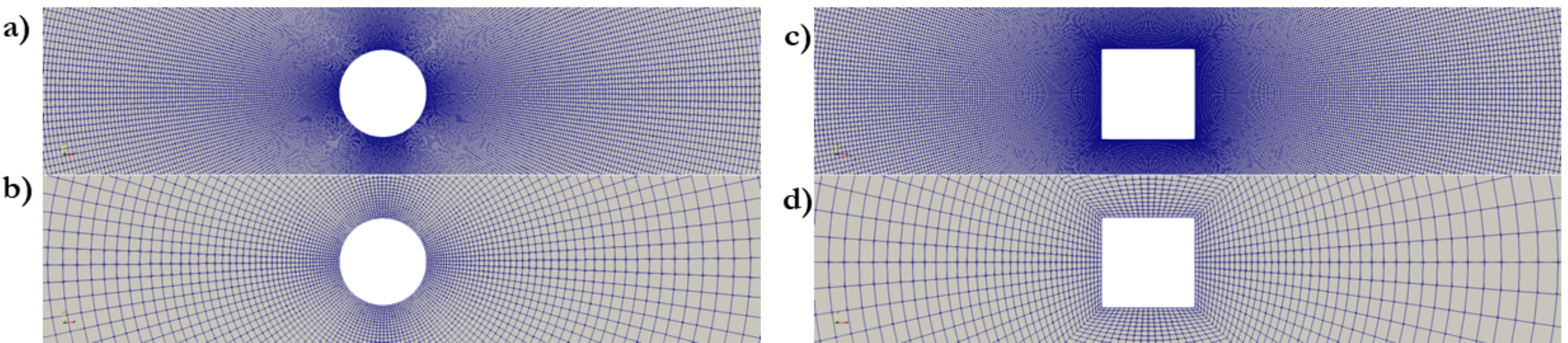}
	\caption{\label{fig:3} a) Circular Cylinder Fluid Mesh , b) Circular Cylinder Acoustic Mesh,  c) Square Cylinder Fluid Mesh , d) Square Cylinder Acoustic Mesh}
\end{figure}

\begin{figure}
	\includegraphics[scale=0.6]{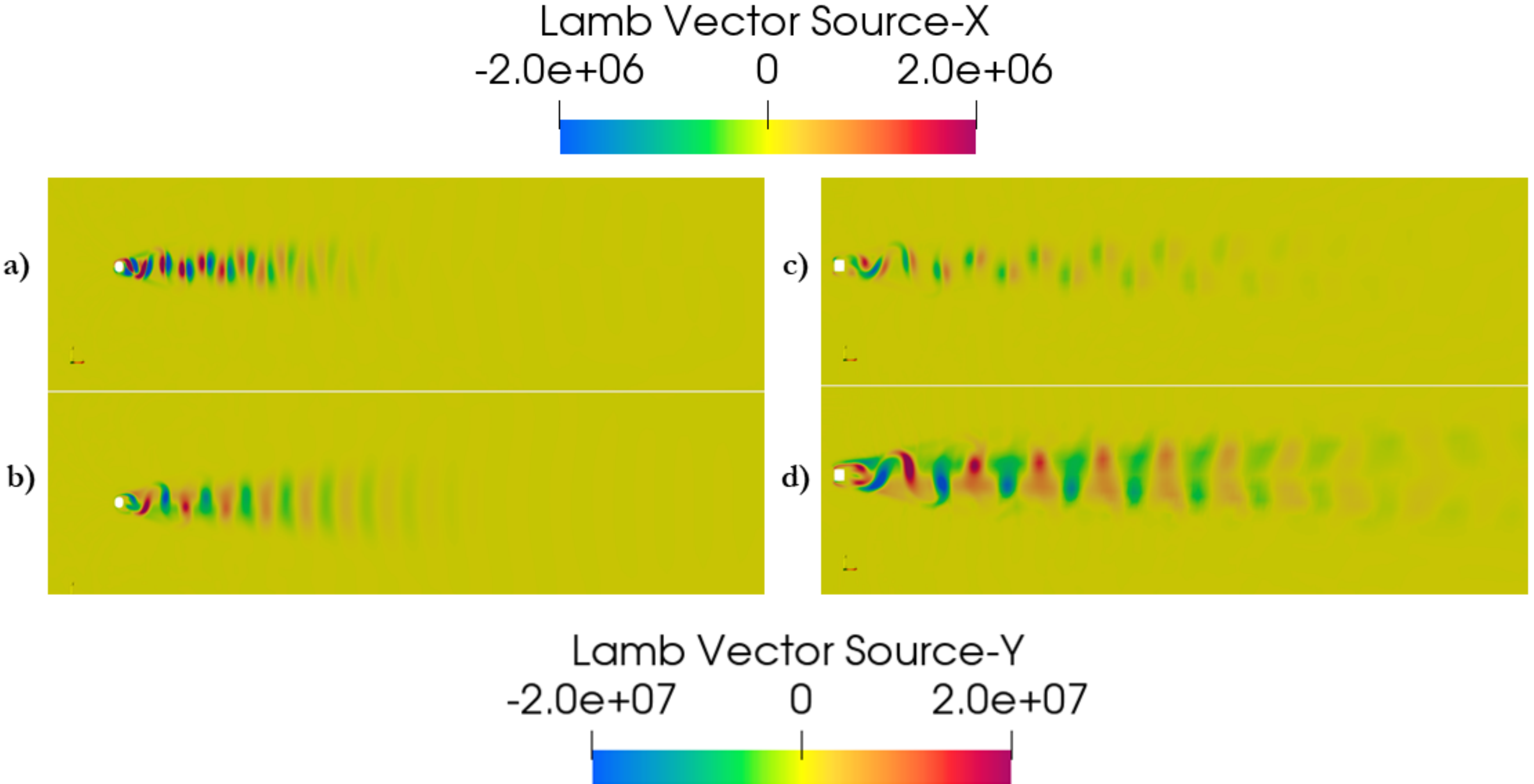}
	\caption{\label{fig:4} Vortex Source terms around the cylinder: a) x-component of  Lamb Vector for Circular Cylinder, b) y-component of Lamb Vector for Circular Cylinder, c) x-component of  Lamb Vector for Square Cylinder, d) y-component of Lamb Vector for Square cylinder}
\end{figure}

\begin{figure}
	\centering
	\includegraphics[scale=0.6]{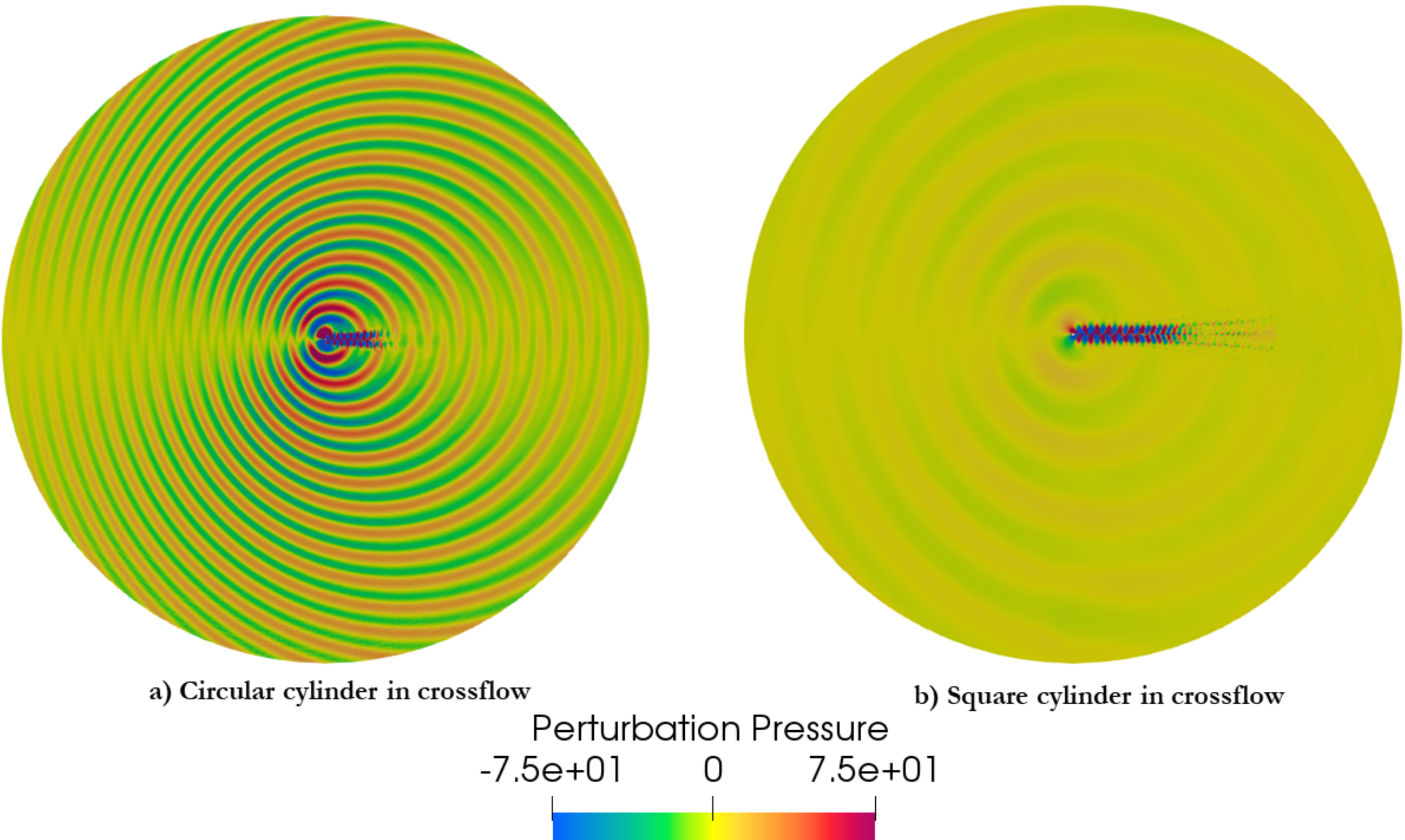}
	\caption{\label{fig:5} Perturbation pressure for cylinder in crossflow: a) Circular cylinder, b) Square Cylinder}
\end{figure}

\begin{figure}
	\centering
	\includegraphics[scale=0.6]{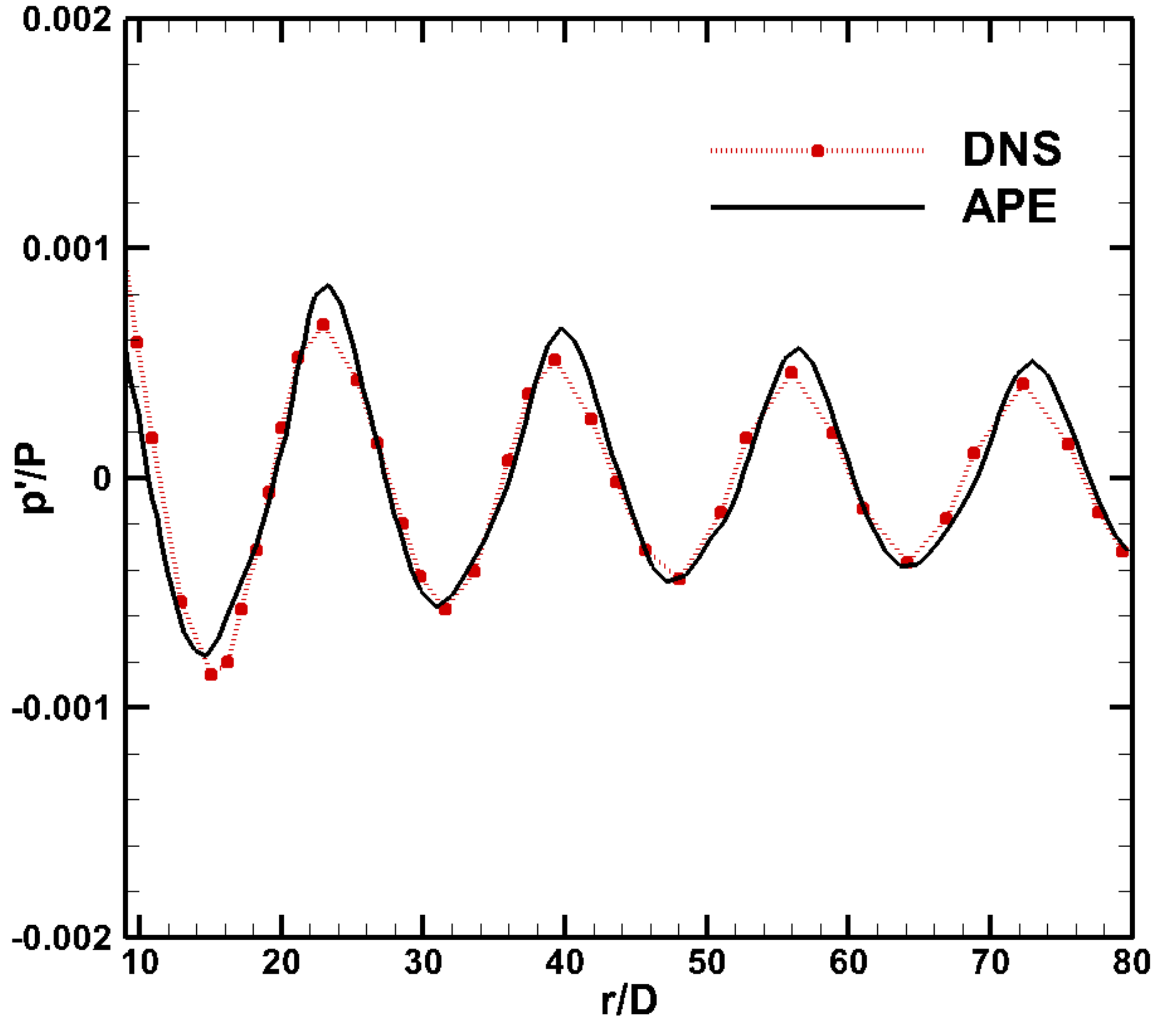}
	\caption{\label{fig:6} Perturbation pressure distribution for circular cylinder along x = 0 compared with DNS \cite{ewert2003acoustic} (reproduced with permission from Journal of Computational Physics 188.2 (2003): 365-398. Copyright 2003 Elsevier Science B.V.) }
\end{figure}

The flow around bluff bodies is known to generate organised structures in the wake due to vortex shedding after a certain critical Reynolds number. The oscillations induced by the vortex shedding generally occur in the direction normal to the crossflow. The maximum amplitude of these oscillations can sometimes be twice the characteristic dimension \cite {bearman1984vortex}. These pressure oscillations are capable of generating acoustic waves and consequently, serve as a perfect test case to validate the hybrid computation of aerodynamic sound. For this, the cases considered are - flow around a circular cylinder (Mach number, $M = 0.3$ and Reynolds number $Re = 200$) and flow around a square cylinder (Mach number, $M = 0.2$ and Reynolds number, $Re = 150$ ) for which Fig. \ref{fig:3} shows the mesh for computations. The used grid fulfills the requirements, as mentioned earlier, to accurately resolve the source terms and, consequently, the generated acoustic waves. We perform these  computations on a 2D domain without invoking any turbulence model. For both cases, the vortex shedding on either side of the cylinder gives rise to fluctuating pressure periodically on the cylinder's surface. This is clear from the contours of the vortex source terms (x-component of Lamb Vector, and y-component of Lamb Vector)from the solver, as depicted in Fig. \ref{fig:4}. The source terms for both cases are slightly dissimilar. This is due to the fact that the difference in geometry results in different separation patterns (and separation points) for both the cases \cite {jiang2020flow}. However, for both cases, the source terms change periodically to give rise to a uniform pattern of acoustic pressure. Fig. \ref{fig:5} reports the contours of pressure perturbation obtained from the simulation. The vortex shedding behind the cylinder results in the differential lift, responsible for the dipole pattern of the acoustic radiation. It is interesting to observe that vortex street in the wake of both the cylinders is also visible in Fig. \ref{fig:5}. This is because the APE system also includes the hydrodynamic pressure fluctuations (but only acoustic pressure fluctuations are convected). Fig. \ref{fig:6} illustrates the perturbation pressure distribution along a line perpendicular to the flow direction (line along which maximum sound radiation occurs) for the circular cylinder while compared with a DNS \cite{ewert2003acoustic}solution. It is clear from the plot that the result obtained from the hybrid approach is in perfect agreement with the DNS result.

\begin{figure}
	\centering
	\includegraphics[scale=0.7]{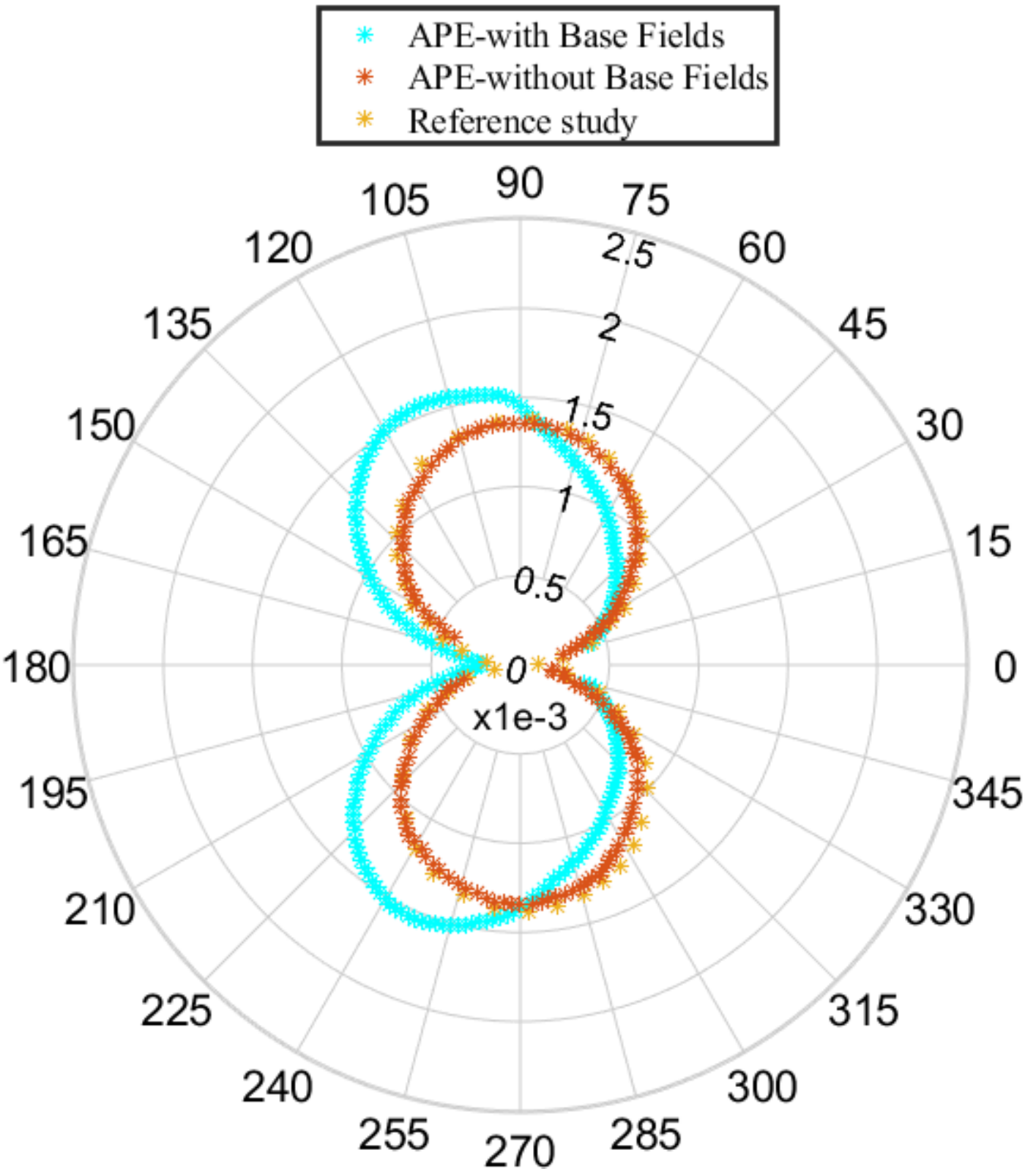}
	\caption{\label{fig:7} Normalised rms perturbation pressure (plotted along radial coordinates) at r/D=80 for Square Cylinder compared with reference study \cite{sukri2013aeolian} (reproduced with permission from AIAA journal 51.2 (2013): 291-301. Copyright 2012 by the American Institute of Aeronautics and Astronautics, Inc.) }
\end{figure}

Though the pressure perturbation plot presented in Fig. \ref{fig:6} provides some confidence in the present hybrid sound computation approach, it does not provide a clear understanding of the effect of the mean flow in this approach. The mean flow for this Mach number does not affect the amplitude of the sound radiation but can affect its directivity. A deeper insight into this notion can be obtained by comparing the results from the present computation from the solution of an acoustic analogy without using any convection effects. Fig. \ref{fig:7} illustrates the root mean square (rms) perturbation pressure obtained from the hybrid simulation (and normalised by the dynamic pressure, $\frac{1}{2}\rho U^2 $) in the far-field region ($r/D=80$) for the square cylinder, compared against a reference study \cite{sukri2013aeolian}. The results obtained by employing APE without the use of base fields are in close agreement with the reference study. It is interesting to observe that for the case employing APE with base fields, the max amplitude of the perturbation pressure remains the same; however, the sound radiation is slightly directed due to the effect of the mean flow. This brings out the advanatage of the present methodology for sound computation.

\subsubsection{\label{sec:level3}Flow over a Forward-Backward Facing Step}

\begin{figure}
	\includegraphics[scale=0.6]{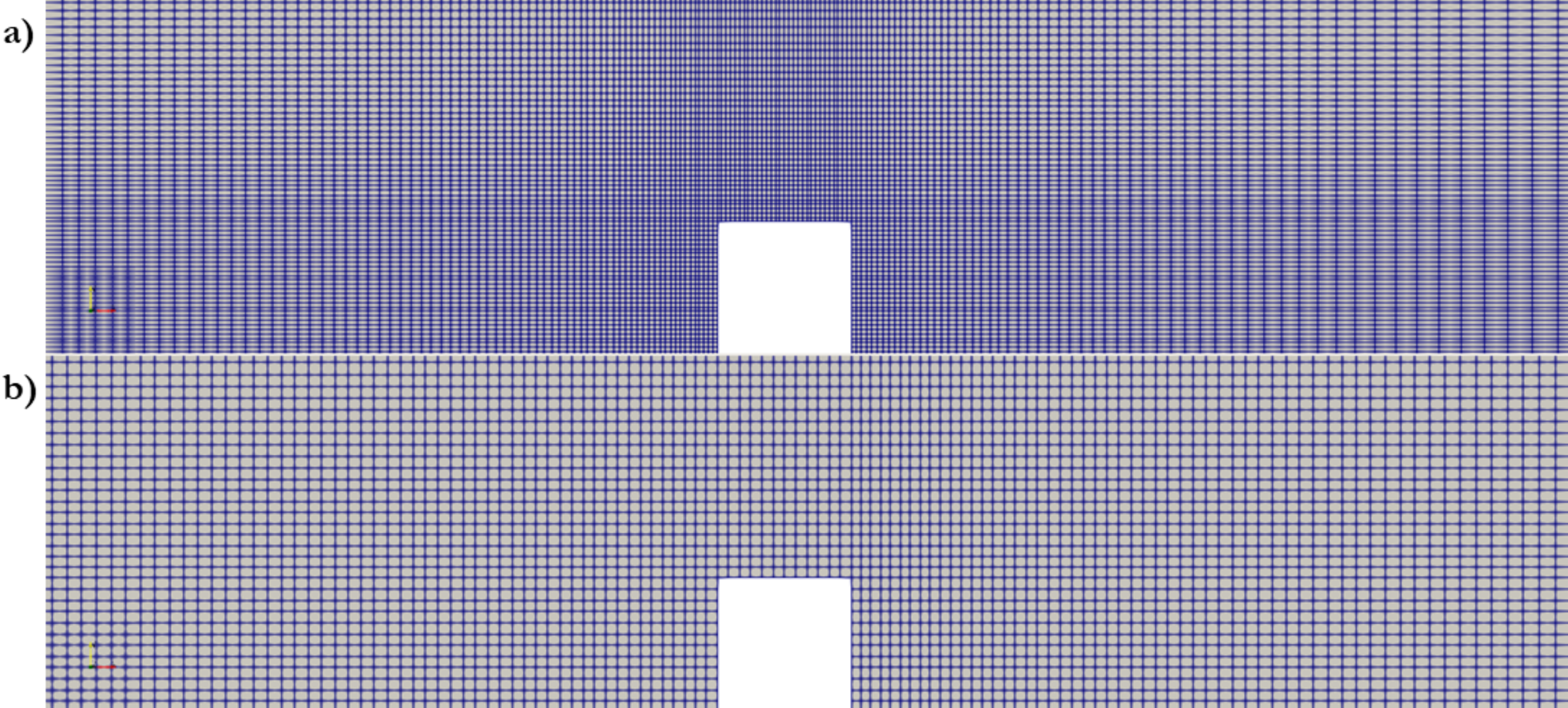}
	\caption{\label{fig:8} a) Fluid Mesh, b) Acoustic Mesh for Forward-Backward Facing Step }
\end{figure}

\begin{figure}
	\includegraphics[scale=0.8]{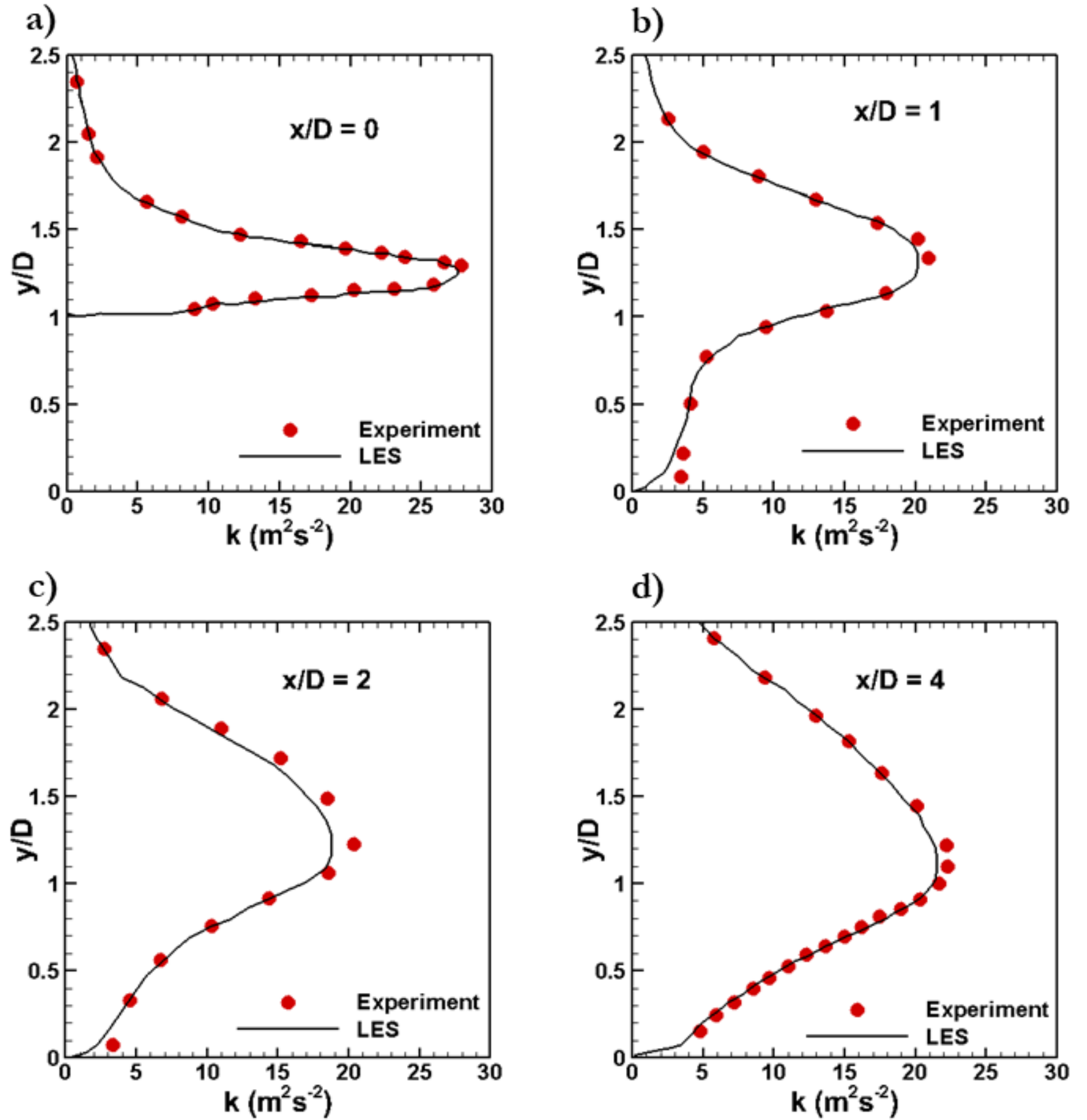}  
	\caption{\label{fig:9}Turbulent Kinetic Energy profiles at different axial locations:a) x/D=0, b) x/D=1, c) x/D=2, d) x/D=4 compared with experiment \cite{springer2016flow}}
\end{figure}

The turbulent flow over a forward-backward facing step is a simple flow configuration that can validate the coupling between the flow solver employing Large Eddy Simulation (LES) and the acoustic solver. For SGS modeling, we have used the Wall Adapting Local Eddy Viscosity (WALE) model \cite {nicoud1999subgrid}. Fig. \ref{fig:8} presents the mesh used in the present computation. The non-dimensional wall distance, $y+$, in the $x$ and $y$ direction is the same and is equal to $10$. In one of the earlier studies, the authors \cite{arya2019effect} have shown that the WALE model provides excellent results for the eddy viscosity profiles near the wall for $y+$ in this range. The incoming flow velocity is $20$ $m/s$. Fig. \ref{fig:9} highlights the quantitative validation of the predicted results obtained from the present LES study compared against the experimental \cite{springer2016flow} profiles of turbulent kinetic energy. As observed, the present LES with the WALE model predicts the turbulent kinetic energy associated with the flow with acceptable accuracy. 

\begin{figure}	
	\includegraphics[scale=0.58]{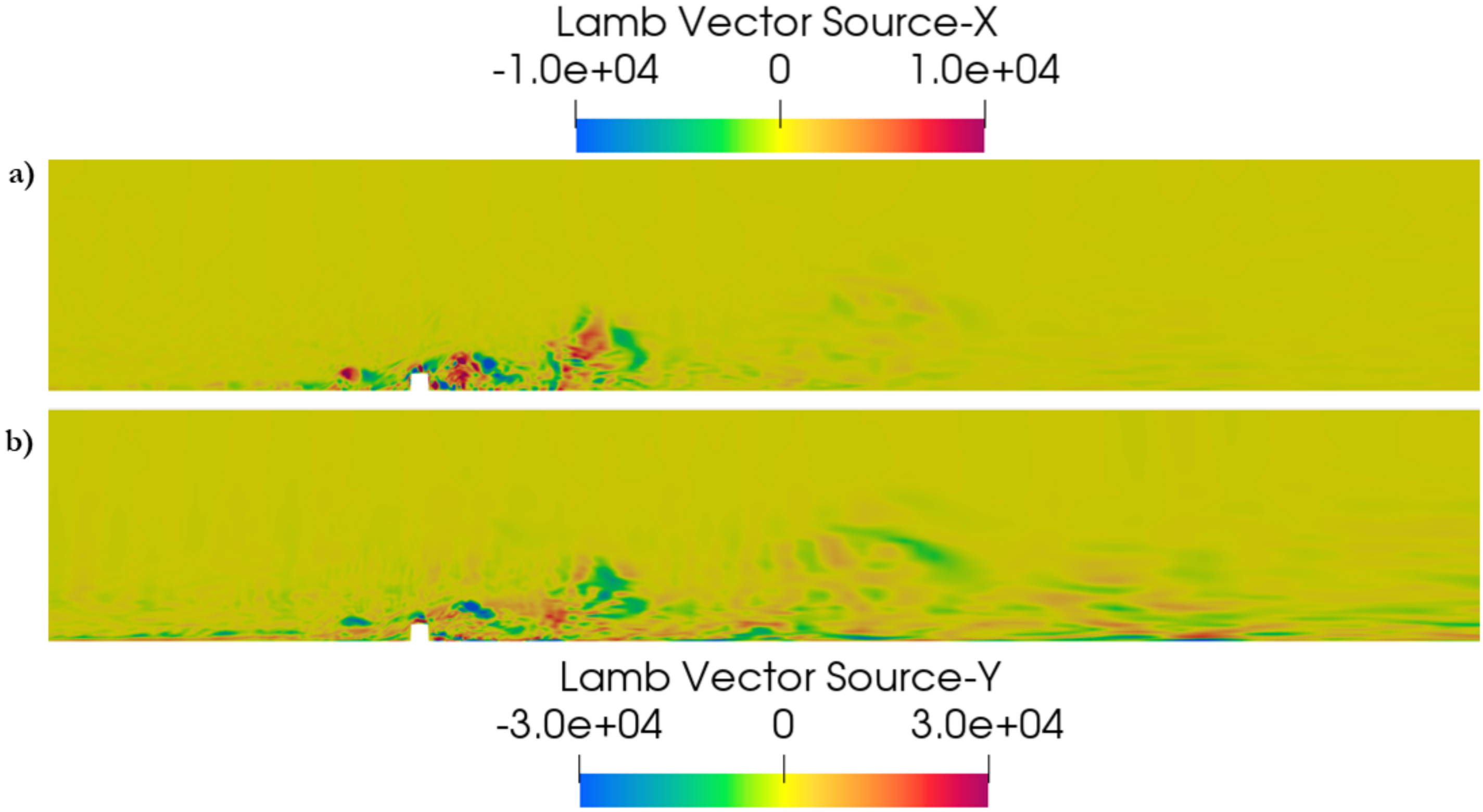}
	\caption{\label{fig:10} Vortex Source terms for Forward-Backward Facing Step: a) x-component of Lamb Vector, b) y-component of Lamb Vector}
\end{figure}

\begin{figure}	
	\includegraphics[scale=0.65]{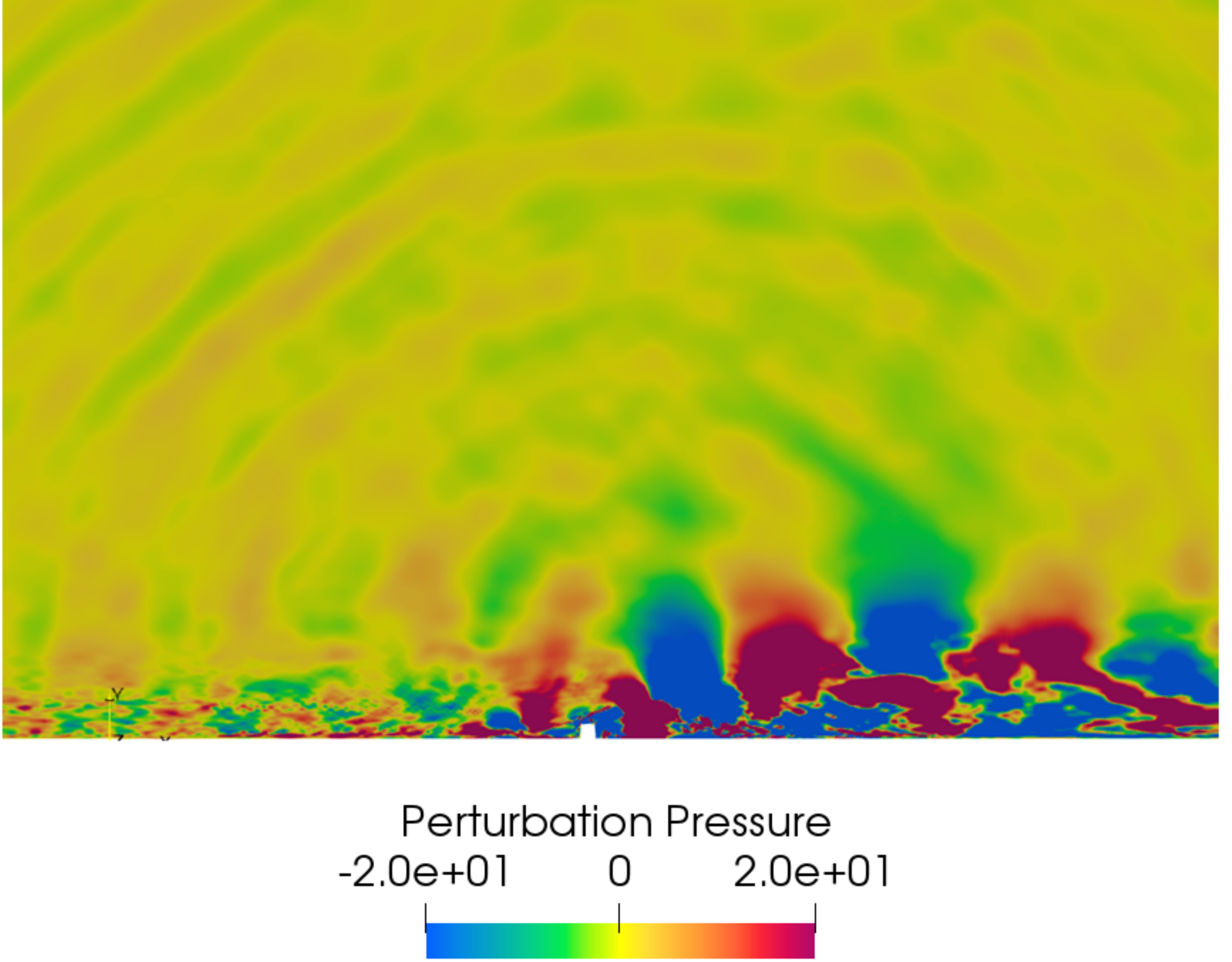}
	\caption{\label{fig:11} Perturbation Pressure Distribution for Forward-Backward Facing Step}
\end{figure}

\begin{figure}	
	\includegraphics[scale=0.6]{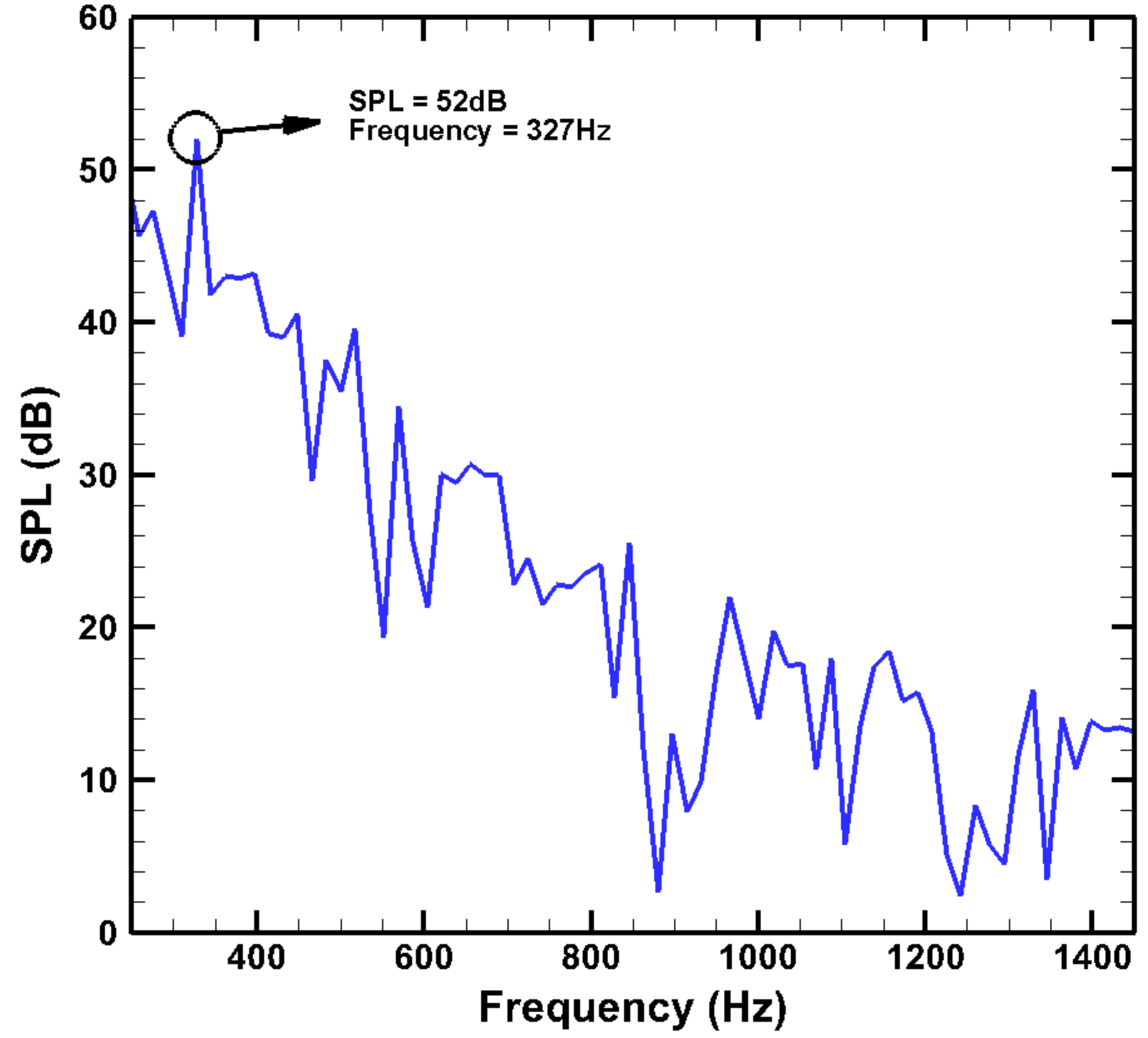}
	\caption{\label{fig:12} SPL at 1m above the step (x/D=0,y/D=50)}
\end{figure}

The source terms obtained from the LES are transferred to the acoustic solver to observe the propagation of the sound. The source terms from the 3D LES are interpolated into the $z=0$ plane for the present case. Fig. \ref{fig:10} describes the Lamb Vector source. It can be observed that the acoustic sources for this flow are distributed randomly near the obstacle. This hints that the acoustic field would consist of broadband noise. Fig. \ref{fig:11} depicts the acoustic pressure in the domain. Notably, the acoustic field consists of some tonal frequencies whose signatures are visible at the far-field in the form of broken circular waves. The domain is also filled with broadband noise. The dominant frequencies can be obtained by an FFT analysis of the pressure signal obtained from the acoustic solver. The perturbation pressure is monitored at a point that is $1$ $m$ above the obstacle. Fig. \ref{fig:12} reports the Sound Pressure Level (SPL) $1$ $m$ above the obstacle obtained by the above analysis. The value obtained is $52$ $dB$ at a frequency of $327$ $Hz$, which is fairly close to the experimental value, $51.4$ $dB$.

\subsubsection{\label{sec:level3}Flow over a cavity with a lip}

The flow over a cavity is relatively simple yet involves very complex flow physics, making it an ideal case for validating the numerical codes. The large-scale pressure fluctuations inside the cavity generate the acoustic waves. There had been some mechanisms suggested for these pressure fluctuations inside the cavity. 

\paragraph{Captive vortex model:}The first mechanism is called the captive vortex model \cite{krishnamurty1955acoustic}, and the motion of this captive vortex system inside the cavity results in the generation of pressure fluctuations.

\paragraph{Shear layer oscillation:}The second type of source mechanism results from the oscillation of the shear layer. It is caused by the interaction of the shear layer shed by the leading edge of the cavity with the trailing edge. The shear layer folds in the cavity to generate a vortex, which grows in size as it moves downstream, causing the shear layer to move upwards. As the vortex ejects out of the cavity, it causes the shear layer to bend into the cavity. This downward position of the shear layer causes the fluid to enter the cavity, and at this time, another vortex begins to form from the leading edge.

There is another mechanism of the shear layer oscillation. At the trailing edge of the cavity, there is a repeated mass intake and mass ejection. This causes regions of higher and lower pressures across the shear layer. This results in the oscillation of the shear layer. Heller and Bliss \cite{heller1975physical} proposed this mechanism.

\paragraph{Fluid-Acoustic Interaction:} The impingement of the shear layer at the trailing edge of the cavity produces intense sound \cite{arya2015investigation}. This sound wave travels upstream and generates the shear layer oscillations at the leading edge. This oscillatory shear layer again impinges on the trailing edge, and the cycle is repeated. Rossiter \cite{rossiter1964wind} has suggested the feedback mechanism. 

\paragraph{Resonant Modes:} This type of mode is purely acoustic and results when the broadband sound generated in the shear layer excites the acoustic modes of the cavity.

\begin{figure}
	\includegraphics[scale=0.6]{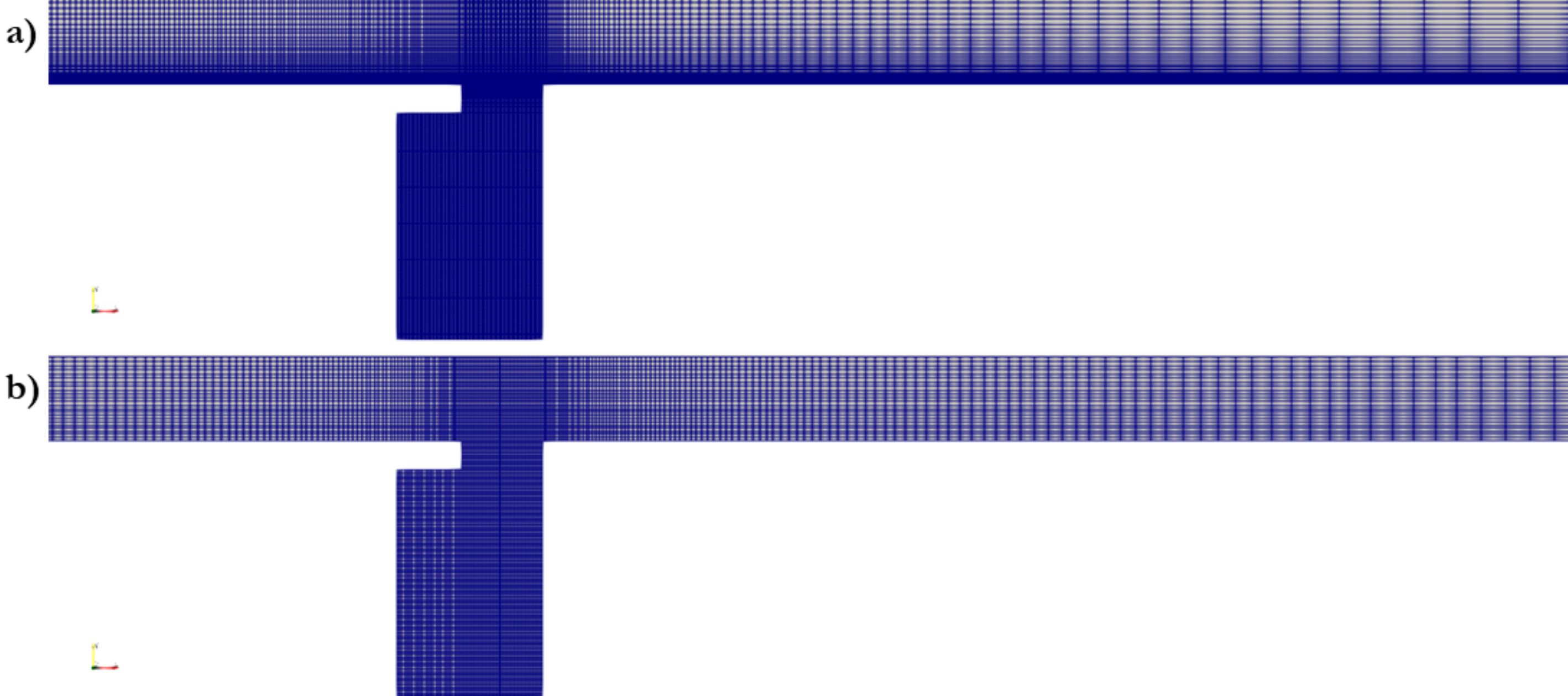}
	\caption{\label{fig:13} a) Fluid Mesh, b) Acoustic Mesh for a cavity with a lip}
\end{figure}

\begin{figure}	
	\includegraphics[scale=0.6]{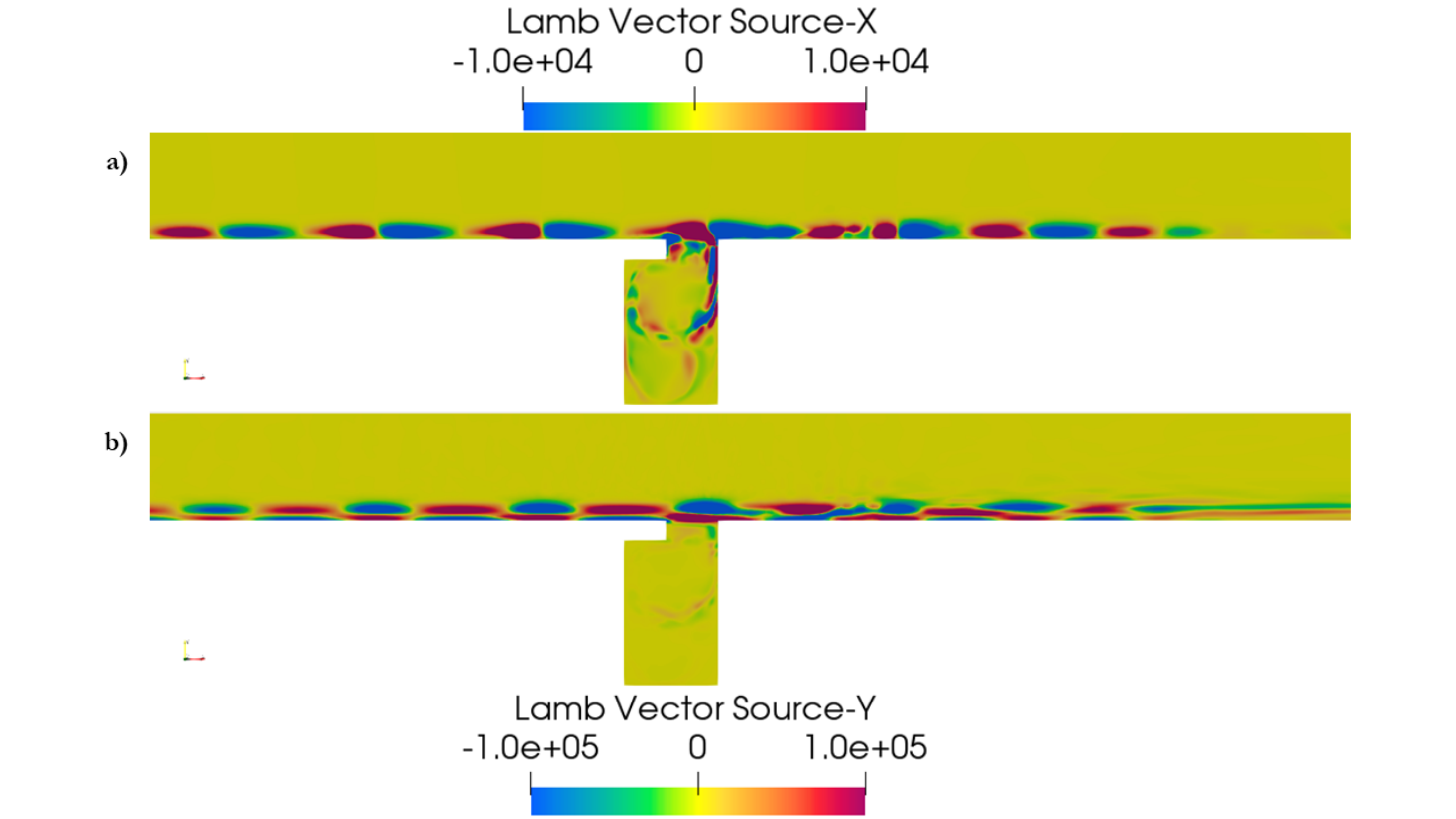}
	\caption{\label{fig:14} Vortex Source terms for a cavity with a lip: a) x-component of Lamb Vector, b) y-component of Lamb Vector }
\end{figure}

A diverse flow physics associated with the sound generation process is witnessed in the cavity, making it an ideal candidate to validate the coupling process. A cavity with a lip is chosen for this purpose \cite{dahl2000third}, for which Fig. \ref{fig:13} illustrates the computational mesh. 
The flow is simulated using an LES with WALE model for SGS modeling. The flow velocity at the inlet is $50$ $m/s$. A turbulent inlet with a power-law profile is prescribed at the inlet boundary. The $y+$ value along the $x$ and $y$ directions is equal to $10$. The boundary layer thickness near the cavity leading edge is $1.25$ $cm$, which is fairly close to the value presented in the reference \cite{dahl2000third}. Fig. \ref{fig:14} shows the source terms for the acoustic simulation. Notably, the source terms are distributed near the shear layer, boundary layer, and inside the cavity.

\begin{figure}	
	\includegraphics[scale=0.6]{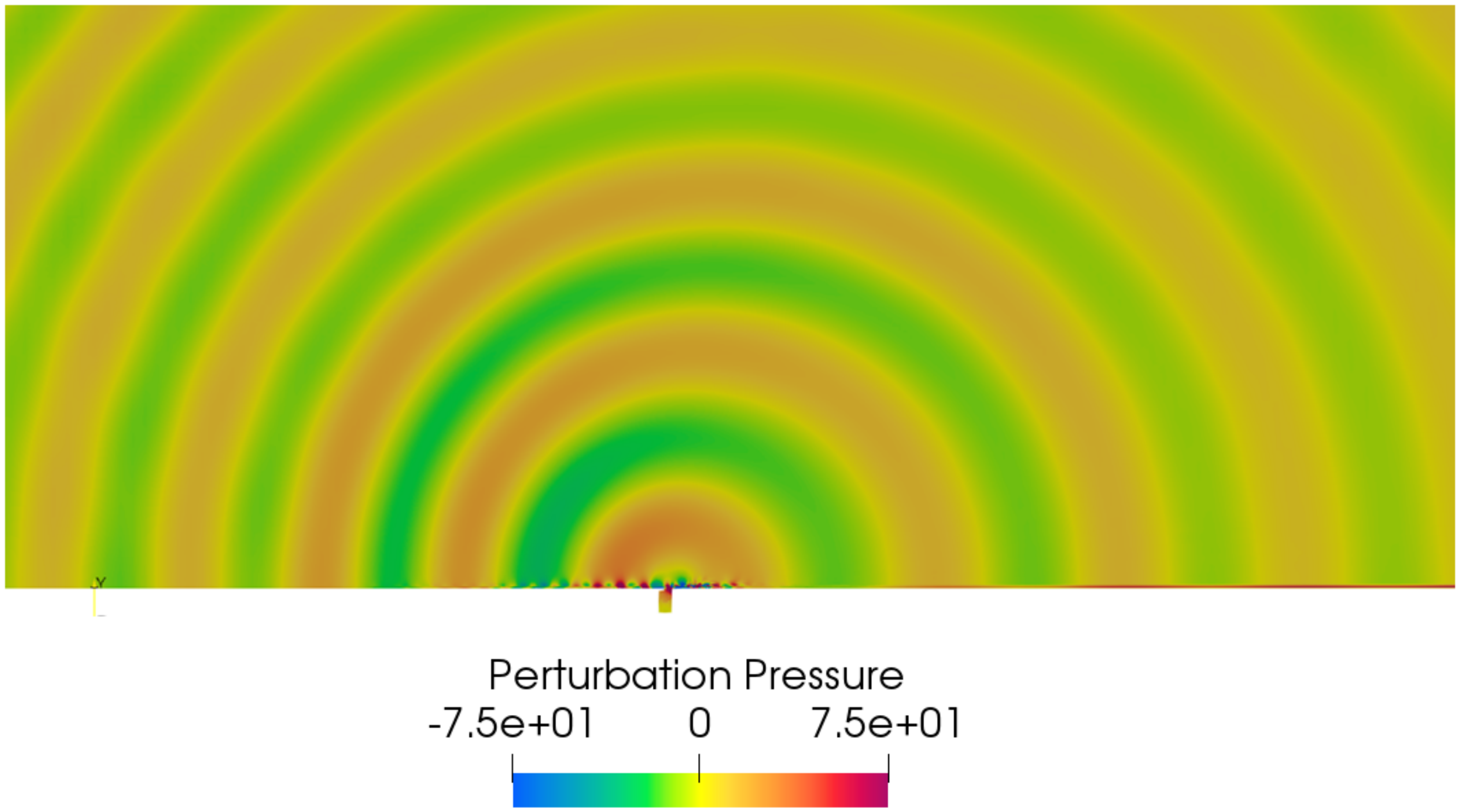}
	\caption{\label{fig:15} Perturbation Pressure for a cavity with a lip }
\end{figure}

\begin{figure}	
	\includegraphics[scale=0.6]{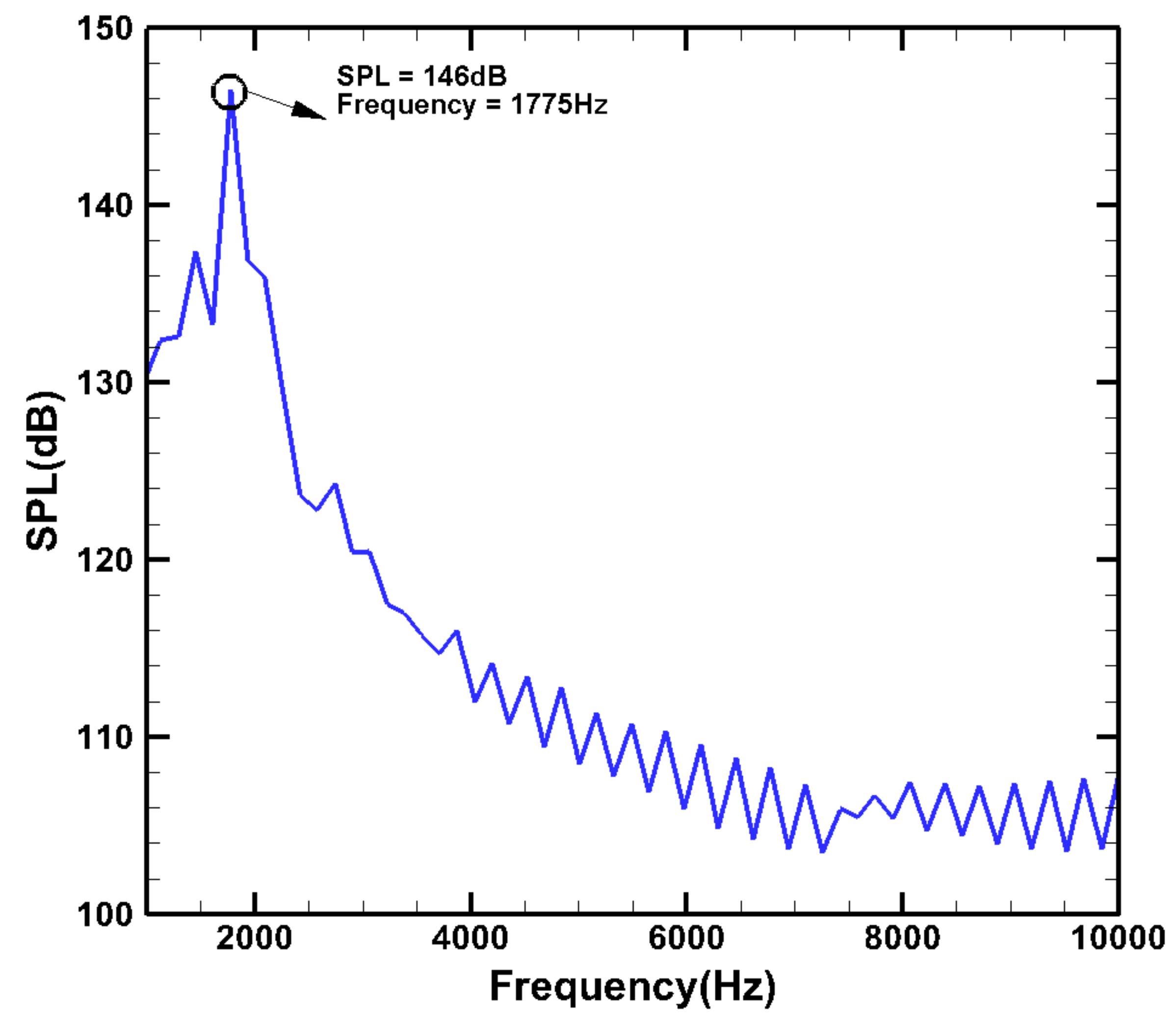}
	\caption{\label{fig:16} Pressure spectrum for a cavity with a lip (x/D=-0.255,y/D=-0.5) }
\end{figure}

Fig. \ref{fig:15} provides the perturbation pressure contours around the cavity. The contours suggest uniform spreading of the single-frequency sound from a directional source. However, the pressure spectrum obtained from a point inside the cavity \cite{dahl2000third} reveals a broadband structure (Fig. \ref{fig:16}) with a peak frequency corresponding to the fluid resonant mode. The frequency at which this peak occurs is $1775$ $Hz$, which is reasonably close to the experimental value \cite{dahl2000third}. The broadband structure is not quite visible in the contours of pressure perturbation due to the smoothening achieved by applying a filter to the received fields, as discussed earlier.


\subsection{\label{sec:level2}Noise Sources in a Rectangular Cavity}

The present section discusses the aeroacoustics analysis of a compressible flow over a cavity. The effect of using the fluctuating entropy source term along with the Lamb Vector source term is also analyzed in the present section.

\subsubsection{\label{sec:level3}Subsonic Cavity}

\begin{figure}	
	\includegraphics[scale=0.6]{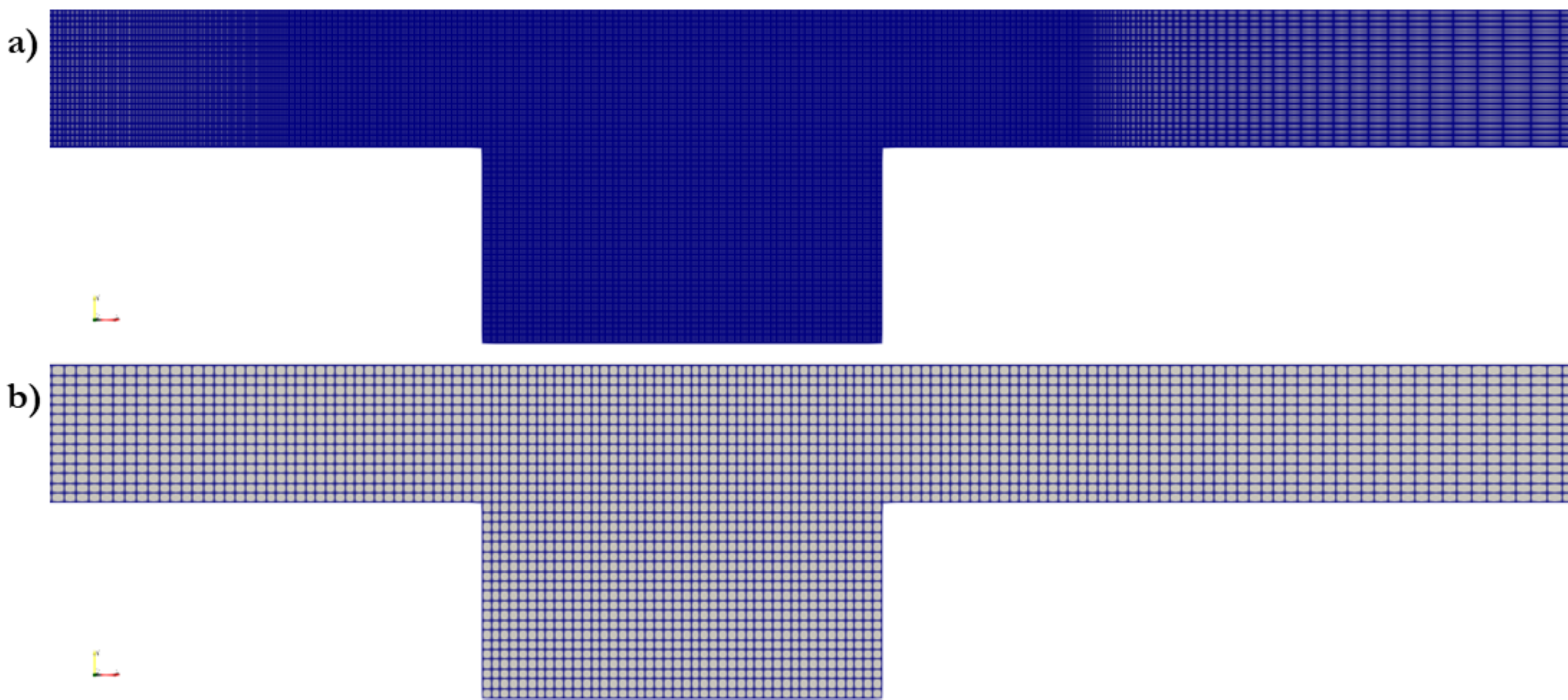}
	\caption{\label{fig:17} a) Fluid Mesh, b) Acoustic mesh for Subsonic Cavity}
\end{figure}

We have simulated four subsonic cavity flow cases corresponding to Mach numbers 0.4, 0.5, 0.6, and 0.7. The mesh has been first generated for cavity flow corresponding to Mach $0.7$ to establish the entropy source term's validity. This same mesh, presented in Fig. \ref{fig:17}, is then used for the other subsonic cases. The flow conditions for all the cases are derived from the experimental work of Krishnamurty \cite{krishnamurty1955acoustic}.

\paragraph{Mach Number 0.7:} The high subsonic Mach number for validating the current methodology is essential in various aspects. The first reason is that at high compressible Mach numbers, the changes in thermodynamic temperature and consequently the entropy fluctuations are significant, and they should, in principle, have some effects on the acoustic characteristics of the flow. Secondly, the frequency of radiation increases with the increase in the Mach number. This means that the associated wavelength becomes comparable to the characteristic dimension of the cavity. Thus, the analysis will also consider when the cavity is no longer compact concerning the acoustic wavelength. Lastly, the mean flow quantities, which are also received from the fluid solver, ensure that the convection effects are considered, which become significant at high Mach numbers. 

\begin{figure}
	
	\begin{subfigure}[]{\textwidth}
		\centering	
		\includegraphics[width=10cm,height=5cm]{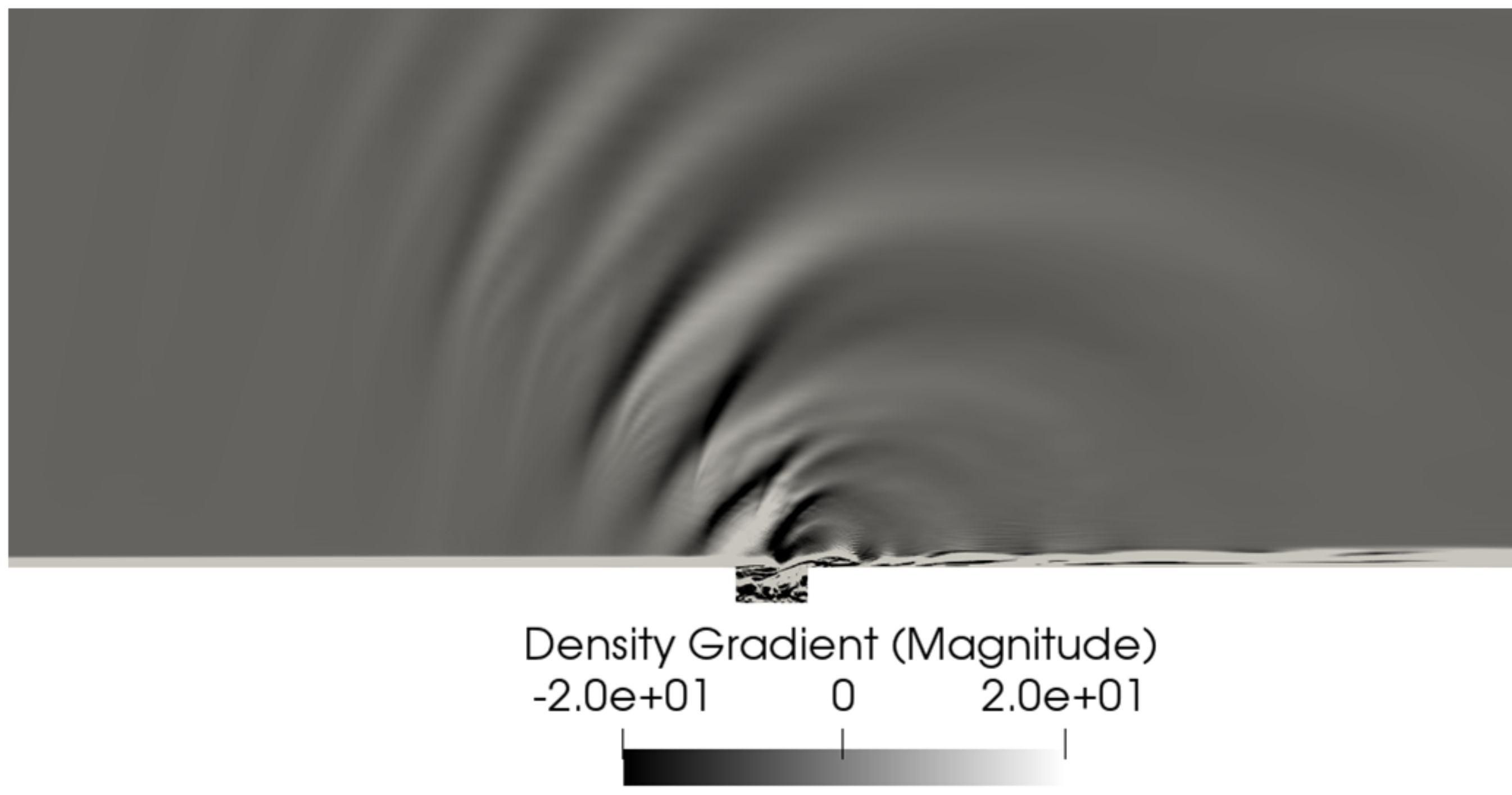}
		\caption{}
		\label{fig:18(a)}
	\end{subfigure}
	
	\begin{subfigure}[]{\textwidth}
		\centering	
		\includegraphics[scale=0.8]{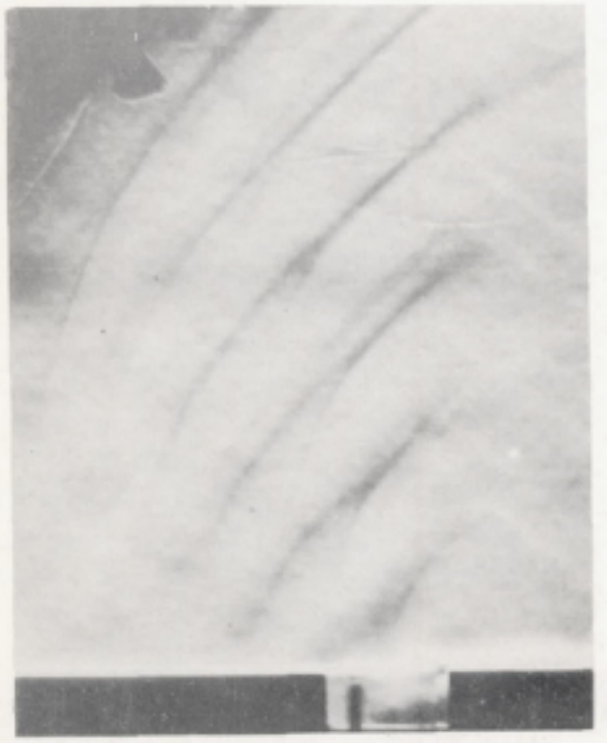}
		\caption{} 
		\label{fig:18(b)}
		\end{subfigure}
		\caption{Direction of Acoustic radiation for M=0.7 cavity a)  Density Gradient along the wall normal direction, b) Acoustic radiation obtained from the experiment \cite{krishnamurty1955acoustic}}
		\label{fig:18}
\end{figure}

In Fig. \ref{fig:18(a)}, the contours of density gradient (obtained from the fluid simulation) along the wall-normal direction are presented. It sheds some light on the nature of the acoustic field associated with the current case. The most striking feature here is the occurence of two wave patterns directed along an angle from the streamwise direction. Such a pattern has been observed in the experiment \cite{krishnamurty1955acoustic} as well which is presented in Fig. \ref{fig:18(b)}. The two wave patterns differ very slightly in terms of their intensities and the angle of propagation, which suggests this pattern is formed due to the direct and reflected sound interaction. The pattern fades away very quickly in the region far from the cavity due to coarser mesh resolution away from the source region. However, these waves are captured efficiently by the acoustic solver, which will be discussed subsequently.

One of the essential features of the flow over a rectangular cavity is the occurrence of self-sustained oscillations. It, therefore, seems tempting to look closely into the generation mechanism of the shear layer oscillations, which would enable us to have a more in-depth insight into the processes that govern the production of acoustic waves. For this, the evolution of vortices over the cavity shear layer during cavity oscillations are studied  with the help of contours of vorticity distribution for one period of self-sustained oscillations, presented in Fig. \ref{fig:19}. The corresponding contours of pressure perturbations at the same time instance are also reported to identify the local low pressure regions, representing vortex cores. In Fig. \ref{fig:19(a)}(i), the vorticity contours clearly depicts the shear layer's rolling resulting in the formation of a vortex. This fact is corroborated by the contours of the fluctuating pressure (Fig. \ref{fig:19(a)}(ii)) which depicts a local low pressure region near the leading edge. The cavity is an area of recirculating flow with a larger recirculation region near the cavity trailing edge. The formation of the vortex at the leading edge is the first step of the feedback process. This vortex then travels downstream along the shear layer and grows in size. This process is observed in Fig. \ref{fig:19(b)} and Fig. \ref{fig:19(c)} . The initial vortex at the leading edge has grown in size and convected near the middle of the shear layer. The low pressure region has also grown at this point which is evident from the contours of perturbation pressure from Fig \ref{fig:19(c)}(ii). This vortex is about to impinge at the trailing edge just when the formation of a new vortex at the leading edge occurs, which is visible in Fig. \ref{fig:19(d)}. The shear layer at the trailing edge at this time is bent downwards due to the ejection of a previous vortex. This is accompanied by mass injection at the trailing edge forming a local high pressure region.The vortex impingement at the trailing edge is visible in Fig. \ref{fig:19(e)}(i). It is also accompanied by the formation of a local low pressure region near the trailing edge. The vortex, after impingement, will be shed out of the cavity, thereby generating a low pressure region again. This is consistent with the mechanism of Heller and Bliss \cite{heller1975physical}, where the shear layer oscillations result from the alternating high and low pressure regions across the shear layer. The mass addition process occurs during the impingement process, where a part of the impinging vortex recirculates inside the cavity while the other part is convected downstream. The recirculating fluid inside the cavity pushes the shear layer up due to a higher pressure inside, thereby ejecting the mass out from the trailing edge. Both the mechanisms of the shear layer oscillations, which are responsible for producing the acoustic waves over the cavity, can be seen at play simultaneously. The formation of acoustic waves due to this process is discussed in the subsequent sections.  

\begin{figure}	
	\includegraphics[scale=0.4]{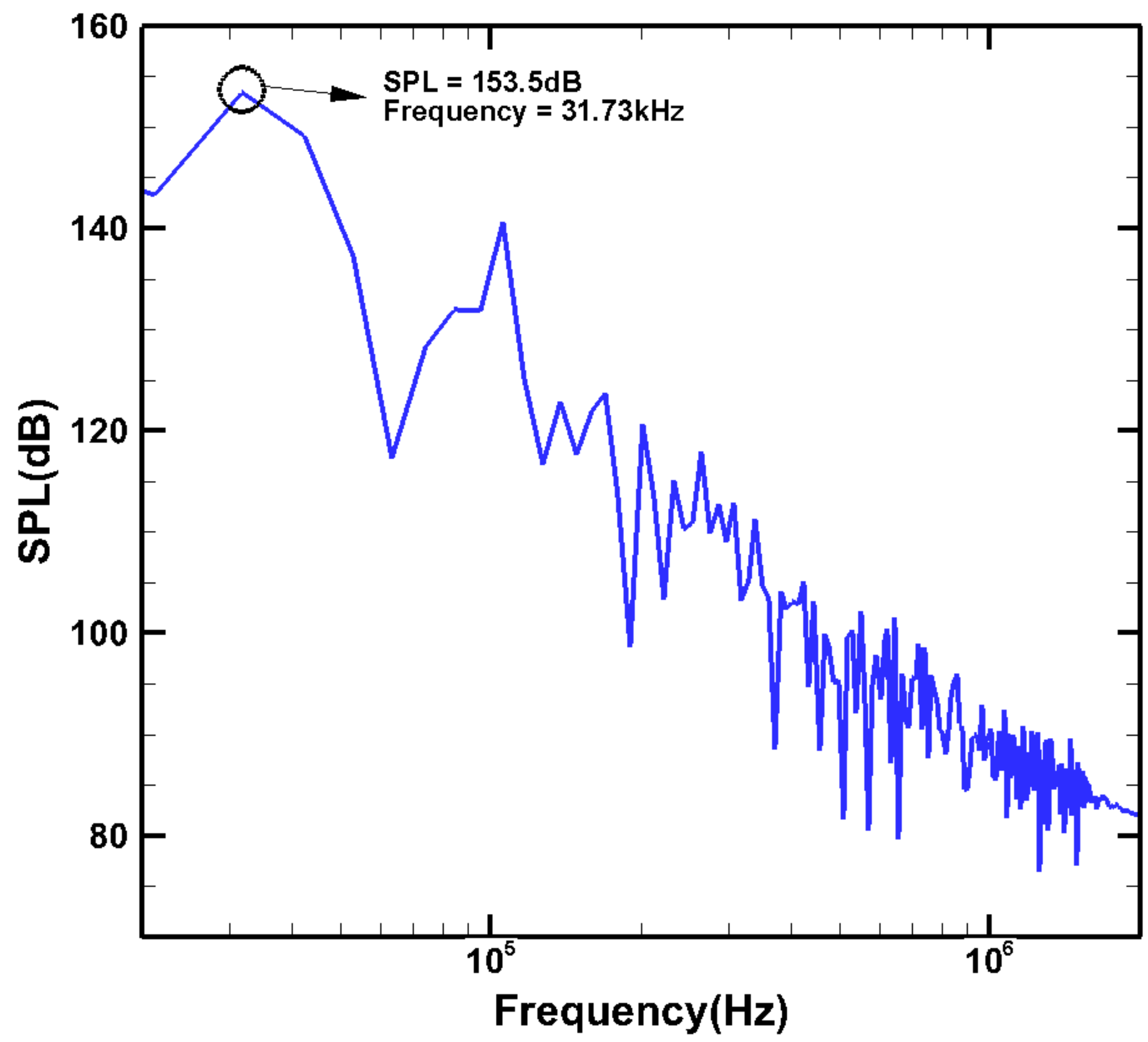}
	\caption{\label{fig:20} SPL for cavity at M = 0.7 (x/D=-0.04,y/D=2) }
\end{figure}

\begin{figure}	
	\includegraphics[scale=0.4]{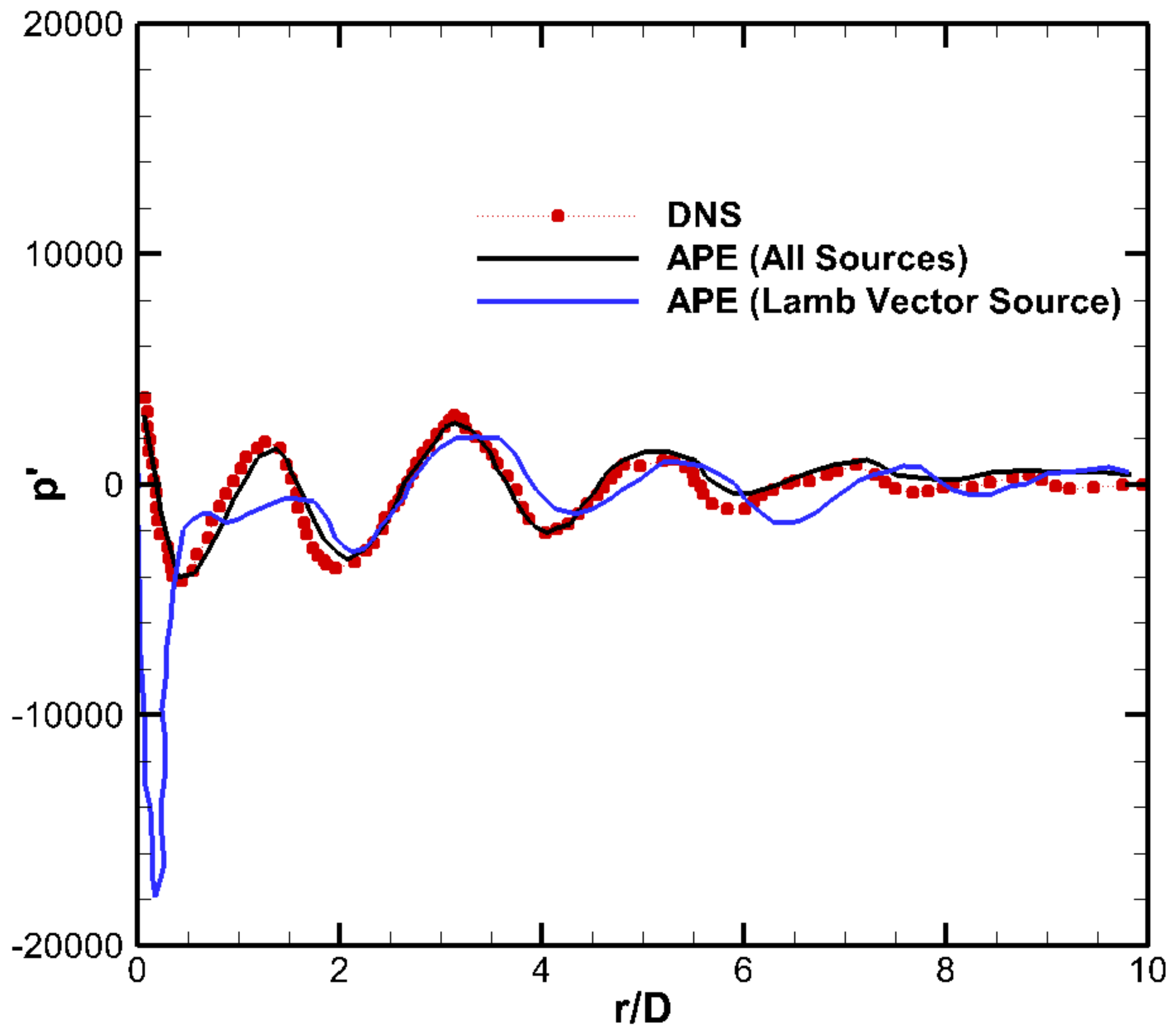}
	\caption{\label{fig:21} Perturbation pressure along the line $x+y=2D$; $r=\sqrt{x^2 + y^2}$ compared with DNS \cite{gloerfelt2003direct} (reproduced with permission from Journal of sound and vibration 266.1 (2003): 119-146. Copyright 2002 Elsevier Ltd.)}
\end{figure}

Fig. \ref{fig:20} reports the SPL obtained at a reference point from the acoustic simulation for the present case employing all the source terms (Lamb Vector and entropy sources) illustrated herein. The SPL $153.5$ $dB$ and the peak frequency at $31.73$ $kHz$ corresponding to Strouhal Number $0.67$ based on the cavity length are fairly close to the experiment's values presented in the experiment \cite{krishnamurty1955acoustic} (Strouhal Number $0.69$) and the reference DNS \cite{gloerfelt2003direct}. The pressure perturbation obtained from the acoustic simulation involving just the Lamb Vector source and another simulation involving all the source terms is compared against the DNS values \cite{gloerfelt2003direct} and is presented in Fig. \ref{fig:21}. One can notice that the perturbation pressure values obtained from employing all the sources are in reasonable agreement with the DNS results, which, in turn, confirms the impact of entropy source in compressible regime. The values obtained by employing just the Lamb Vector source terms display sharp variation near the cavity source region. This region corresponds to the region where the shear layer impinges on the cavity trailing edge and is generally associated with high vorticity. The large pressure perturbation predicted by the application of Lamb Vector sources gets canceled by the entropy sources. Cancellations of different source terms have also been observed earlier in the case of Lighthill Tensor for jet aeroacoustics \cite{lew2007investigation}. Thus, it is clear from the presented results that fluctuating entropy also plays a crucial role in generating acoustic waves in a compressible cavity flow.

\begin{figure}
	\includegraphics[scale=0.6]{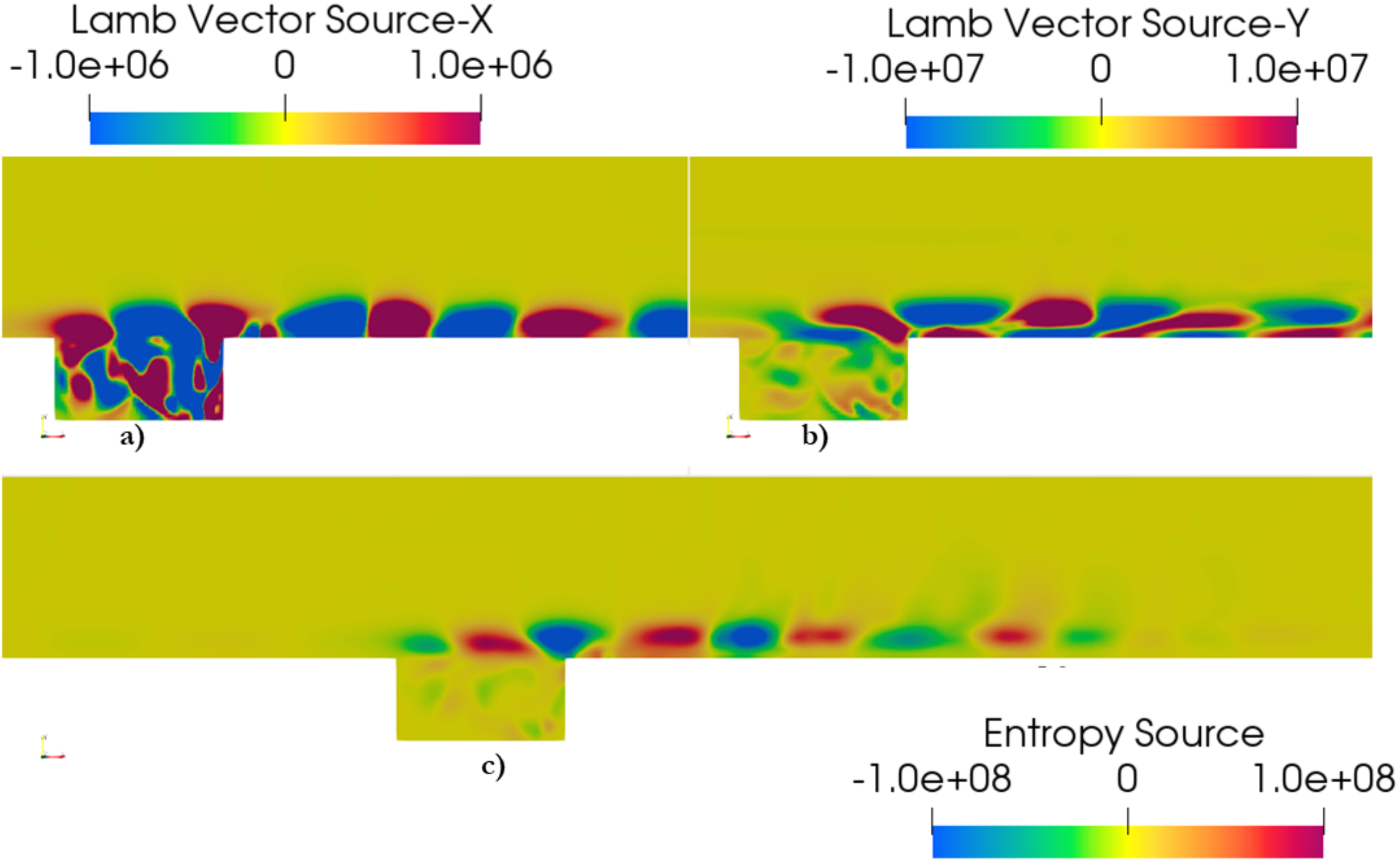}
	\caption{\label{fig:22} Source terms for M = 0.7; a) x-component of Lamb Vector, b) y-component of Lamb Vector, c) Entropy Source term}
\end{figure}

\begin{figure}
	\includegraphics[scale=0.6]{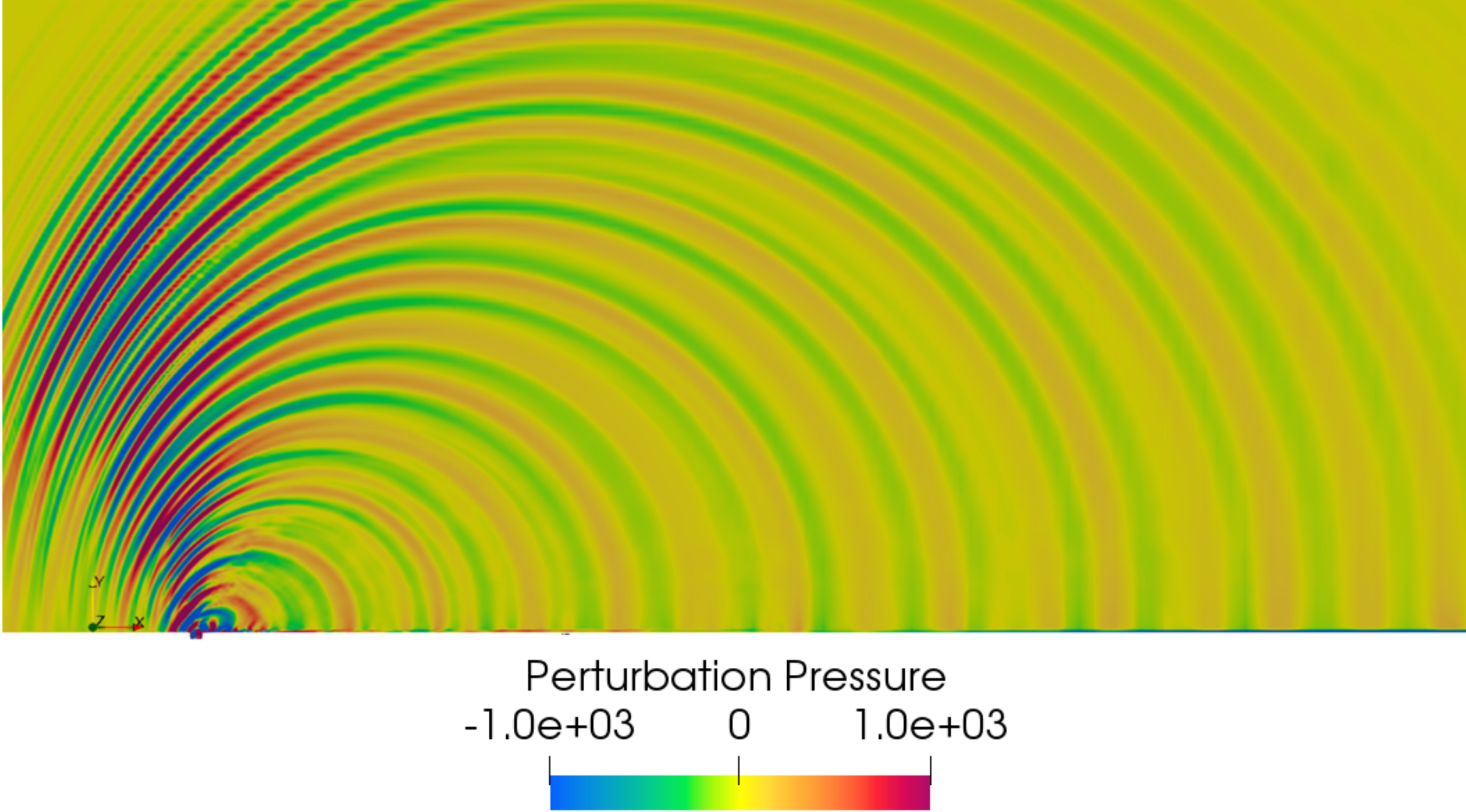}
	\caption{\label{fig:23} Perturbation pressure contours for M = 0.7}
\end{figure}

Fig. \ref{fig:22} shows the source terms in which the entropy source term is also included. As expected, all the source terms are distributed along the shear layer and inside the cavity. Fig. \ref{fig:23} depicts the pressure perturbations generated due to the sources. The observation suggests that the resulting acoustic field is highly directional, with the majority of radiation along an upstream direction. The convection effects are distinctly visible in the contours with the upstream traveling waves being slightly compressed due to incoming flow.

\begin{figure}	
	\includegraphics[scale=0.6]{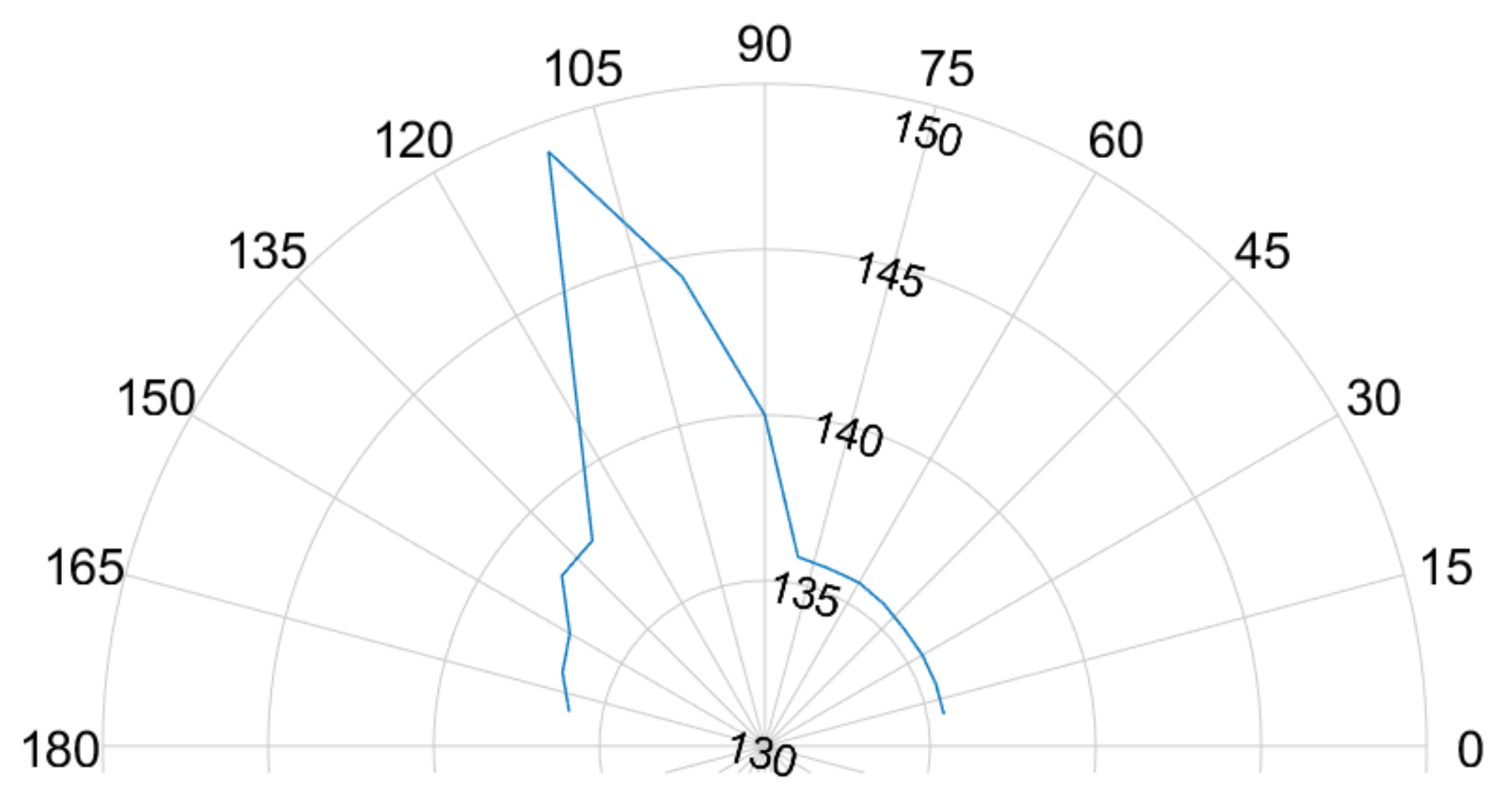}
	\caption{\label{fig:24} Farfield (r=100D) Sound directivity for M = 0.7, $f=31.73$ $kHz$, with $SPL(dB)$}
\end{figure}

The far-field acoustic directivity along a semi-circle of radius $100D$ and center at the cavity leading edge is presented in Fig. \ref{fig:24}. There is a sharp peak at an azimuthal angle of $120$ degrees from the downstream direction. The reference DNS \cite{gloerfelt2003direct} and experiment \cite{krishnamurty1955acoustic} also observed a similar value of this peak directivity.

\begin{figure}
	\includegraphics[scale=0.6]{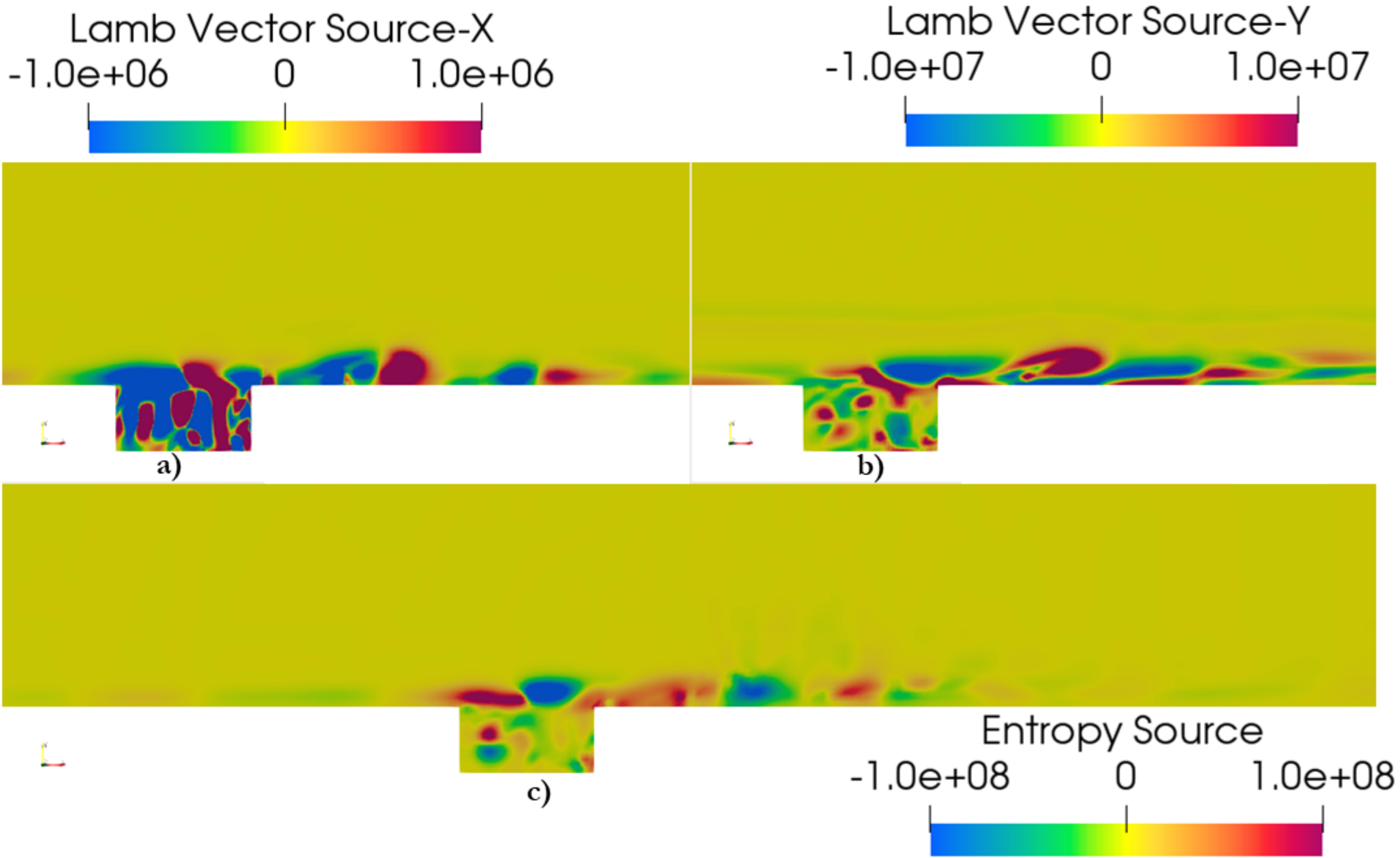}
	\caption{\label{fig:25} Source terms for M = 0.6; a) x-component of Lamb Vector, b) y-component of Lamb Vector, c) Entropy Source term}
\end{figure}

\begin{figure}	
	\includegraphics[scale=0.6]{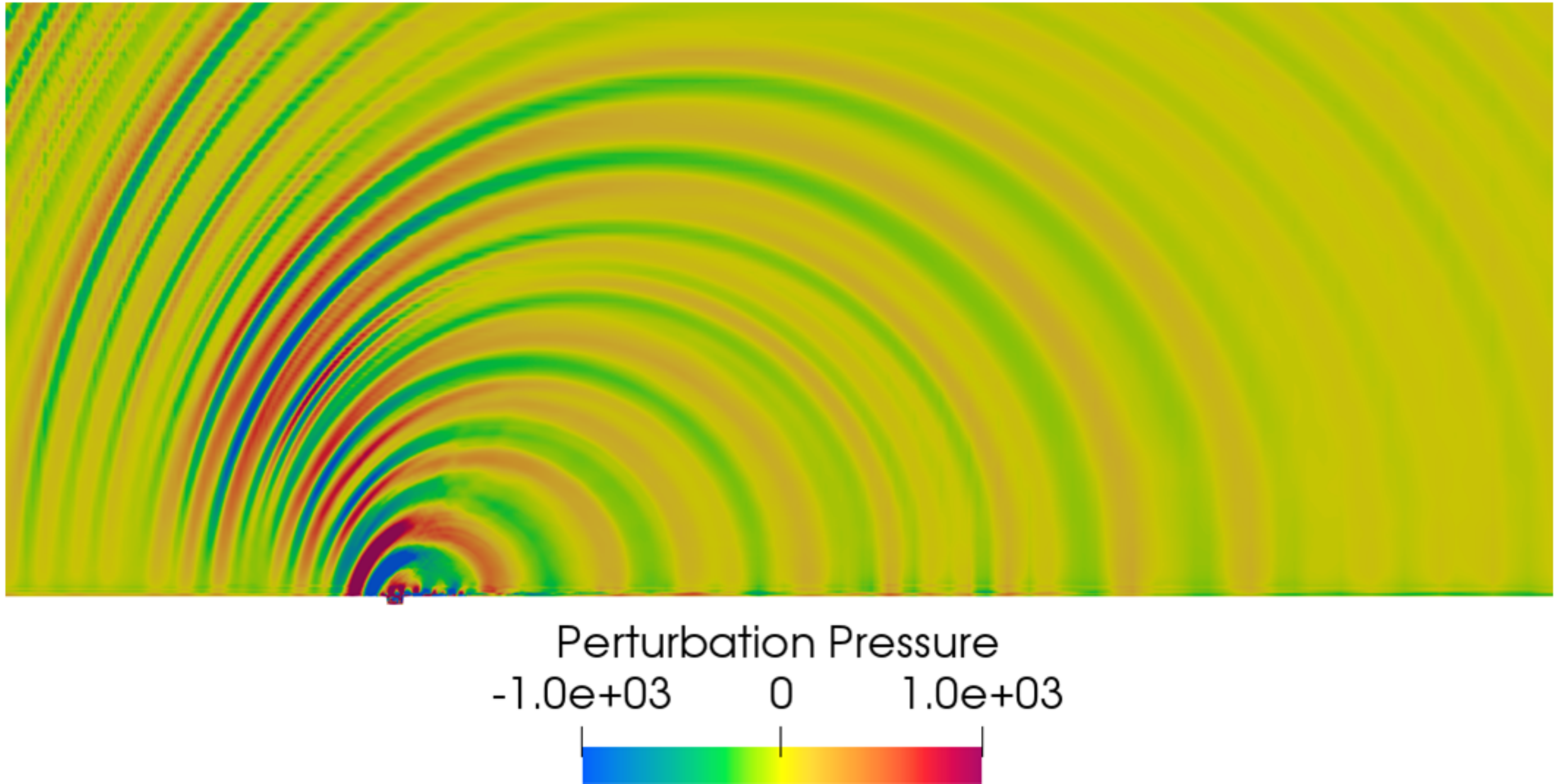}
	\caption{\label{fig:26} Perturbation pressure contours for M = 0.6}
\end{figure}

\begin{figure}	
	\includegraphics[scale=0.6]{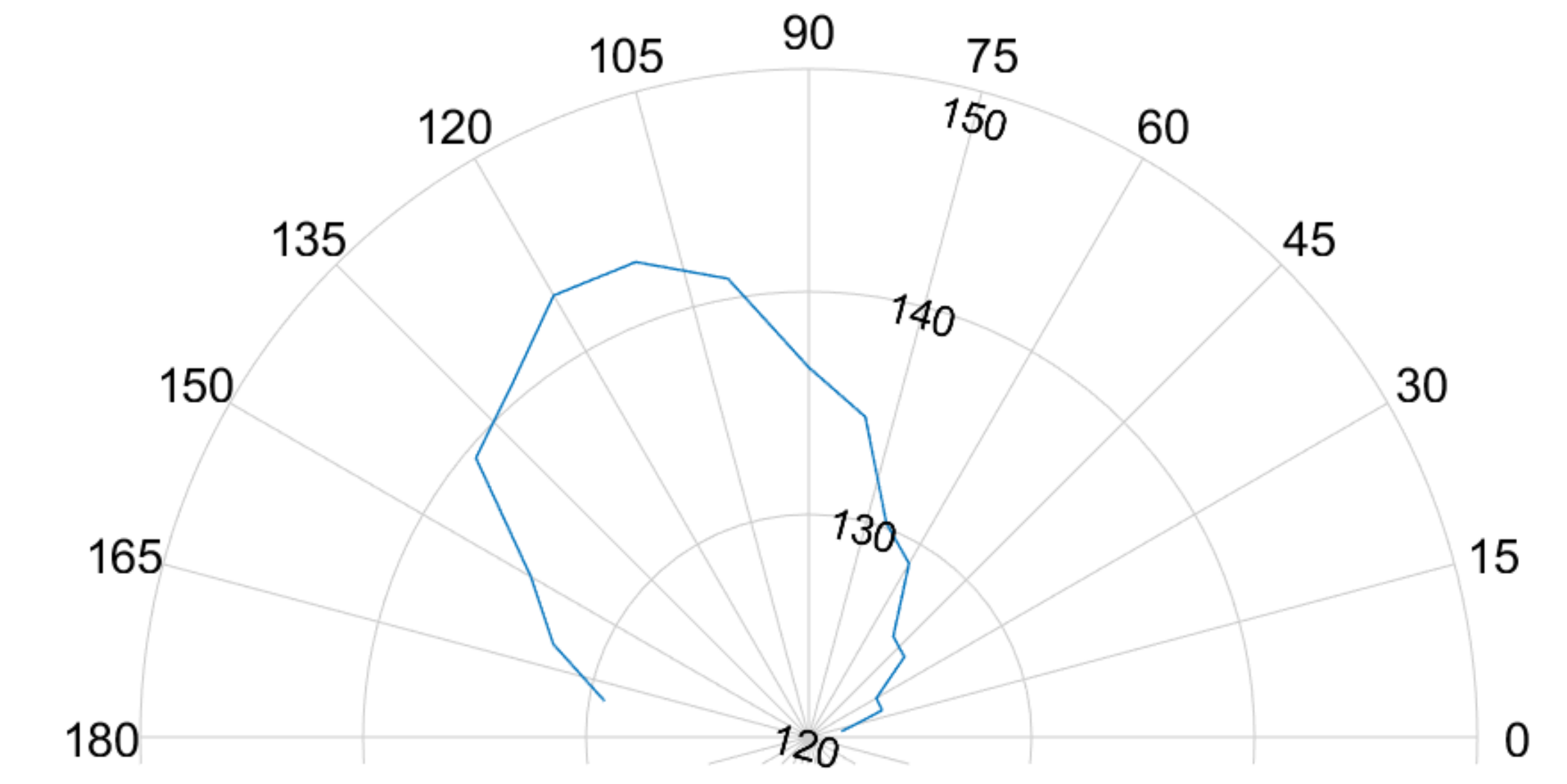}
	\caption{\label{fig:27} Farfield (r=100D) Sound directivity for M = 0.6, $f=28.1$ $kHz$, with $SPL(dB)$}
\end{figure}

\paragraph{Mach Number 0.6:}Fig. \ref{fig:25} presents the source terms for the cavity flow at Mach $0.6$, and Fig. \ref{fig:26} shows the contours of pressure perturbation. As observed, though the sound radiation is directed at an angle, the compression in the upstream waves is not as intense as for Mach $0.7$, as illustrated in the directivity plot in Fig. \ref{fig:27}. Notably, the peak directivity occurs nearly at the same angle as for Mach $0.7$. However, the peak is not as sharp as before. The curve obtained around the peak is smoother, which means that the SPL values are similar around the peak.

\begin{figure}	
	\includegraphics[scale=0.6]{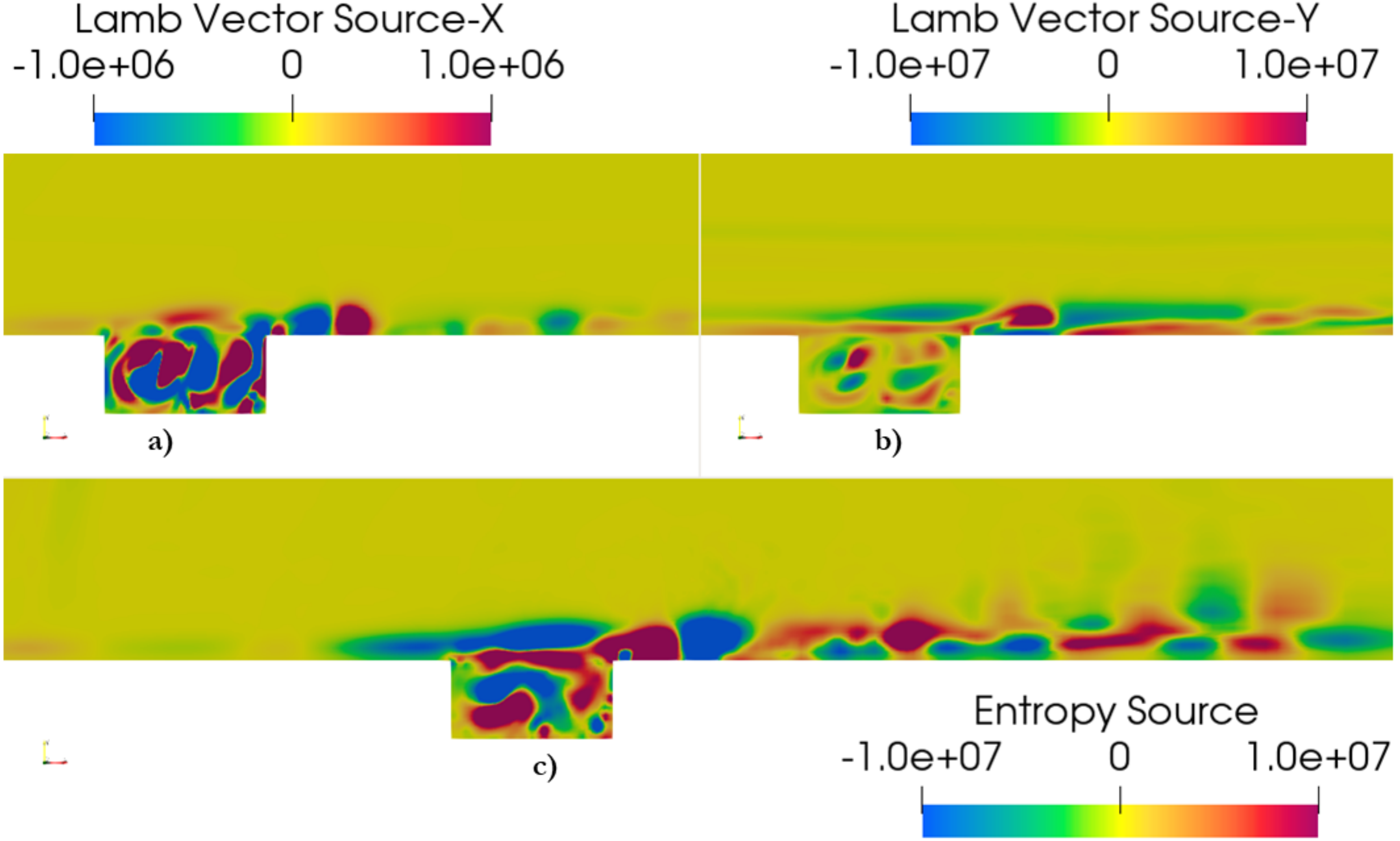}
	\caption{\label{fig:28} Source terms for M = 0.5; a) x-component of Lamb Vector, b) y-component of Lamb Vector, c) Entropy Source term}
\end{figure}

\begin{figure}	
	\includegraphics[scale=0.6]{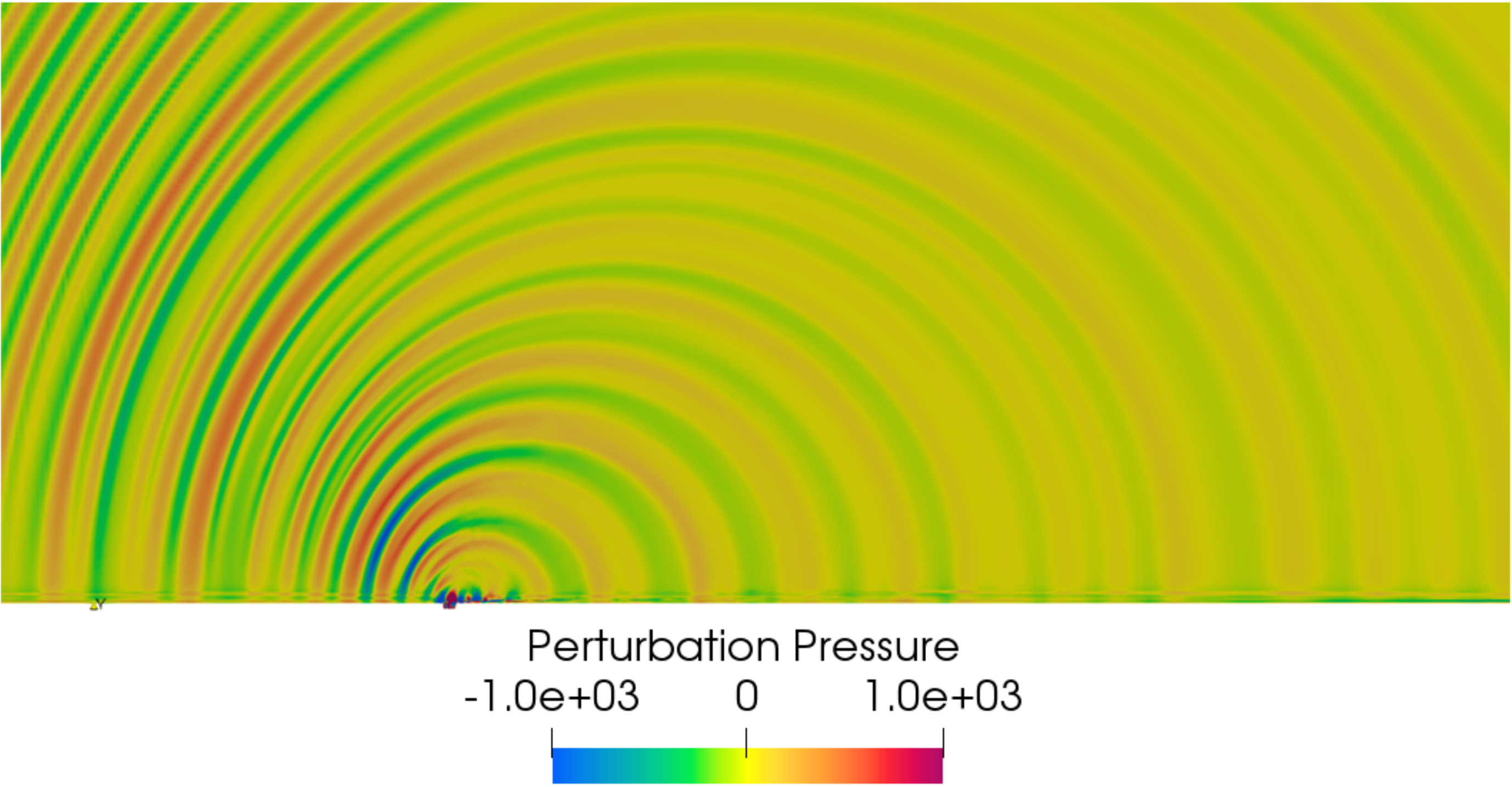}
	\caption{\label{fig:29} Perturbation pressure contours for M = 0.5}
\end{figure}

\begin{figure}	
	\includegraphics[scale=0.6]{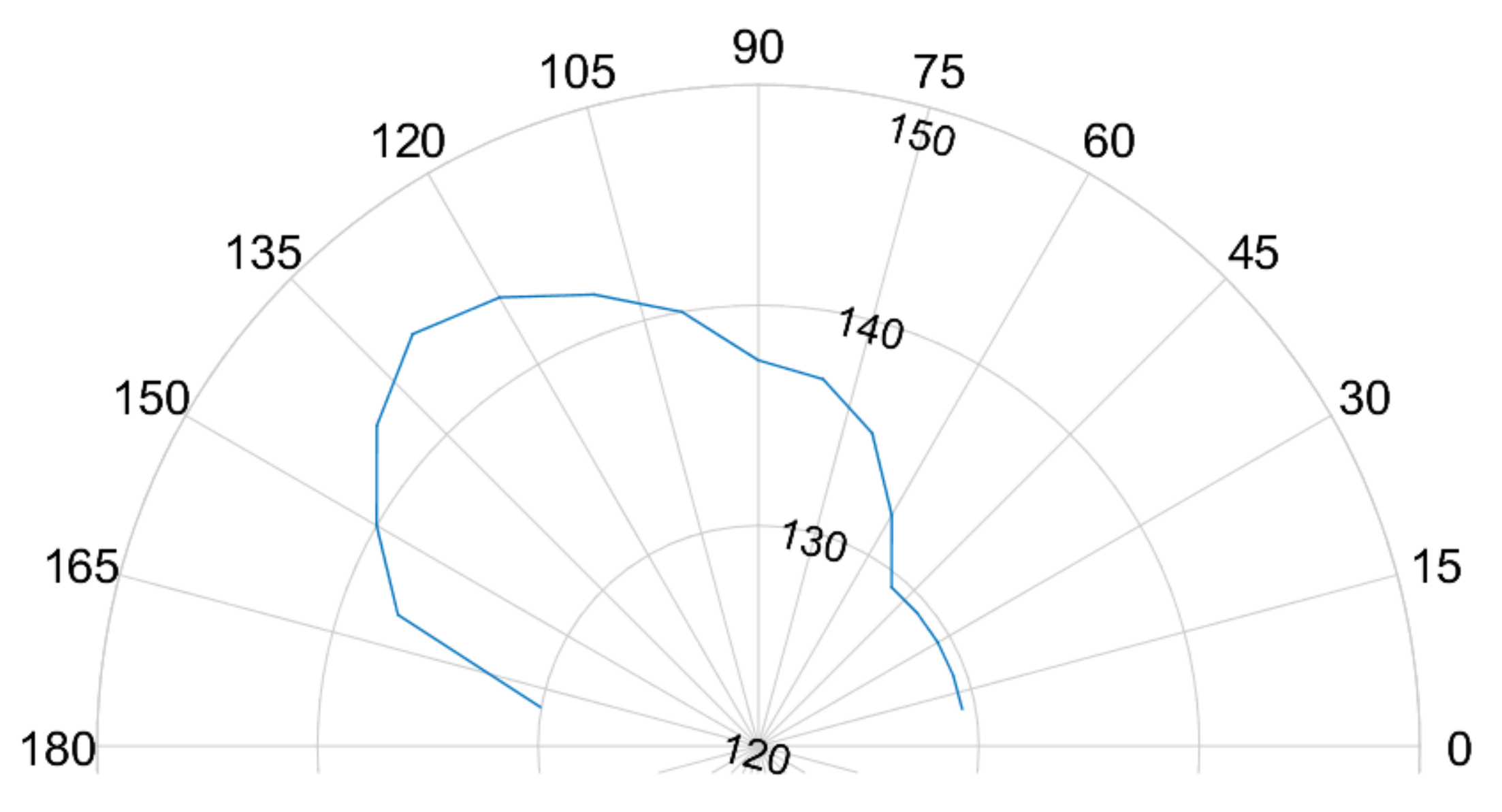}
	\caption{\label{fig:30} Farfield (r=100D) Sound directivity for M = 0.5, $f=25.4$ $kHz$, with $SPL(dB)$}
\end{figure}

\paragraph{Mach Number 0.5:} Fig. \ref{fig:28} depicts the source terms for the cavity flow at Mach $0.5$, and Fig. \ref{fig:29} provides the contours of pressure perturbation. The convection effects on the perturbation pressure contours are relatively less even though the upstream directivity is still visible (Fig. \ref{fig:30}). The peak directivity is spread across various azimuthal angles, with a peak occurring somewhere near $110$ degrees from the downstream direction.

\begin{figure}	
	\includegraphics[scale=0.6]{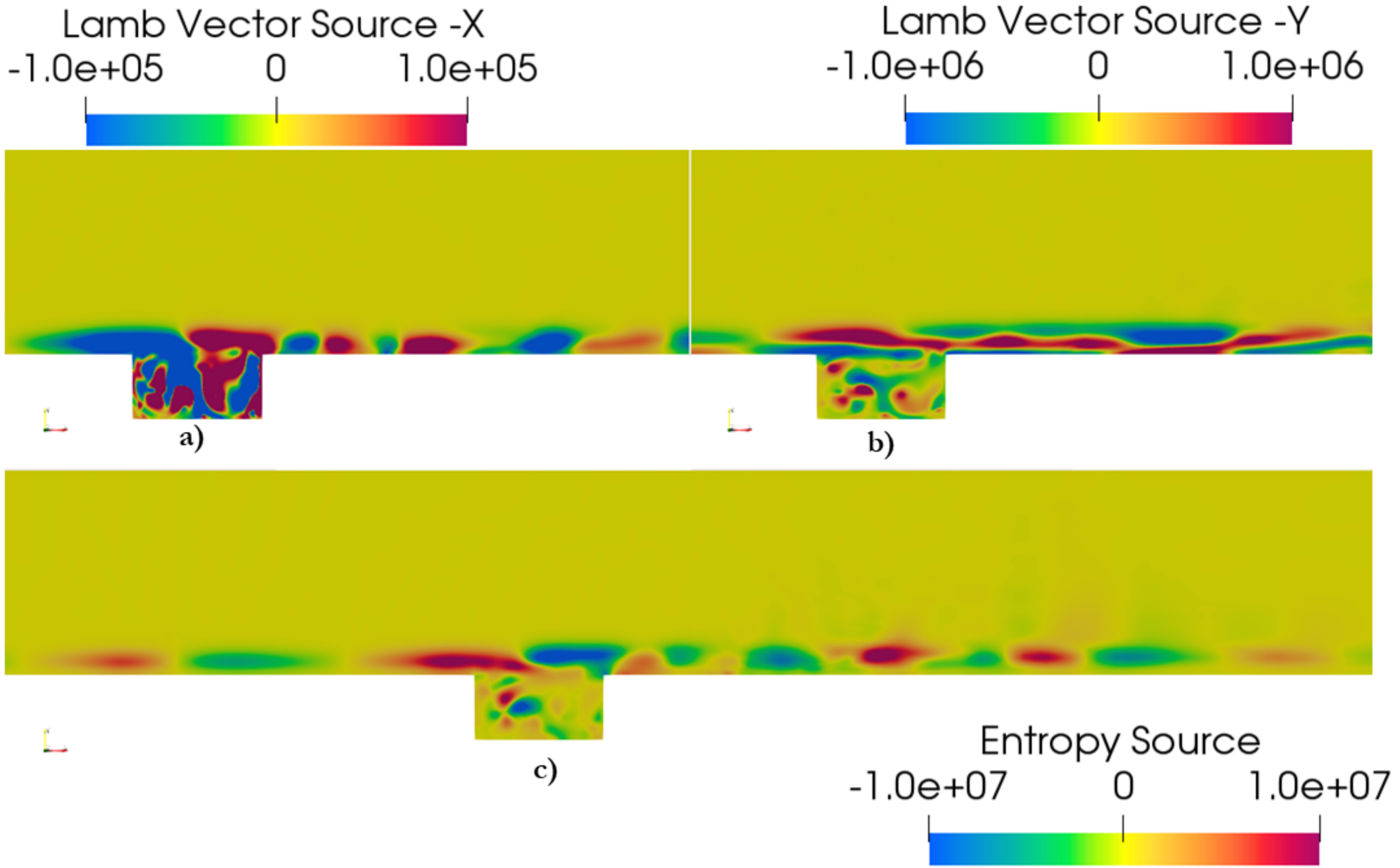}
	\caption{\label{fig:31} Source terms for M = 0.4; a) x-component of Lamb Vector, b) y-component of Lamb Vector, c) Entropy Source term}
\end{figure}

\begin{figure}
	\includegraphics[scale=0.6]{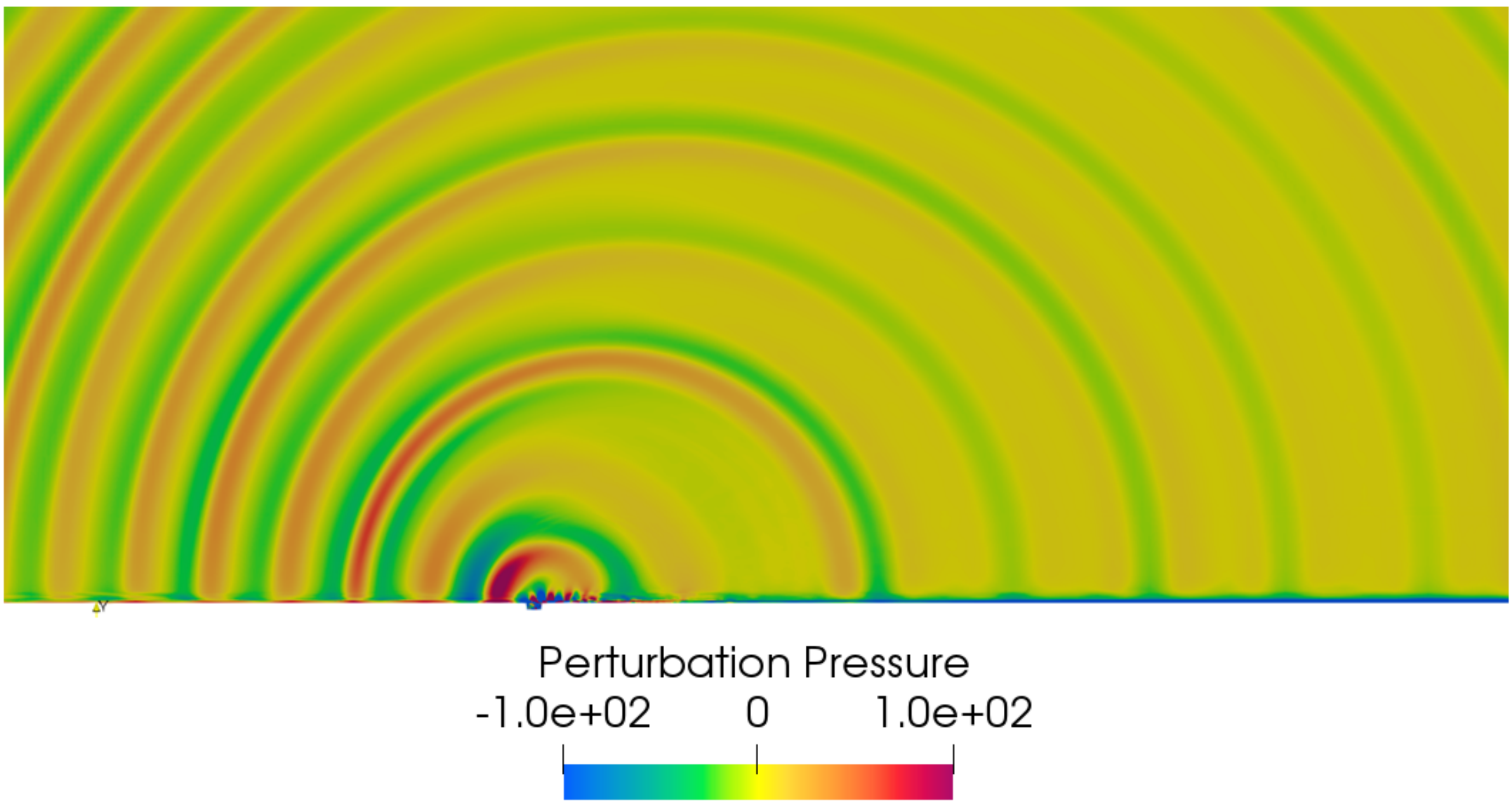}
	\caption{\label{fig:32} Perturbation pressure contours for M = 0.4}
\end{figure}

\begin{figure}	
	\includegraphics[scale=0.6]{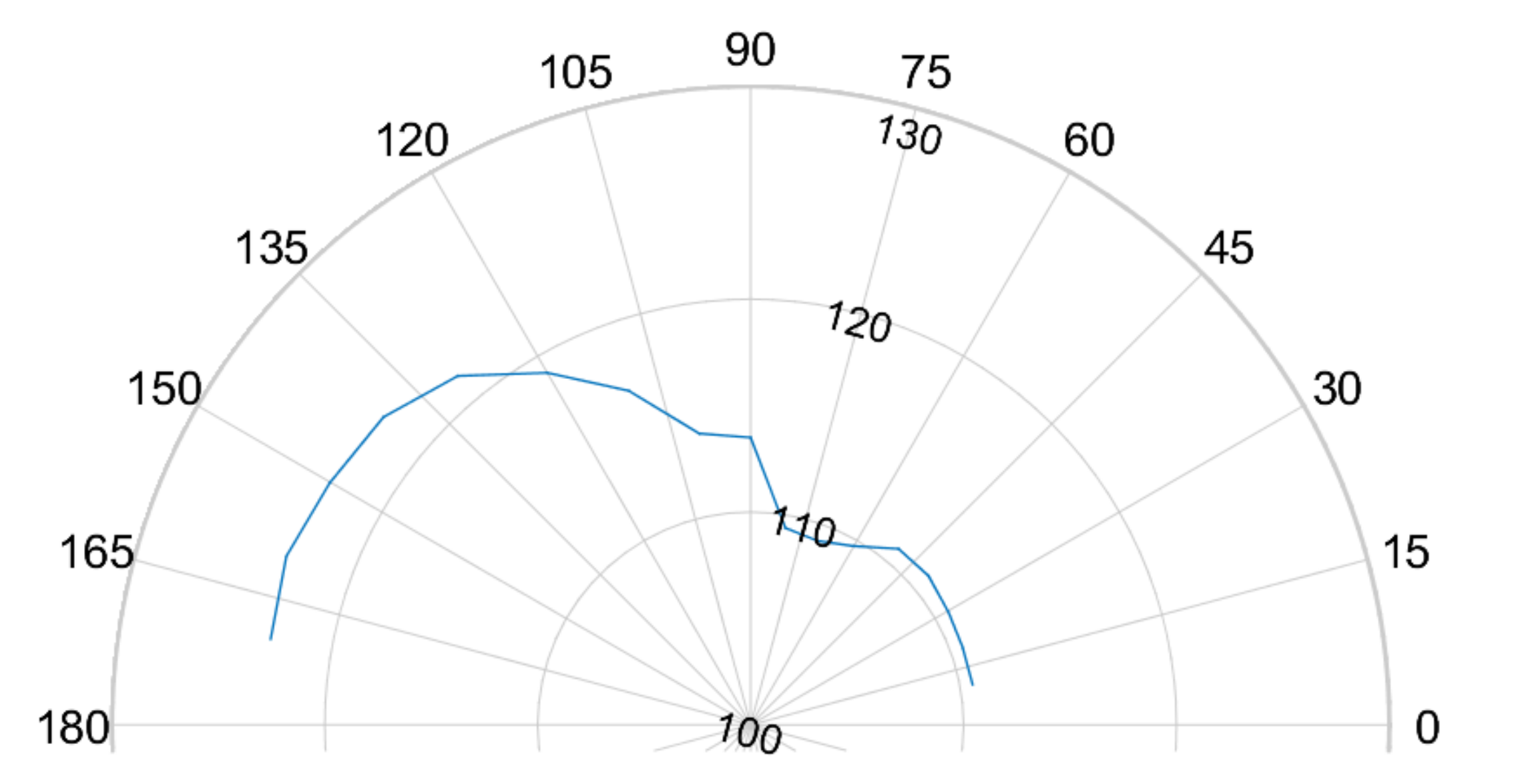}
	\caption{\label{fig:33} Farfield (r=100D) Sound directivity for M = 0.4, $f=22.0$ $kHz$, with $SPL(dB)$}
\end{figure}

\paragraph{Mach Number 0.4:} The source terms for the cavity flow at Mach 0.4 are presented in Fig. \ref{fig:31}, while \ref{fig:32} shows the pressure perturbation contours for the same. The entropy source term is relatively less inside the cavity and is distributed in the shear layer. The convection effects are not significant, and there are slight differences between the upstream and downstream traveling waves. One can notice from the directivity plot (Fig. \ref{fig:33}), that the SPL at the points along the upstream direction is relatively higher than that at the downstream points. Still, there is no significant directionality in the propagation of the acoustic waves.

\subsubsection{\label{sec:level3}Supersonic Cavity} 

\begin{figure}	
	\includegraphics[scale=0.6]{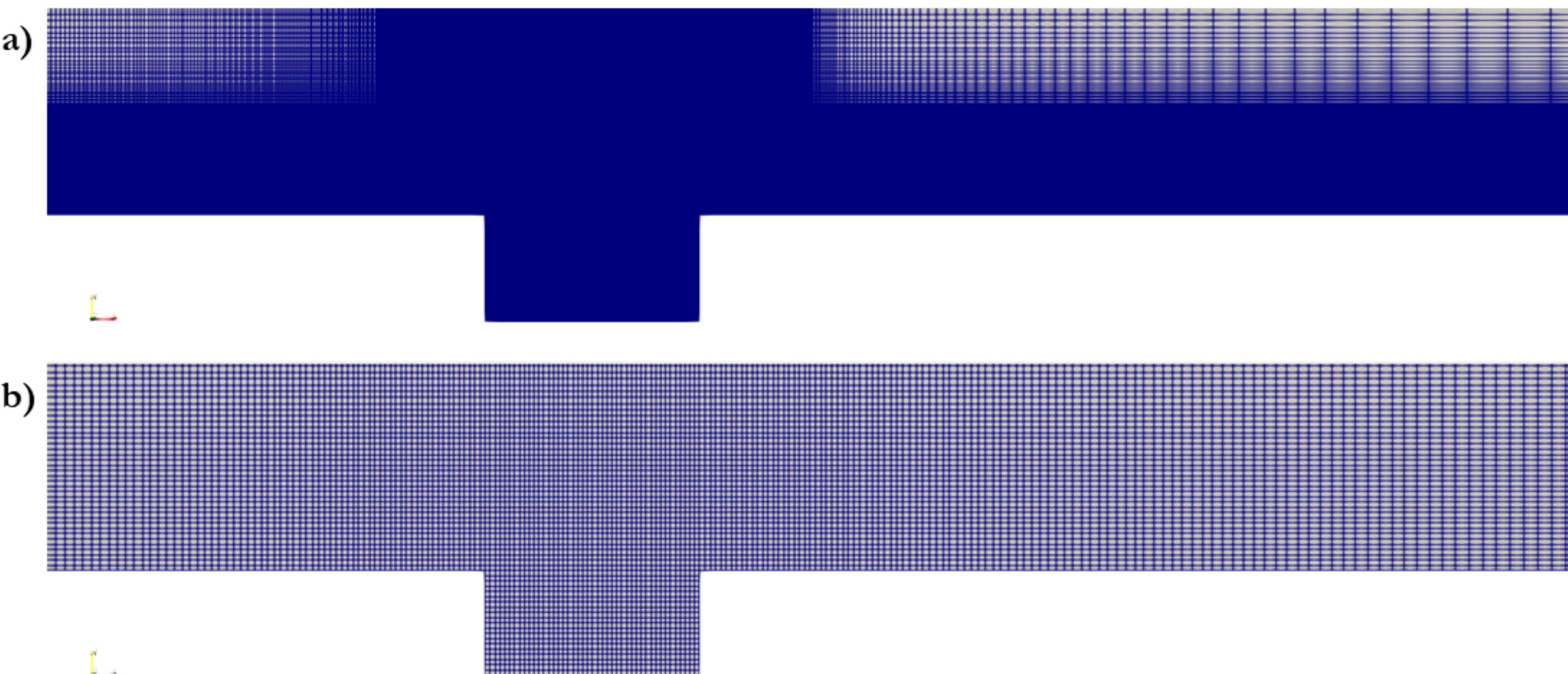}
	\caption{\label{fig:34} a) Fluid Mesh, b) Acoustic Mesh for M = 1.5}
\end{figure}

We now extend our analysis to a supersonic cavity flow at Mach number, $M=1.5$ (mesh shown in Fig. \ref{fig:34}) to  assess the noise sources using the hybrid method. The polynomial order for the acoustic simulation is reduced to keep the computational cost under acceptable limits and reduce the oscillatory behavior and larger time in the solver convergence. The mean velocity and density fields in the fluid simulation are discontinuous across the shock generated over the cavity, which is detrimental to the stability of the APE system. Therefore, to avoid any numerical instability in the acoustic solver while at the same time considering the convection effects associated with the flow, we have used the reference base field from a shock free flow simulation ($M=0.85$) for the supersonic study.

\begin{figure}
	\includegraphics[scale=0.55]{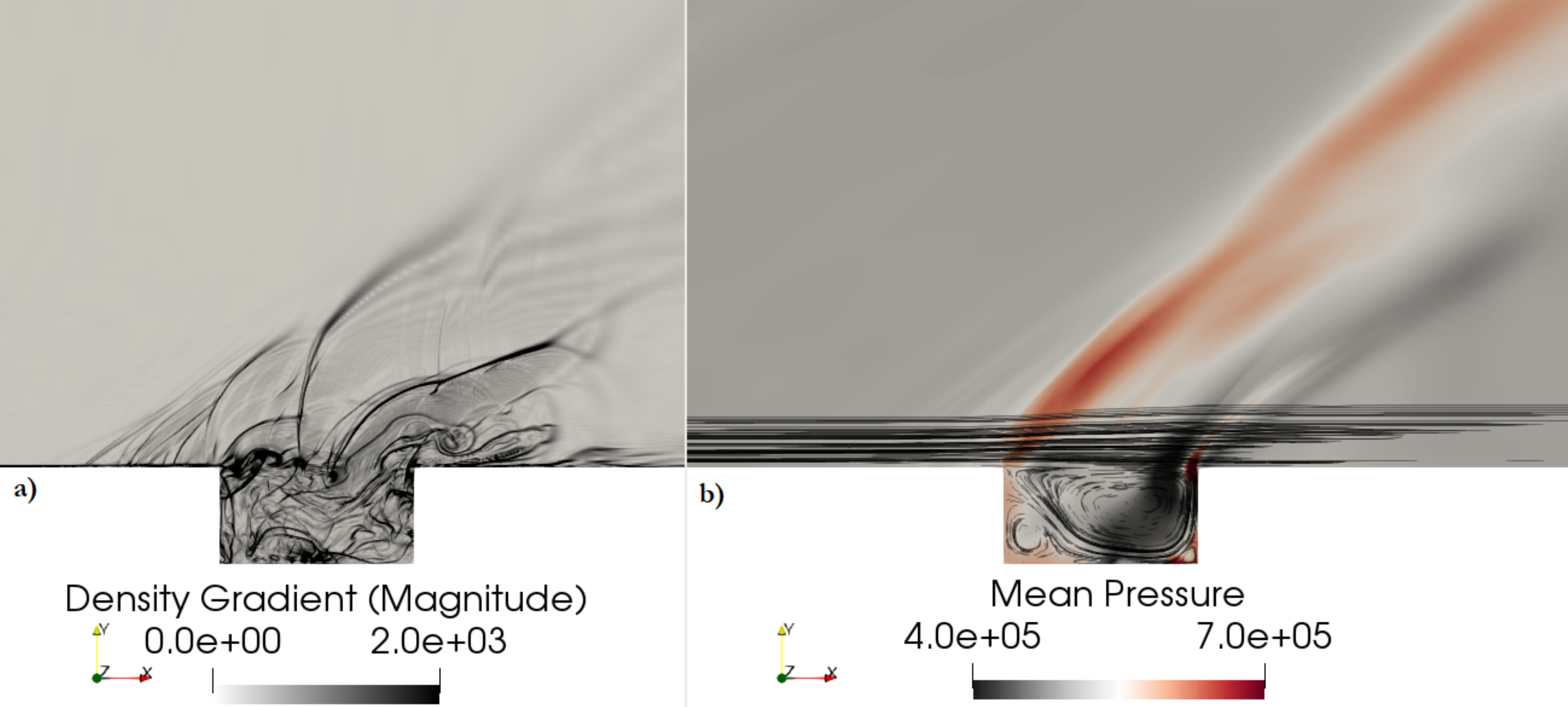}
	\caption{\label{fig:35} Contours of (a) Density Gradient Magnitude and (b) Mean Pressure for M=1.5 cavity}
\end{figure}

Numerical schlieren (density gradient magnitude) for the present case are illustrated in Fig. \ref{fig:35}(a). The region near the cavity consists of supersonic as well as subsonic flow regimes that are wave dominated containing complex flow structures. When the incoming supersonic flow encounters the shear layer at the leading edge of the cavity, it forms an oblique shock visible in Fig. \ref{fig:35}(b). This shock front serves as the boundary of the information propagation from the cavity. Therefore, any acoustic waves generated inside the cavity and travelling upstream are restricted at this boundary and are convected downstream with the flow. This is visible in Fig. \ref{fig:35}(a) where some wave fronts are restricted and forced to move along the shock boundary. Minor shocks upstream of the leading edge are also present due to the interaction of the flapping shear layer with the incoming flow. It can also be noted from Fig. \ref{fig:35}(b) that a compression corner is formed at the trailing edge due to the impingement of incoming supersonic flow.
Fig. \ref{fig:36} reports the vorticity distribution and pressure fluctuations around the supersonic cavity. It is evident that for the supersonic case, the cavity undergoes violent oscillations. The shear layer oscillation mechanism with mass addition and ejection is quite similar to the subsonic case. However, the structure of the shear layer itself is slightly different. The vortex pairing in the shear layer is visible through the vorticity plots in Fig \ref{fig:36(a)}(i) and Fig \ref{fig:36(b)}(ii). This vortex pair is convected subsequently in the downstream direction by the flow. The shear layer for the supersonic case is composed of discrete vortices, which is clear by the pressure perturbation plots (Fig \ref{fig:36(c)}(ii) and Fig \ref{fig:36(d)}(ii))having many small locally low pressure regions. The effect of these structures on the acoustic field around a supersonic cavity is discussed in the next section. 

\begin{figure}
	\includegraphics[scale=0.6]{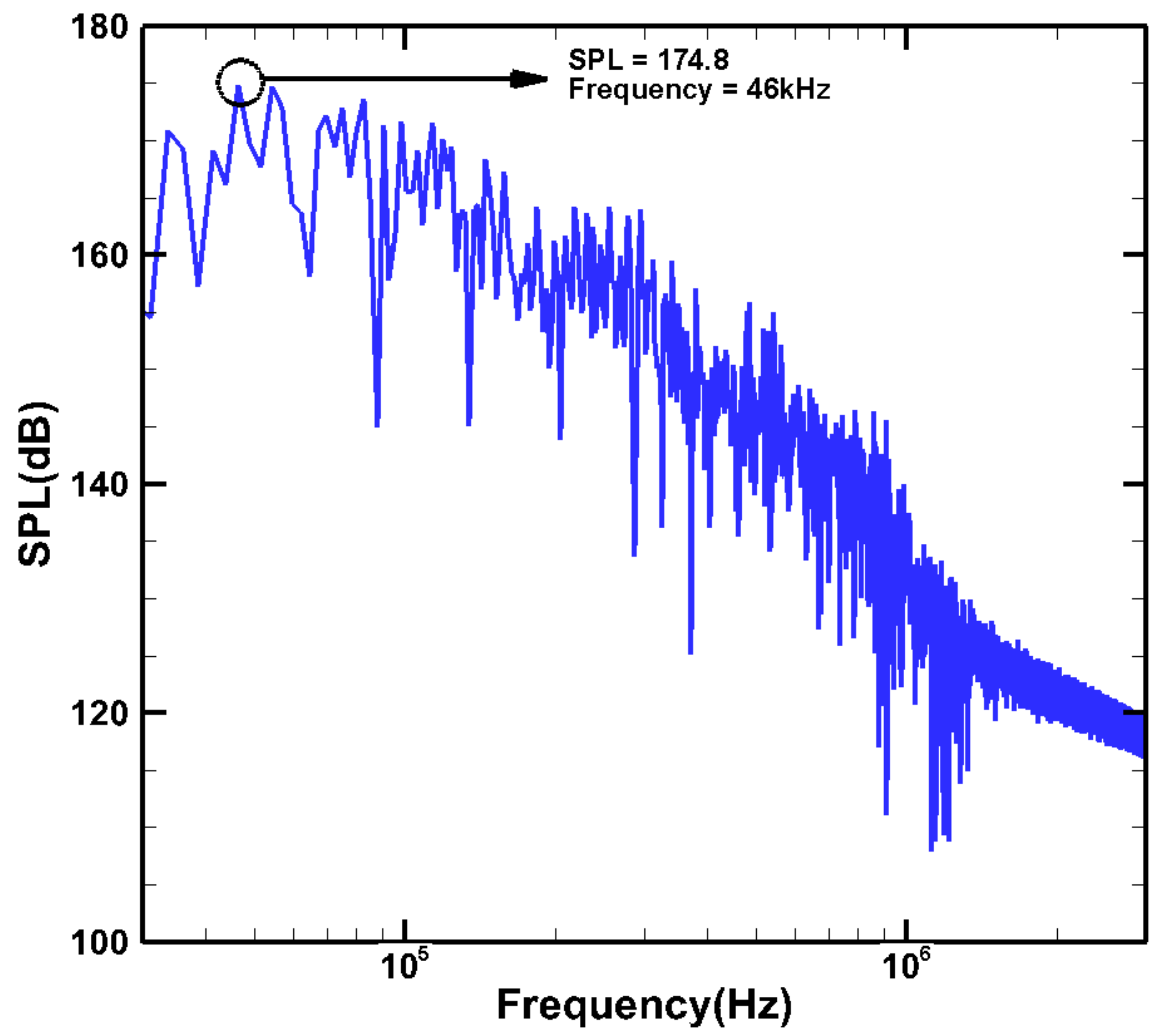}
	\caption{\label{fig:37} SPL for the cavity at M = 1.5 ($x/D=-0.04, y/D=2$)}
\end{figure}

\begin{figure}	
	\includegraphics[scale=0.6]{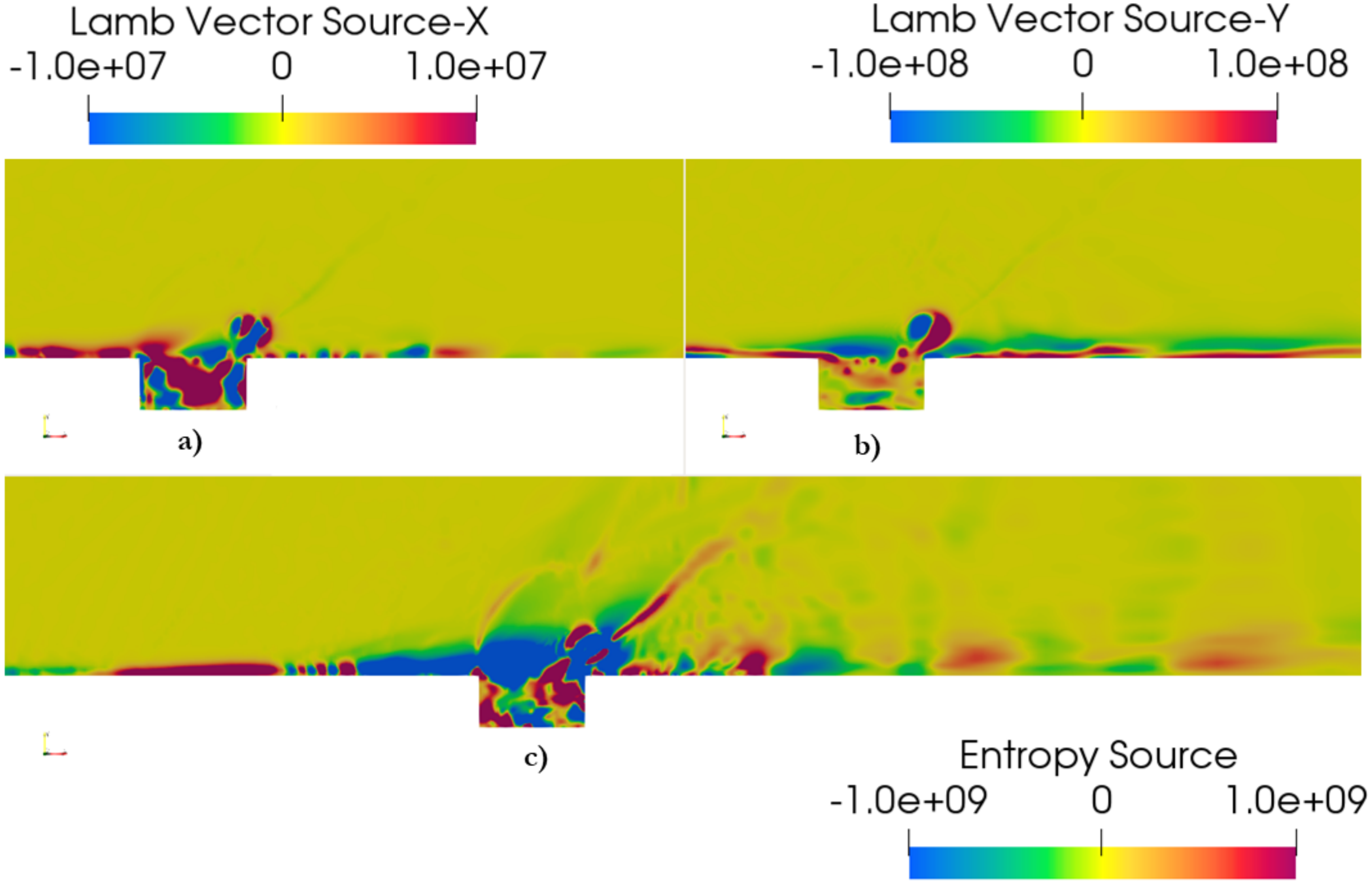}
	\caption{\label{fig:38} Source terms for M = 1.5; a) x-component of Lamb Vector, b) y-component of Lamb Vector, c) Entropy Source term}
\end{figure}

Fig. \ref{fig:37} reports the SPL at the reference point, while Fig. \ref{fig:38} depicts the source terms. The peak SPL is encountered at $46$ $kHz$, which is in accordance with the experiment \cite{krishnamurty1955acoustic}.
It is interesting to note that for this case, while the Lamb Vector sources are concentrated in the shear layer and inside the cavity, the entropy source has significant values above the cavity shear layer as well. These regions correspond to the shock waves' location in the cavity flow across which the temperature and entropy gradients occur. Thus, the acoustic field is expected to be broadband with multiple frequencies, evident from the spectrum.

\begin{figure}	
	\includegraphics[scale=0.9]{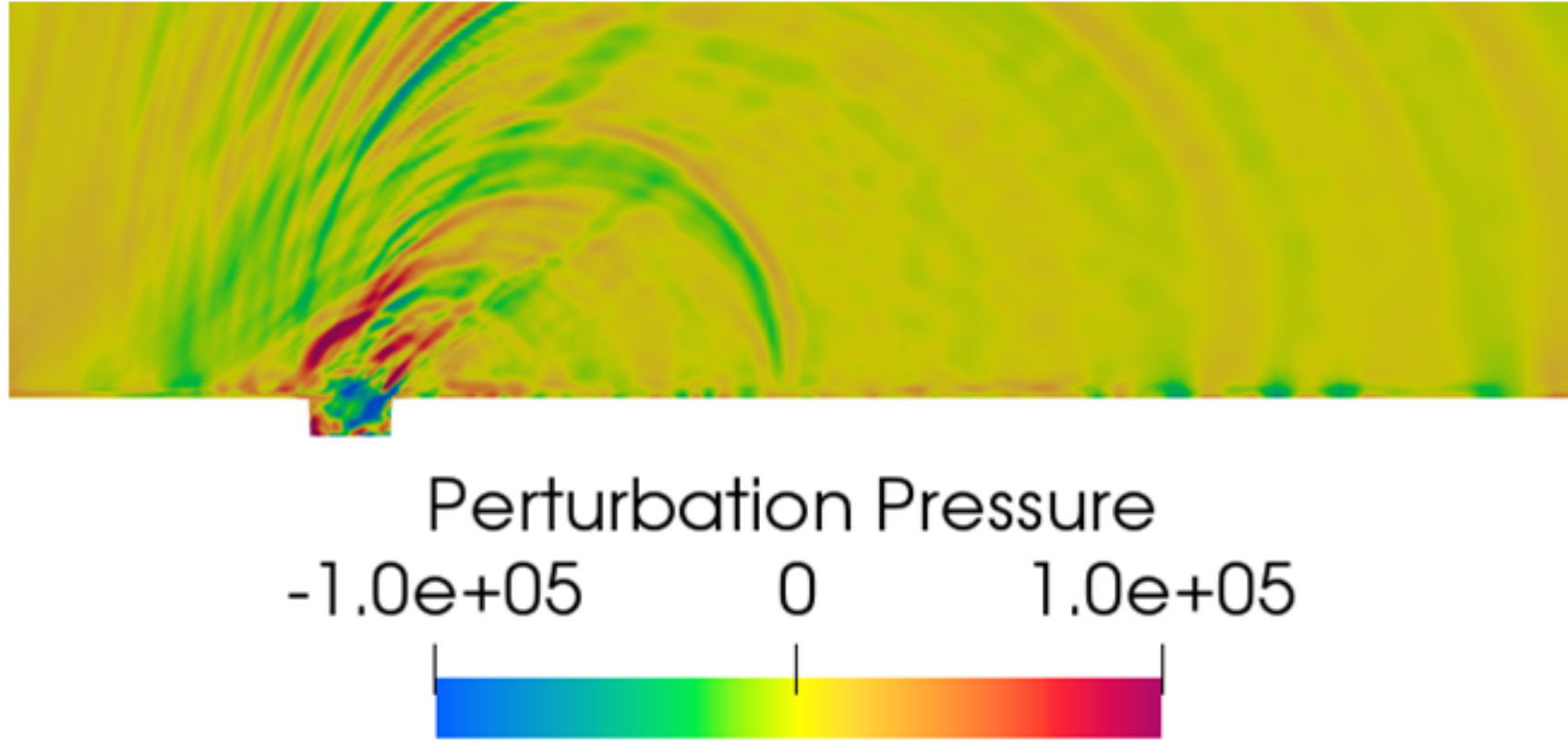}
	\caption{\label{fig:39} Perturbation pressure for M = 1.5}
\end{figure}

For this, Fig. \ref{fig:39} illustrates the contours of pressure perturbation.
Intense acoustic radiation around the cavity is observed. The acoustic waves should ideally have a direction of propagation in the downstream direction. In experiment \cite{krishnamurty1955acoustic}, the wave patterns are restricted for upstream travel near the leading edge's shock boundary. The perturbation pressure contours presented also follow the same pattern. Some upstream traveling waves are present in the system because of the nature of the base fields, as discussed earlier. However, these are insignificant compared to the primary propagation direction, which is in accordance with the experimental observation.

\subsection{\label{sec:level2}Interpretation of the acoustic source terms}

Since the Lamb Vector sources by formulation inherently describe the vortex dynamics, the noise generation mechanism can be described by the evolution of the vortices. The vorticity distribution obtained from the CFD simulation might be helpful in such a case. The limitation, however, is that it should be coupled with the evolution of the pressure perturbation. This limitation is overcome by the present analysis in which the vorticity source terms are directly employed to generate the sound. By analyzing the source terms, one can observe the underlying vortex dynamics responsible for generating the acoustic waves.

A general recirculation region exists inside the cavity consisting of one small recirculating region near the leading edge and one larger region near the trailing edge. The shear layer of a cavity consists of discrete vortices \cite{ziada1982vortex} due to the resonant vortex shedding. The source terms can be interpreted in terms of these discrete vortices. In Fig. \ref{fig:40}, the evolution of discrete vortices in the shear layer and their interaction with the cavity is presented. The vortex (Vortex-1) visible in the $y-$Lamb Vector source term near the trailing edge is clipped as observed in Fig.\ref{fig:40(a)}(ii). A part of this clipped vortex convects in the downstream direction. Simultaneously, the other part is trapped by the cavity walls and is recirculated inside the cavity due to its inertia and the geometric constraint. This vortex becomes a part of the larger recirculation region inside the cavity. While this process happens, a new vortex (Vortex-2) is shed from the leading edge, as observed in Fig. \ref{fig:40(b)}(ii). As the Vortex-2 is convected in the downstream direction, the edge of the larger recirculating region pushes the shear layer up (and consequently, Vortex-2). The bending of this vortex results in a positive wavefront generation, as depicted in Fig. \ref{fig:40(c)}(iv). As soon as this bent vortex reaches the trailing edge and the clipping process starts, the new vortex, formed at the leading edge, slides inside the cavity generating the negative wavefront as shown in Fig. \ref{fig:40(d)} as well as in Fig. \ref{fig:40(b)}. This new vortex then again bends due to the recirculation regions's upward push, and the cycle continues. The mechanism mentioned above is presented for M = 0.4 cavity flow, for which the convection effects are not large. The wavefronts are formed at different positions above the cavity; however, the spreading of these wavefronts happens uniformly. There are other minor sources at the trailing edge as well, the strength and contribution of which is relatively less.

The same mechanism is also followed for high subsonic Mach numbers ($M > 0.6$) as well. The bending and clipping processes for high Mach numbers are dominated by convection, which results in several wavefronts of comparable intensities during the bending and clipping phases giving rise to a broadband structure. . For example, for $M = 0.7$, the dense wavefronts (Fig. \ref{fig:23}) along the direction of maximum sound radiation result from wavefronts generated by the process mentioned above and their rapid convection along the shear layer while generating these wavefronts. 
 
The underlying source mechanism is found to be similar for supersonic flows as well. It is observed, however, that there are some distinct differences between the two in some respects. The vortex shedding and the excitation due to the cavity recirculation region, the shear layer may give rise to a vortex pair as observed in Fig. \ref{fig:41(a)}(ii) and Fig. \ref{fig:41(b)}(ii). The bending of this vortex pair and further clipping at the trailing edge also produce the same wavefronts as before. Still, this process produces sudden "bursts" of acoustic radiation instead of when a regular vortex undergoes the same process. The discrete vortices at the leading edge sometimes undergo bending as before, but at times they merge, giving rise to a larger vortex, which follows the same bending giving rise to a sharp wavefront. This whole process is illustrated in Fig. \ref{fig:41(c)} to Fig. \ref{fig:41(e)}. Due to the generation of a compression corner at the cavity trailing edge, the entropy fluctuations at this point are pronounced, which also gives rise to acoustic radiation traveling at the same angle as the leading edge shock makes with the downstream direction. The overall noise generation mechanism for the supersonic cavity flow is somewhat complicated, but the current analysis highlights certain major aspects of this process.

It is interesting to observe that the study of the source terms for the APE system brings forward the hidden vortex dynamics associated with the sound generation process. The present case shows that the sound generation process for the cavity flow can be broken down into studying the dynamics and evolution of discrete vortices along the cavity shear layer. The present analysis also reveals vortices on either side of the shear layer. Such a vortex train has also been demonstrated by Soni et al. \cite{soni2019modal} by applying modal decomposition techniques on a supersonic cavity flow. They have also been studied by Sridhar et al. \cite{sridhar2018mach} for the generation of Mach waves over a supersonic cavity. Thus, the study of source terms for sound generation using vortex sources uses the coupling between hydrodynamics and acoustics to reveal the underlying vortex dynamics, which are essential for the evolution of the flow itself.

\section{\label{sec:level1}Conclusion}   

The present work employs a hybrid CAA technique to study the noise sources in a compressible cavity flow for different Mach numbers. The effect of base fields (mean velocity and density) are taken into consideration. The hybrid methodology uses an innovative approach for the data transfer between fluid mesh and acoustic mesh. In the first step, base fields are transferred from the fluid solver to the acoustic solver. In the next step, unsteady sources (Lamb Vector and Entropy Source) are calculated and sent to the acoustic solver. The approach ensures that no damping function is required for the current study. This methodology is first validated against a few benchmark cases involving laminar and turbulent flow regimes. The obtained results are in close agreement with the existing literature. The coupled methodology efficiently predicts the tonal as well as the broadband sound. The effect of the base flow field on the sound directivity and the maximum amplitude of the perturbation pressure is demonstrated by comparing it with existing results obtained through an acoustic analogy without the effect of base flow. The results suggest that even though the base flow does not affect the amplitude of the pressure perturbations, it imparts some directivity to the acoustic waves. The perturbation pressure obtained from the hybrid approach in the near-field as well as in the far-field is in excellent agreement with the existing DNS study. The methodology is used to validate the importance of entropy source term for sound generation for a Mach $0.7$ cavity flow. The impact of entropy and vortex source terms on perturbation pressure is found to be in excellent agreement with the DNS results. Only use of vortex source over-predicts the perturbation pressure values in the near field region. The method is then used to study the noise sources in subsonic, and supersonic cavity flow. The obtained results using the current method are in good agreement with the experimental results in terms of acoustic spectra and sound directivity. Further, the evolution of the source terms in the shear layer, as well as inside the cavity explains the production of acoustic waves over the cavity. The results aslo report the directivity pattern for different Mach numbers. For high subsonic Mach numbers, the acoustic waves are highly directional. The analysis of the source terms highlights the key features involved in the sound generation process. These features are in accordance with the existing models. The generation of vortices at the leading edge, their subsequent convection resulting in the bending of the shear layer and eventually their clipping at the trailing edge is clearly demonstrated from the analysis of the source terms. This complete cycle results in the generation of acoustic waves over the cavity and the direction of their propagation depends on the incoming Mach number of the flow.

\section*{Acknowledgement}
Financial support for this research is provided through IITK-Space Technology Cell (STC). The authors would also like to acknowledge the High-Performance Computing (HPC) Facility at IIT Kanpur (www.iitk.ac.in/cc).

\section*{Data Availability}

The data that support the findings of this study are available from the corresponding author upon reasonable request.

\section*{References}

\bibliography{aipsamp}
\clearpage

\begin{figure}[]
	\begin{subfigure}[]{\textwidth}	
		\centering
	\includegraphics[width=0.5\textwidth]{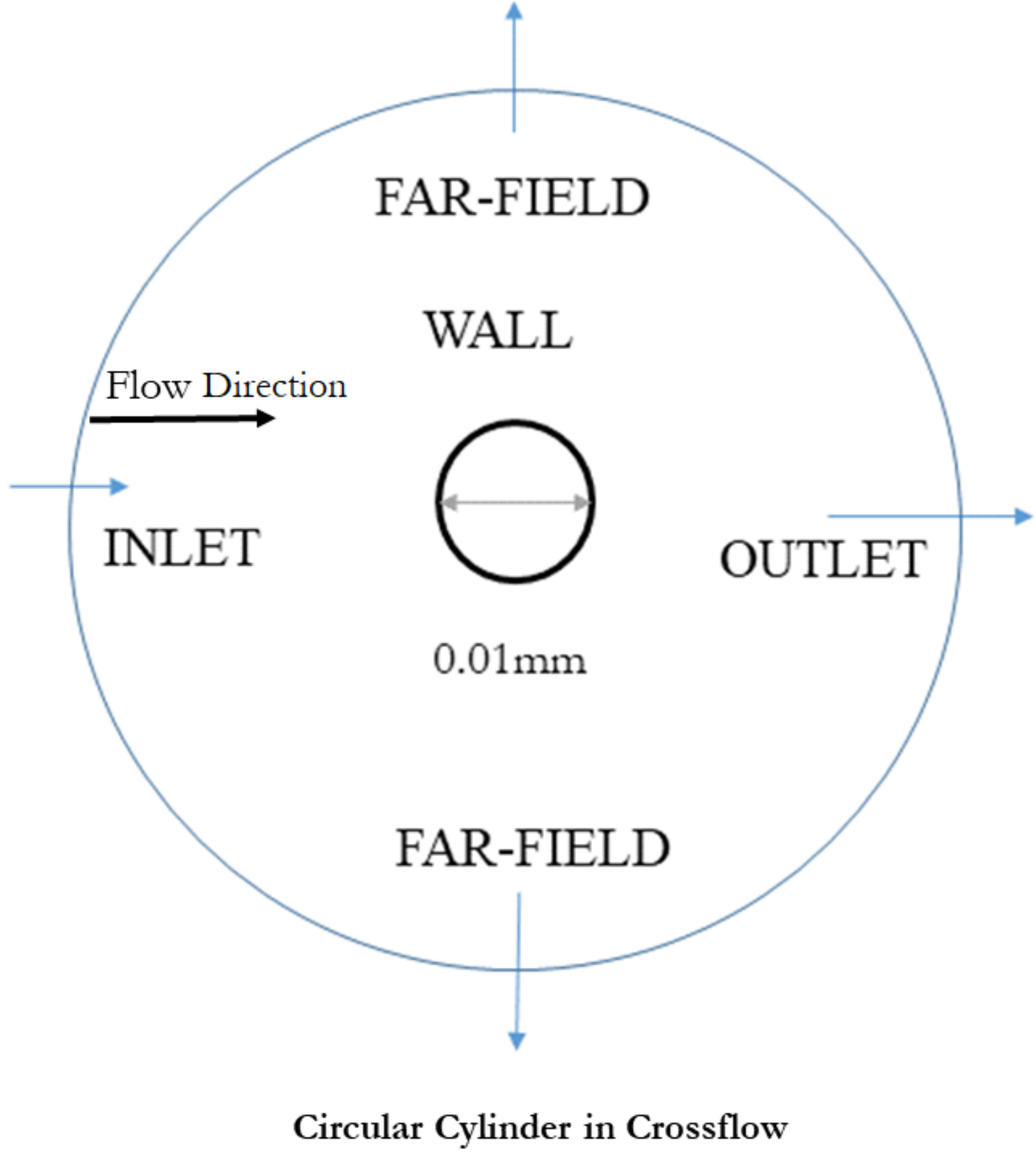}
	\caption{\label{fig:2(a)}}
	\end{subfigure}

  \begin{subfigure}[]{\textwidth}
  	\centering
  	\includegraphics[width=0.5\textwidth]{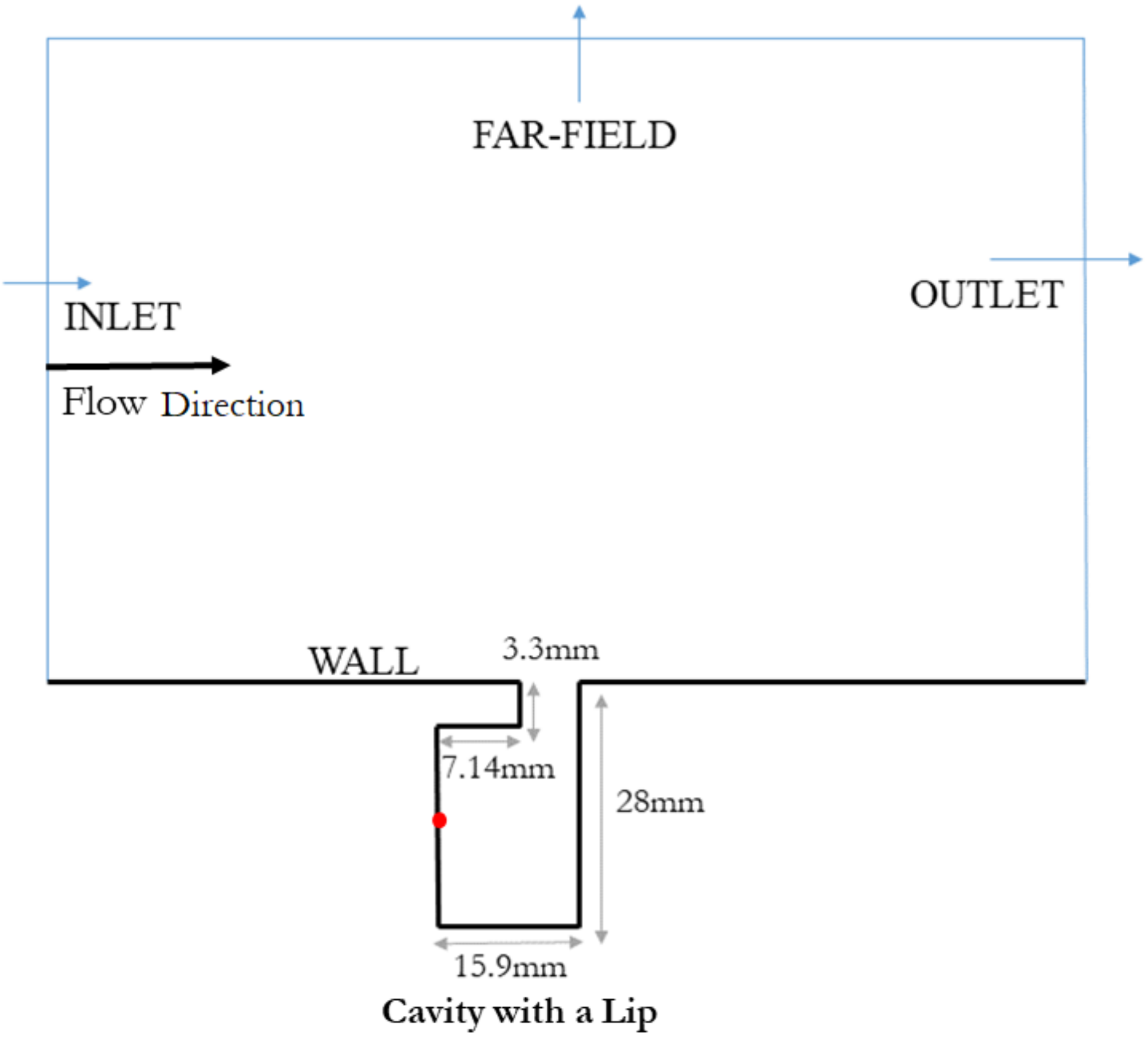}
  	\caption{\label{fig:2(b)} }
  \end{subfigure}
 \end{figure}

 \begin{figure}
 	\ContinuedFloat
  \begin{subfigure}[]{\textwidth}
  	\includegraphics[width=0.5\textwidth]{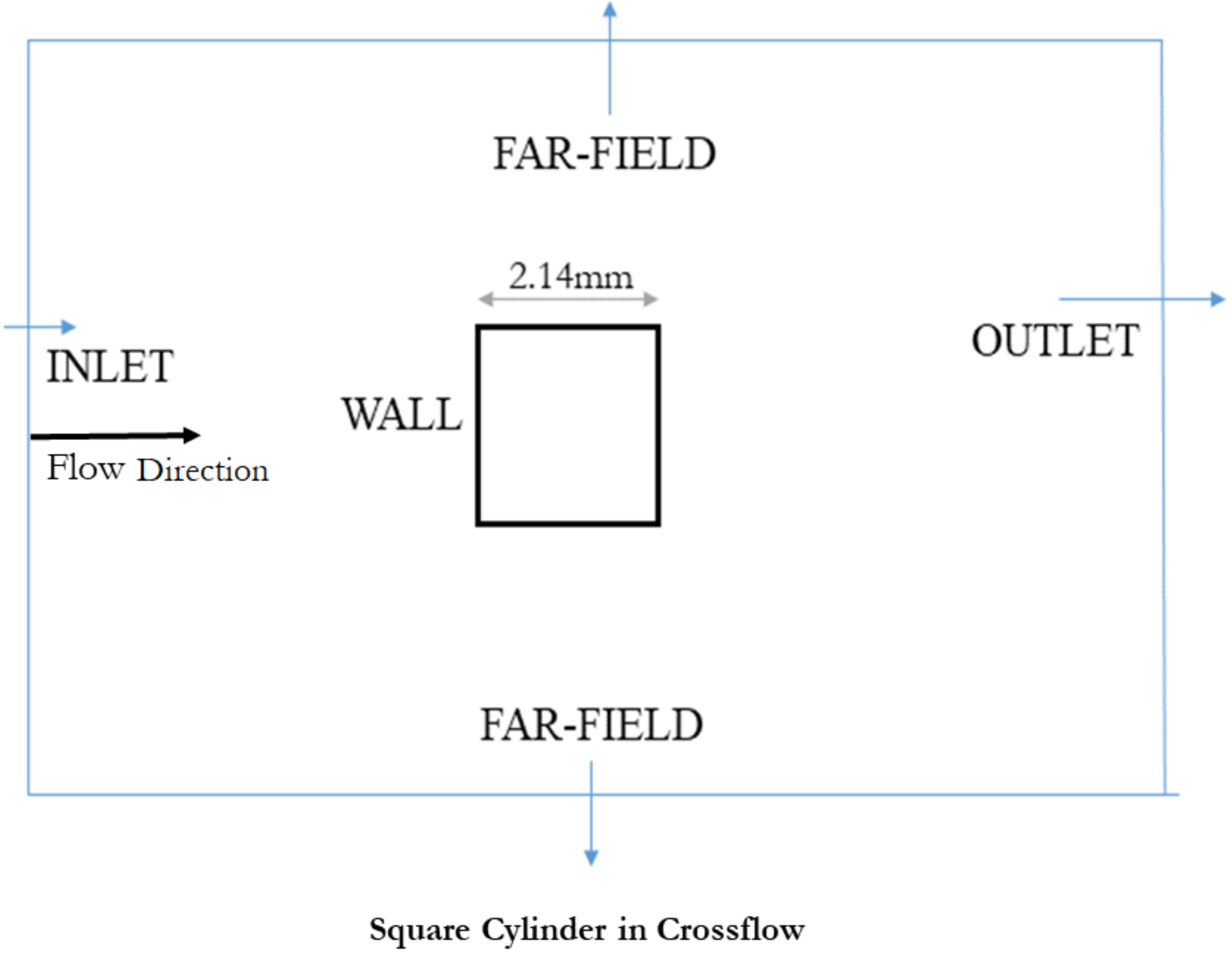}
  	\caption{\label{fig:2(c)} }
  \end{subfigure}

   \begin{subfigure}[]{\textwidth}
  	\includegraphics[width=0.5\textwidth]{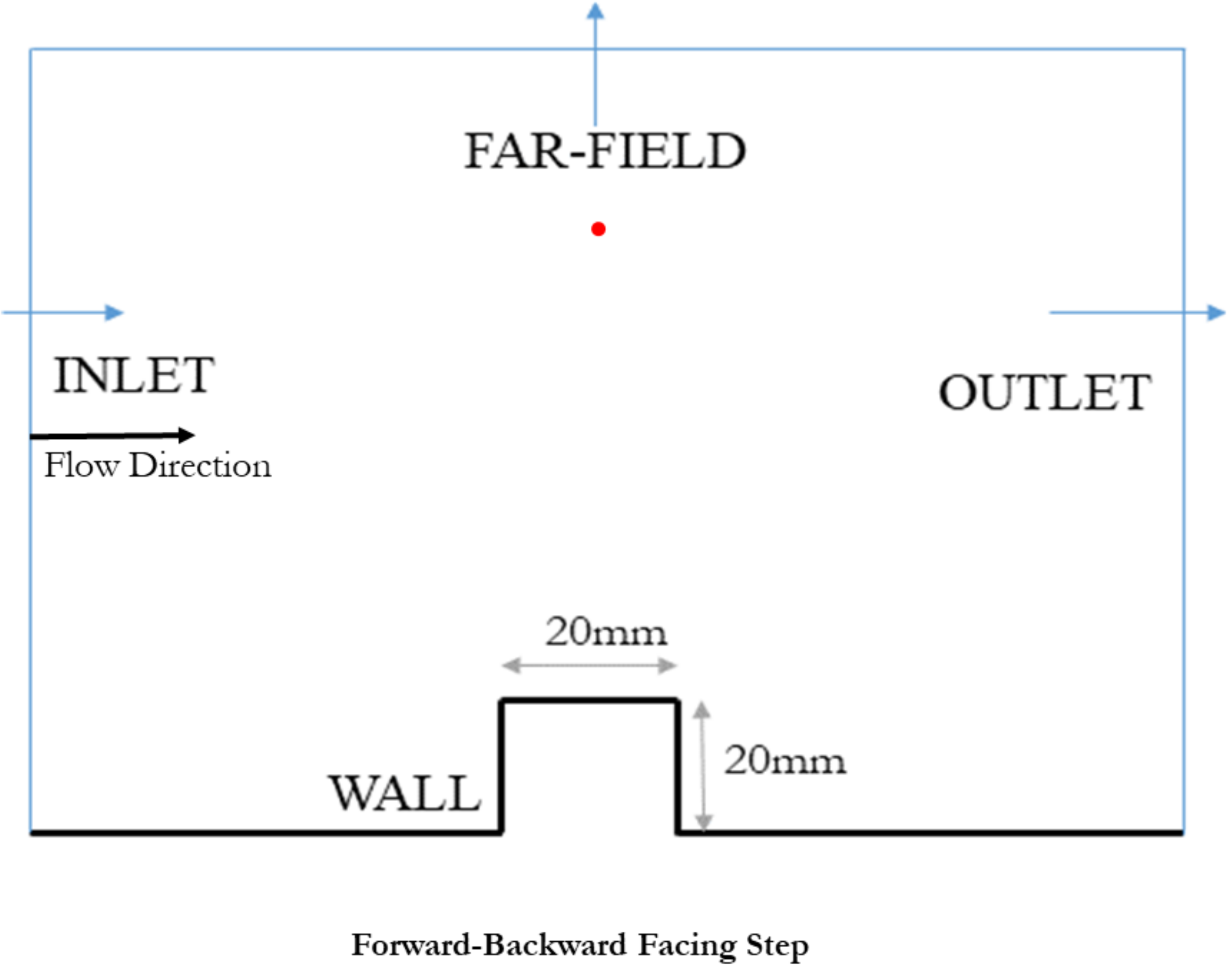}
  	\caption{\label{fig:2(d)}}
  \end{subfigure}

  \begin{subfigure}[]{\textwidth}
	\includegraphics[width=0.5\textwidth]{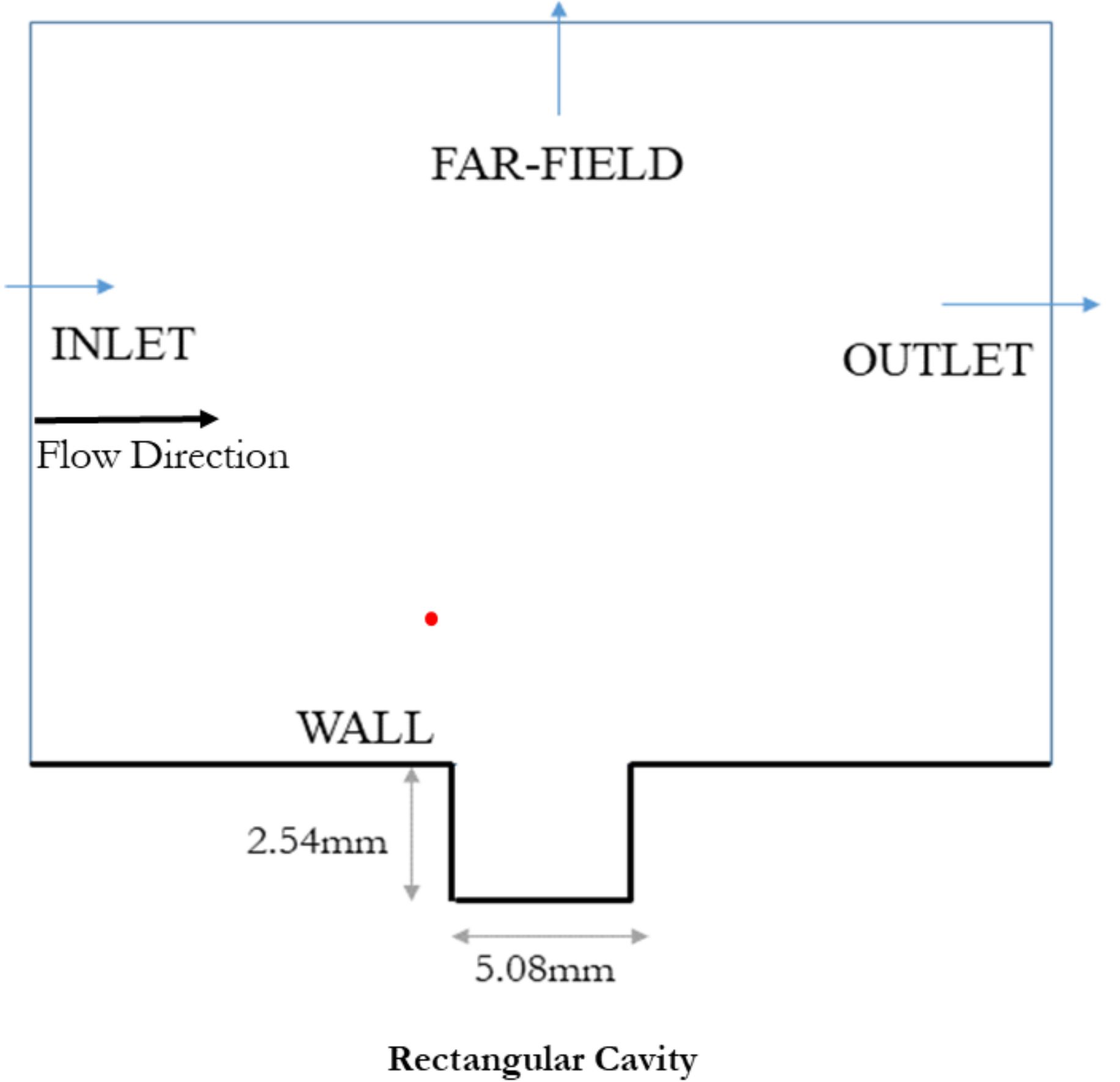}
	\caption{\label{fig:2(e)} }
\end{subfigure}
  	\caption{Geometry and Domain for all cases: a) Circular Cylinder in crossflow, b) Square Cylinder in crossflow c) Forward Backward Facing Step, d) Cavity with a Lip, e) Rectangular Cavity (The experimental values are obtained from the literature at the point marked in red)}
  \label{fig:2}
\end{figure}

\begin{table}
	\caption{Fluid Domain size for all the cases}
	\phantom{~}\noindent
	\begin{tabular}{p{0.2\textwidth}  p{0.1\textwidth}  p{0.07\textwidth} b{0.07\textwidth}  b{0.07\textwidth} b{0.07\textwidth}}
		Geometry&Characteristic length(D)mm&Inlet&Outlet&Farfield&Span \\
		\hline\hline \\
		Cylinder & 0.01 & 510D & 510D & 510D & 0 \\
		\hline \\
		Forward-Backward Step & 20 & 510D & 510D & 510D & 5D \\
		\hline \\
		Cavity with Lip & 28 & 510D & 510D & 510D & 10D \\
		\hline \\
		Cavity subsonic & 2.54 & 160D & 510D & 510D & 0 \\
		\hline \\
		Cavity supersonic & 2.54 & 160D & 510D & 510D & 0 \\
		\hline \\
	\end{tabular}
	\label{table:1}
\end{table}
\begin{table}
	\caption{Acoustic Domain size for all the cases}
	\phantom{~}\noindent
	\begin{tabular}{p{0.2\textwidth}  p{0.1\textwidth}  p{0.07\textwidth} b{0.07\textwidth}  b{0.07\textwidth} b{0.07\textwidth}}
		Geometry&Characteristic length(D)mm&Inlet&Outlet&Farfield&Span \\
		\hline\hline \\
		Cylinder & 0.01 & 500D & 500D & 500D & 0 \\
		\hline \\
		Forward-Backward Step & 20 & 500D & 500D & 500D & 0 \\
		\hline \\
		Cavity with Lip & 28 & 500D & 500D & 500D & 0 \\
		\hline \\
		Cavity subsonic & 2.54 & 150D & 500D & 500D & 0 \\
		\hline \\
		Cavity supersonic & 2.54 & 150D & 500D & 500D & 0 \\
		\hline \\
	\end{tabular}
	\label{table:2}
\end{table}
\begin{table}
	\caption{Mesh Details}
	\phantom{~}\noindent
	\begin{tabular}{p{0.2\textwidth}  p{0.1\textwidth}  p{0.1\textwidth} p{0.2\textwidth} }
		Geometry&Fluid Mesh&Acoustic Mesh&Polynomial Order \\
		\hline\hline \\
		Cylinder & 1.5e+05 & 2.4e+04 & 5 \\
		\hline \\
		Forward-Backward Step &1.0e+05  & 1.5e+04 & 5 \\
		\hline \\
		Cavity with Lip & 2.0e+05 & 3.0e+04 & 5 \\
		\hline \\
		Cavity subsonic & 7.5e+05 & 6.0e+04 & 5  \\
		\hline \\
		Cavity supersonic & 1.2e+06 & 1.0e+05 & 3  \\
		\hline
	\end{tabular}
	\label{table:3}
\end{table}


\begin{figure}
	\begin{subfigure}{1\textwidth}
		\includegraphics[scale=0.55]{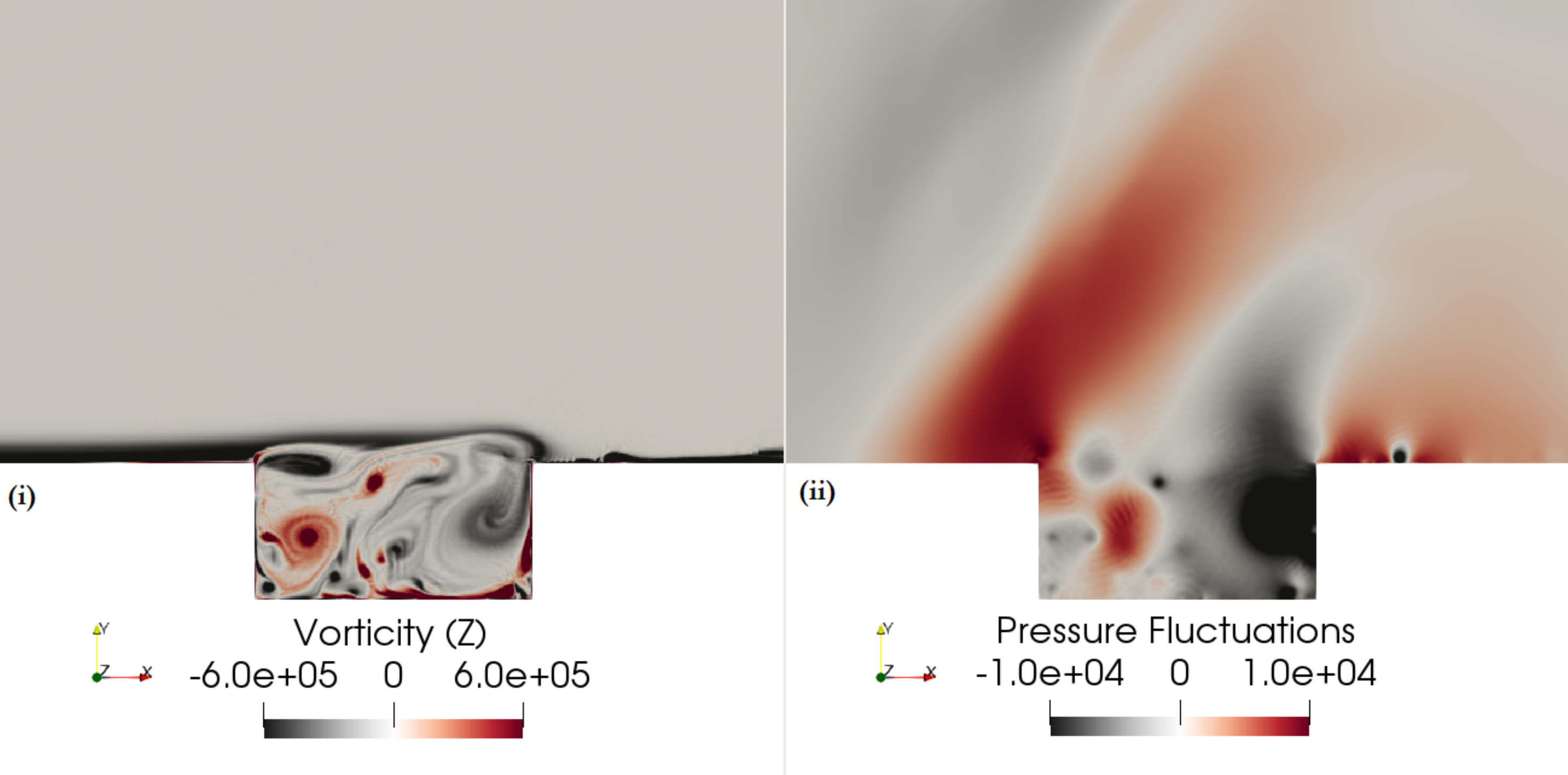}  
		\caption{}
		\label{fig:19(a)}
	\end{subfigure}
	
	\begin{subfigure}{1\textwidth}
		\includegraphics[scale=0.55]{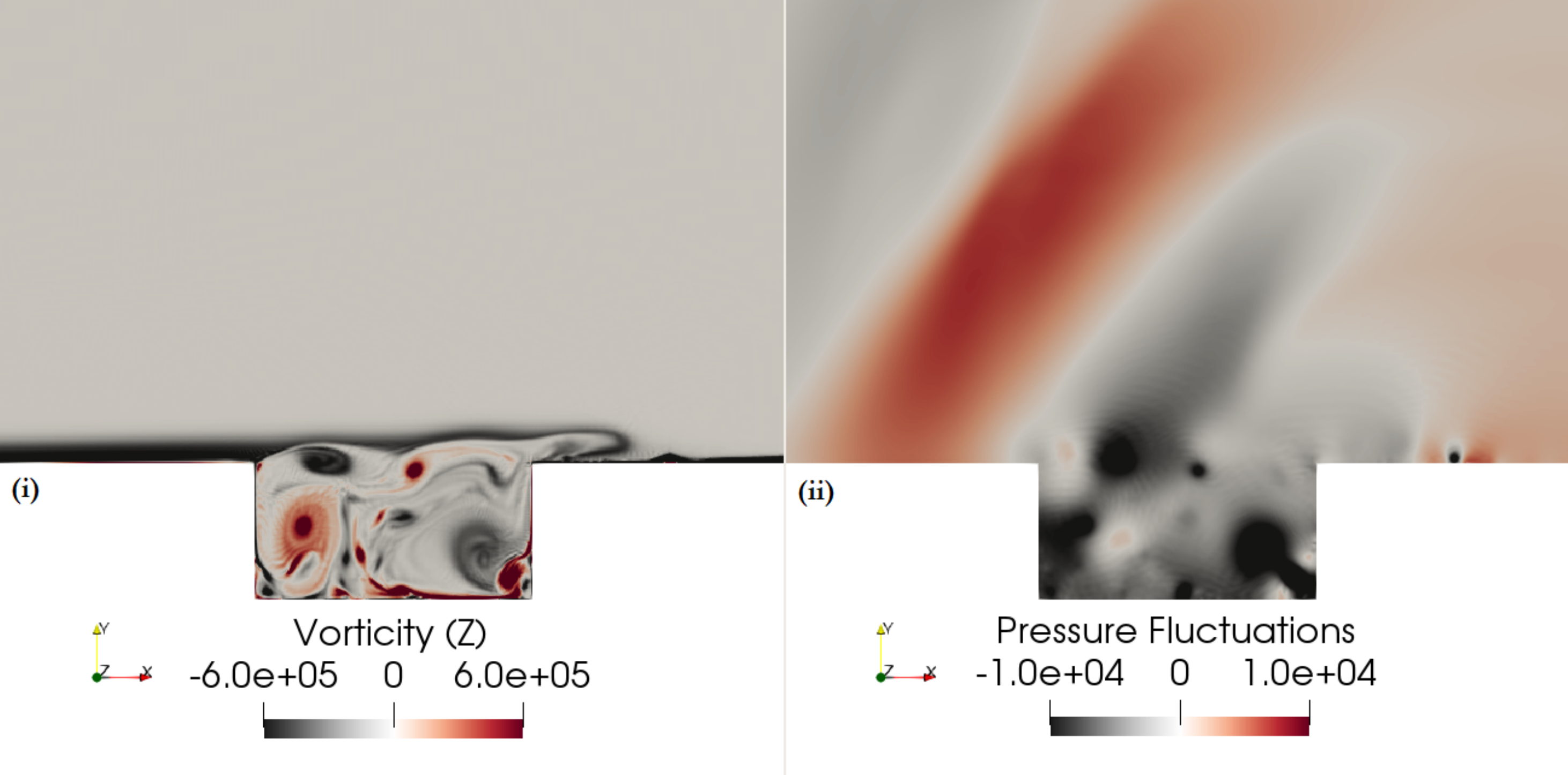}  
		\caption{}
		\label{fig:19(b)}
	\end{subfigure}
\end{figure}
\begin{figure}
	\ContinuedFloat
	\begin{subfigure}{1\textwidth}
		\includegraphics[scale=0.55]{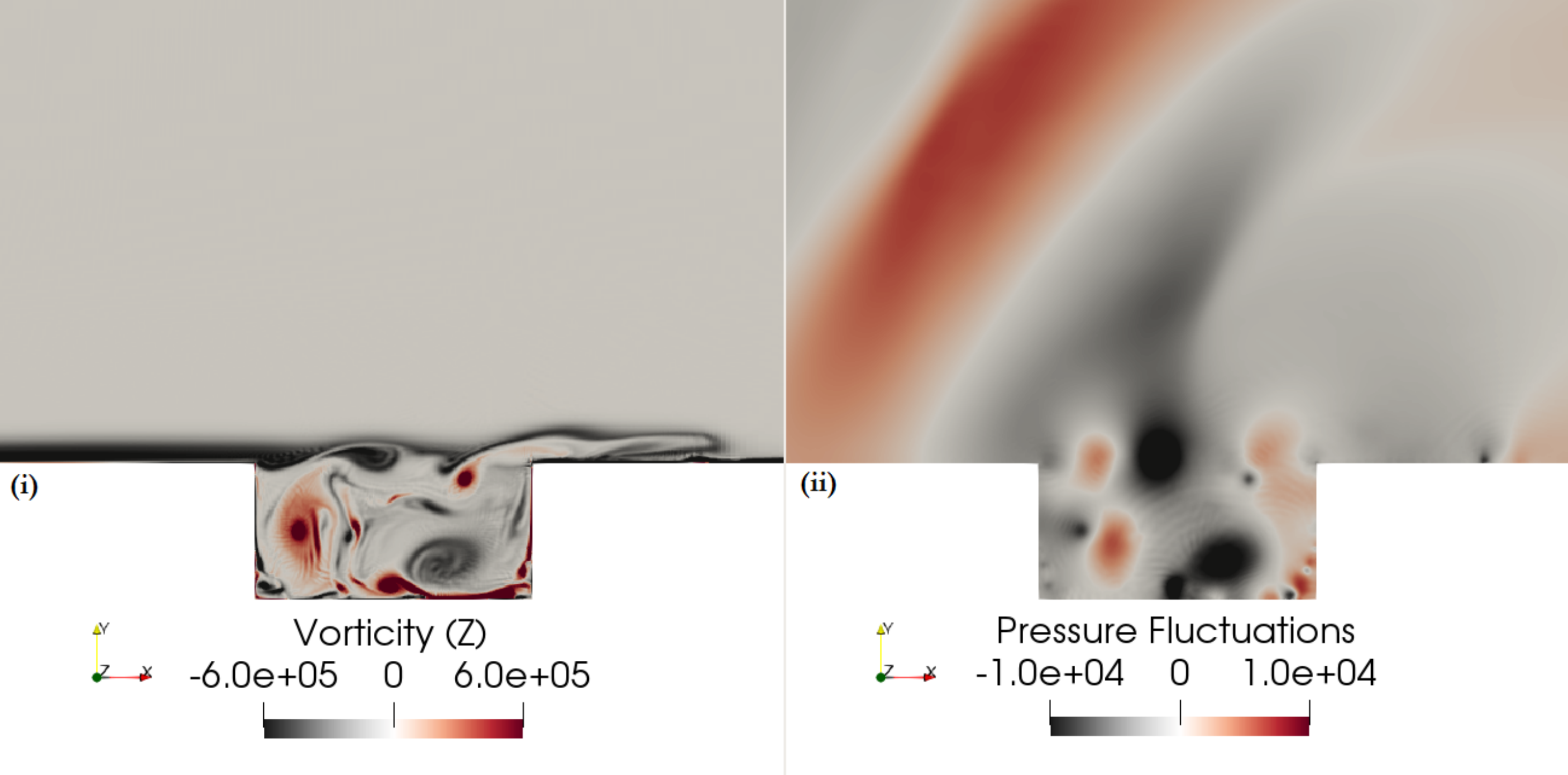}  
		\caption{}
		\label{fig:19(c)}
	\end{subfigure}
	\begin{subfigure}{1\textwidth}
	\includegraphics[scale=0.55]{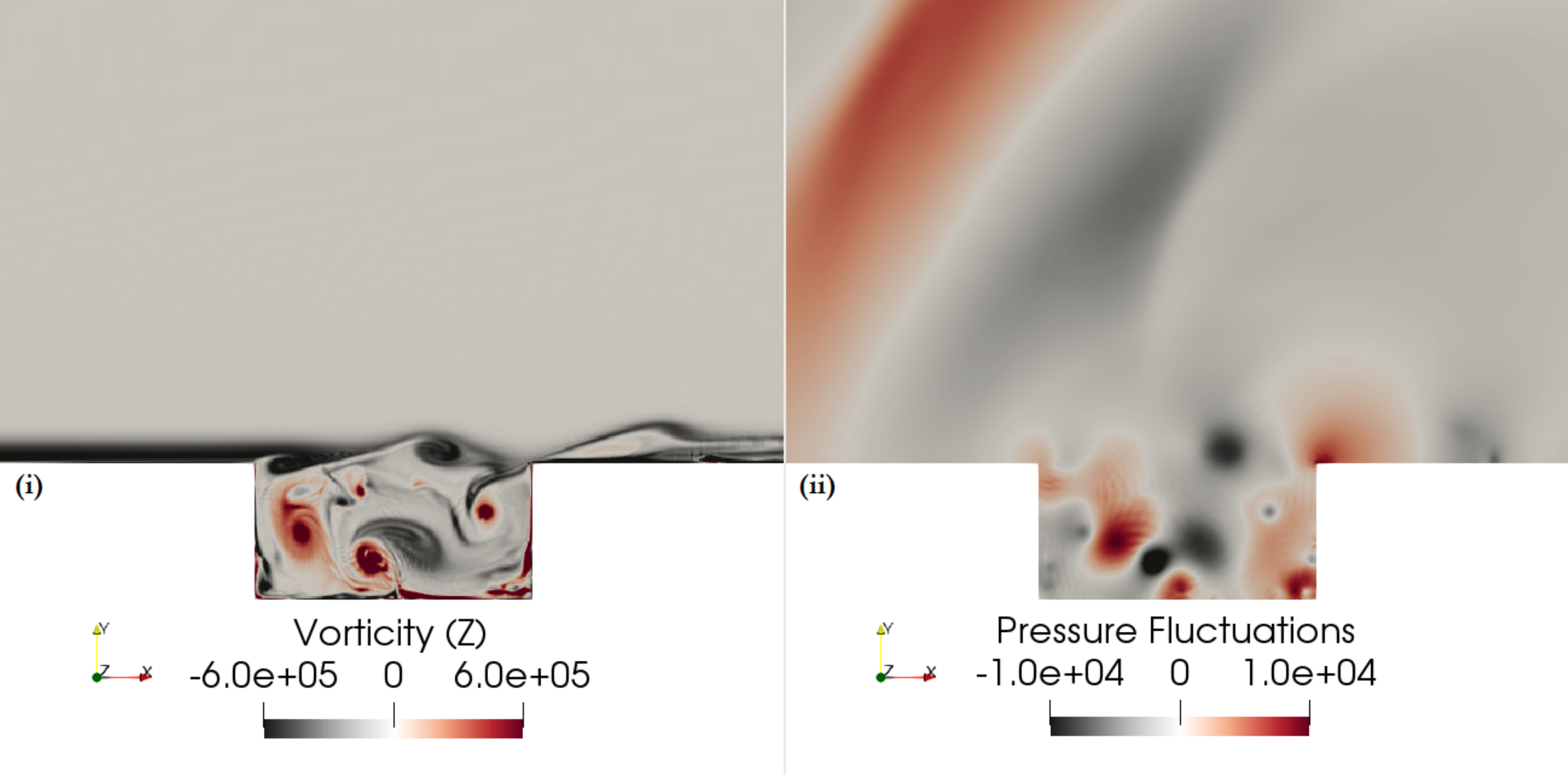}  
	\caption{}
	\label{fig:19(d)}
    \end{subfigure}
\end{figure}

\begin{figure}
	\ContinuedFloat
	\begin{subfigure}{1\textwidth}
		\includegraphics[scale=0.55]{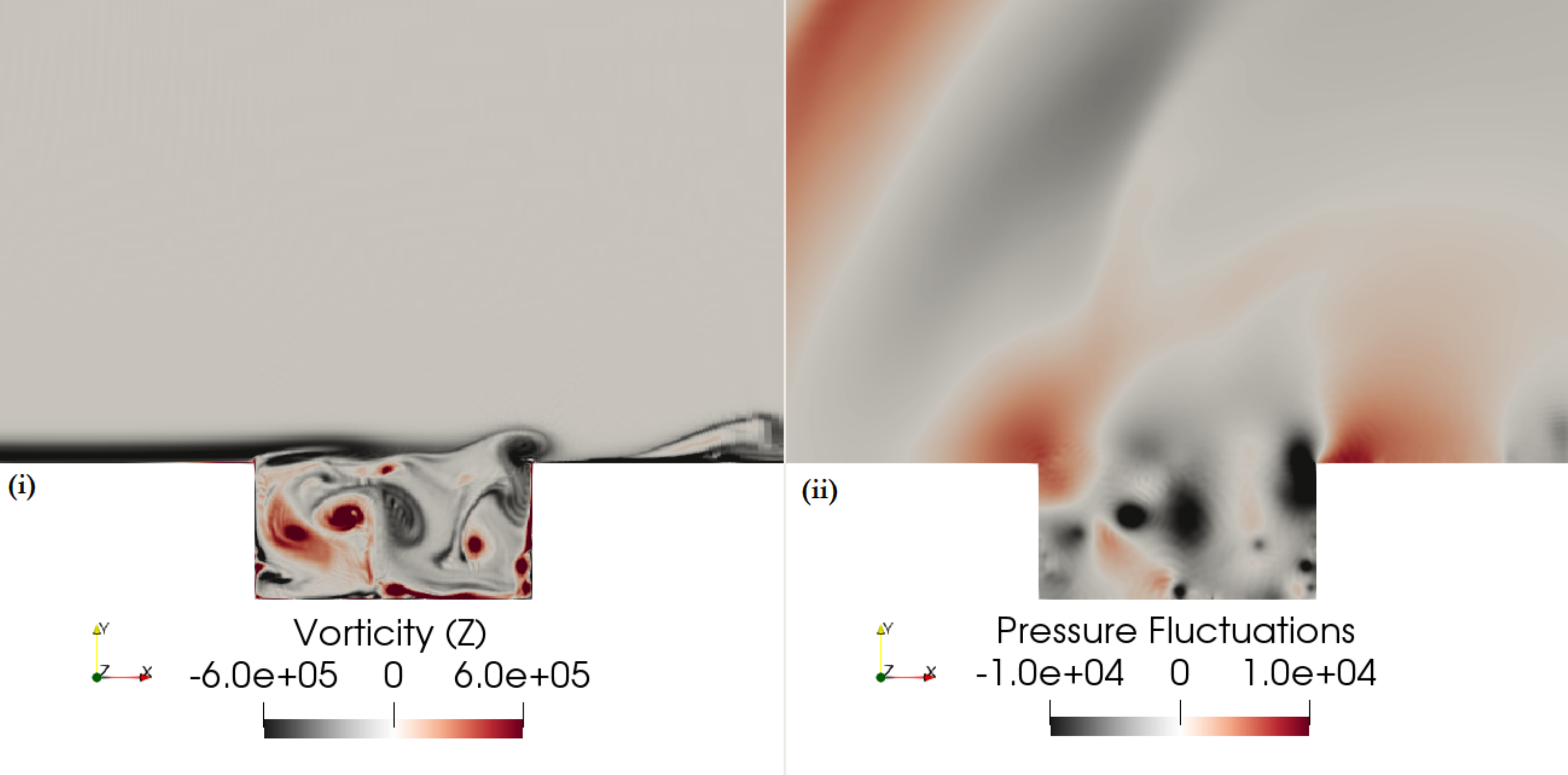}  
		\caption{}
		\label{fig:19(e)}
	\end{subfigure}
	\caption{(i) Vorticity and (ii) Perturbation pressure over the cavity (M=0.7) for one cycle a) Shedding of vortex from the leading edge, b)Vortex convection downstream c) Growth of vortex and bending of the shear layer, d) Vortex Impingement and generation of new vortex at the leading edge, e) Convection of the vortex downstream of trailing edge }
\label{fig:19}
\end{figure}


\begin{figure}
	\begin{subfigure}{1\textwidth}
		\includegraphics[scale=0.55]{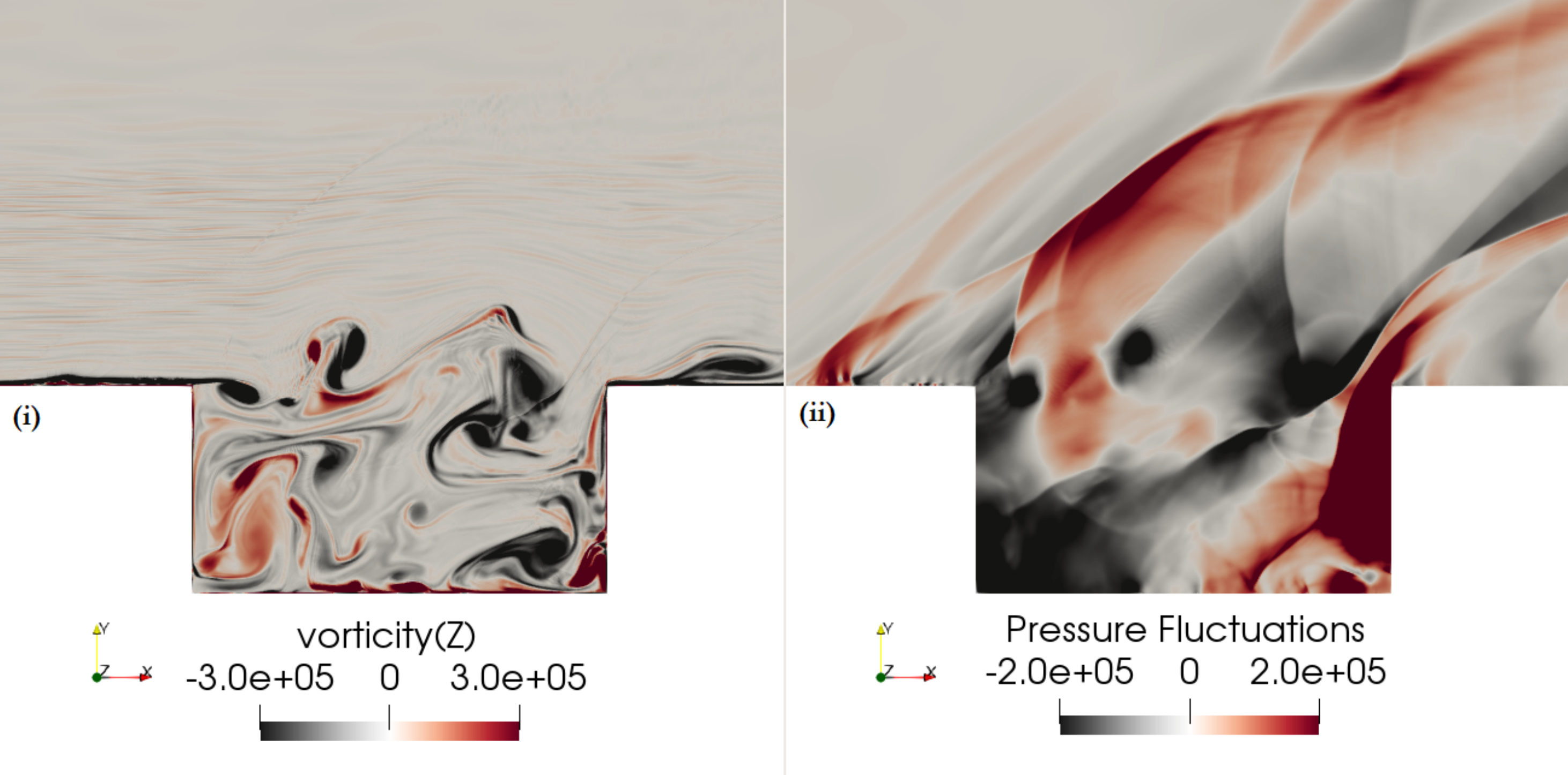}  
		\caption{}
		\label{fig:36(a)}
	\end{subfigure}
	
	\begin{subfigure}{1\textwidth}
		\includegraphics[scale=0.55]{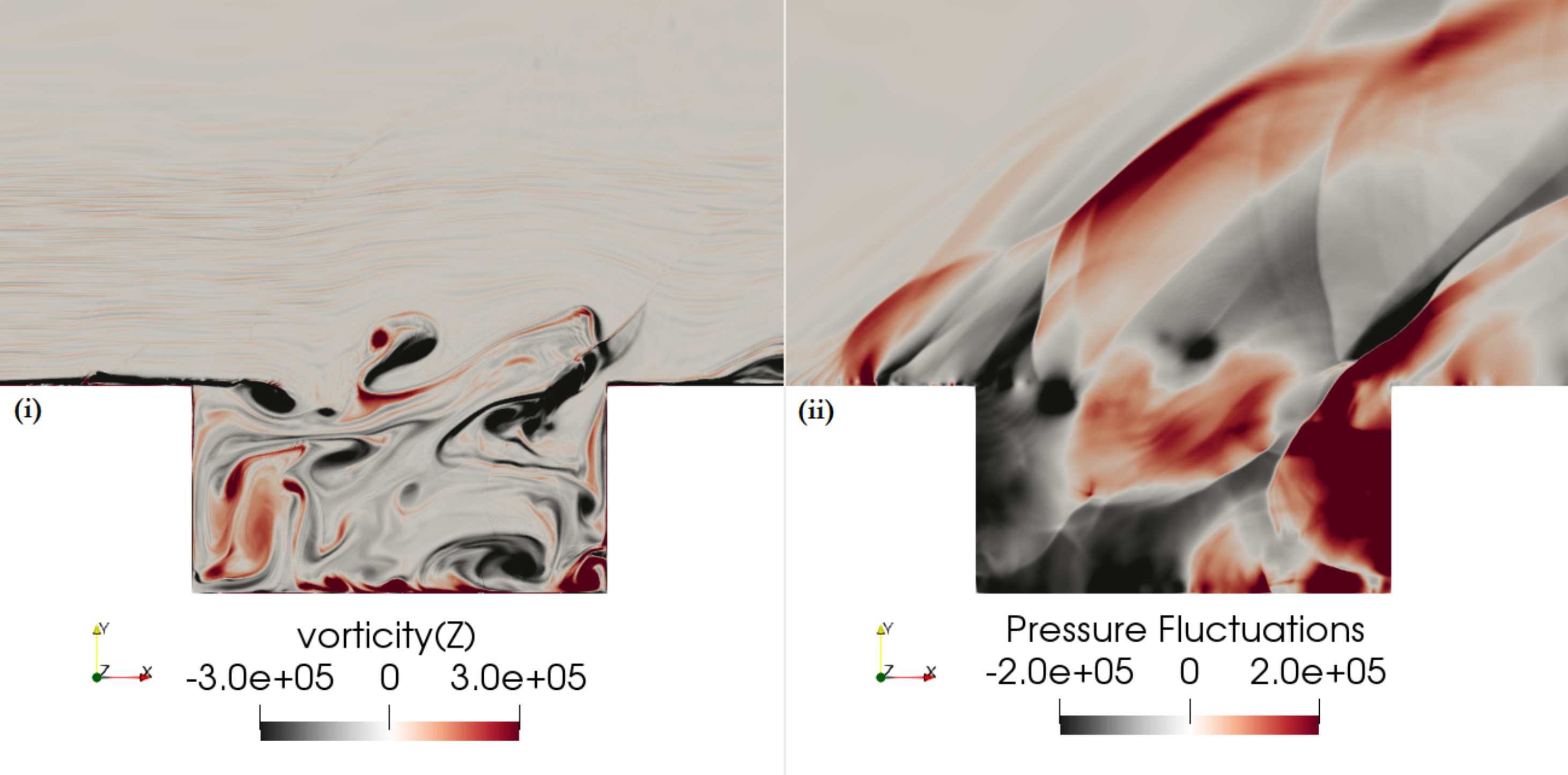}  
		\caption{}
		\label{fig:36(b)}
	\end{subfigure}
\end{figure}
\begin{figure}
	\ContinuedFloat
	\begin{subfigure}{1\textwidth}
		\includegraphics[scale=0.55]{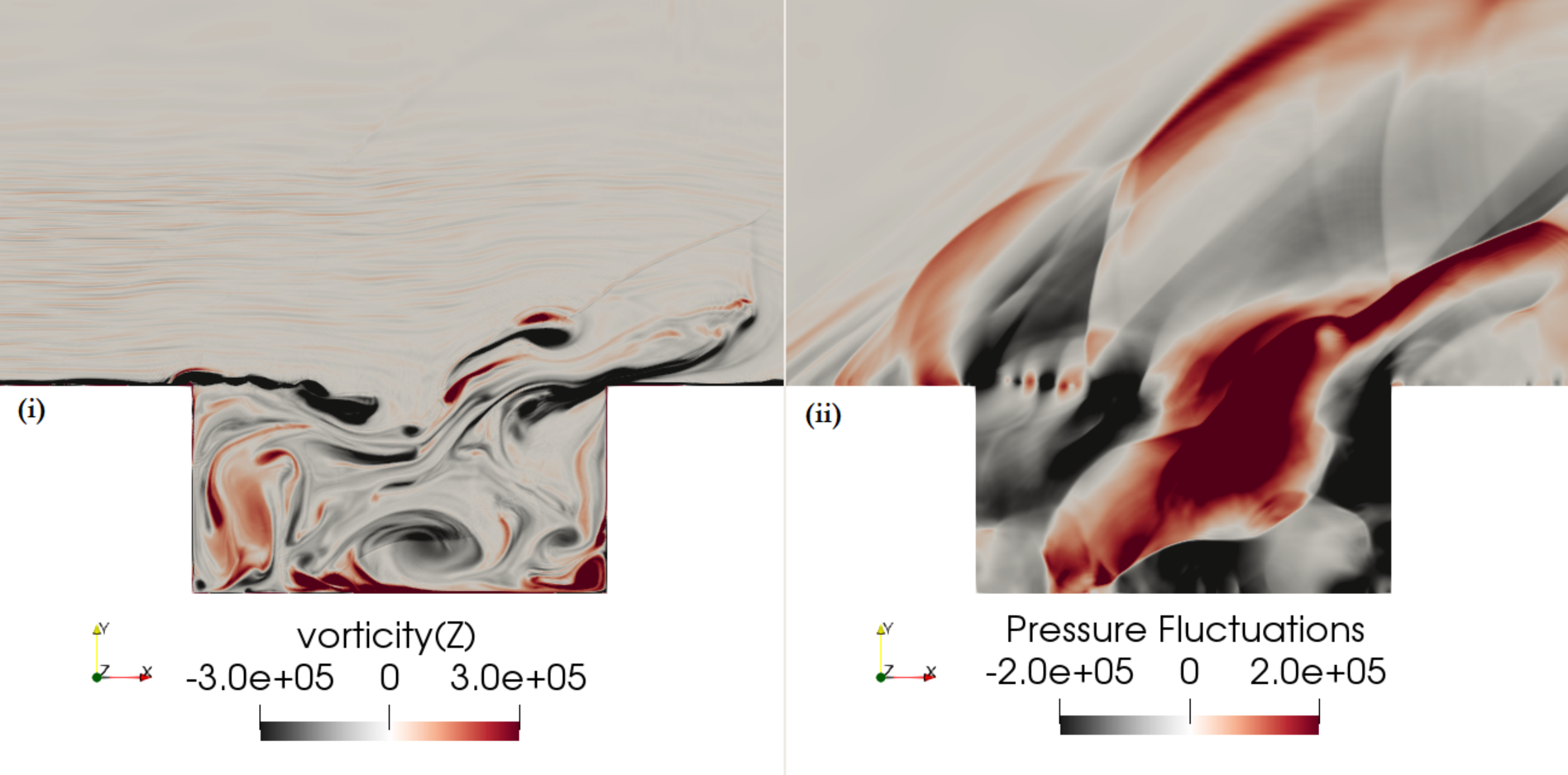}  
		\caption{}
		\label{fig:36(c)}
	\end{subfigure}
	\begin{subfigure}{1\textwidth}
		\includegraphics[scale=0.55]{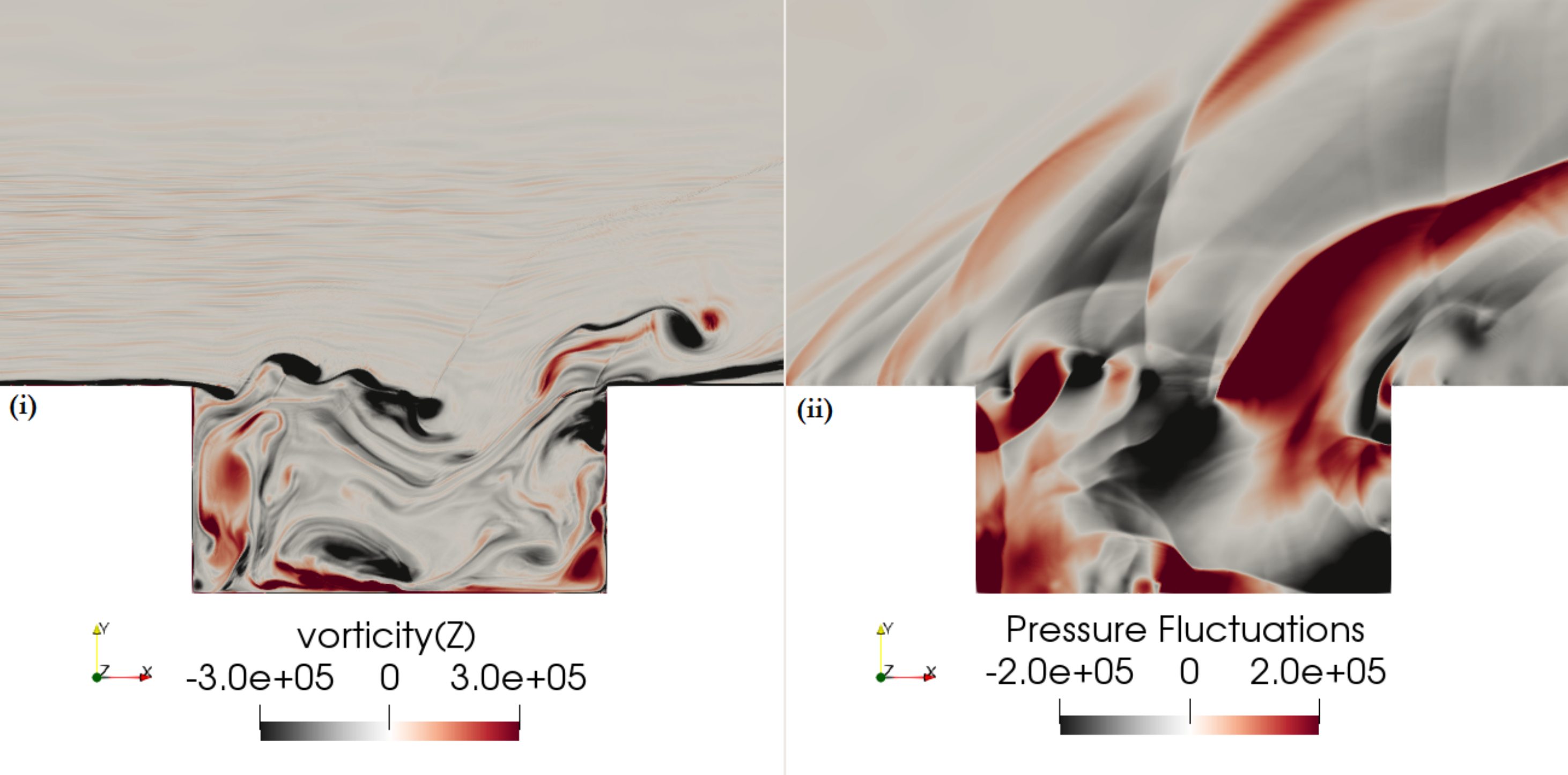}  
		\caption{}
		\label{fig:36(d)}
	\end{subfigure}
	\caption{(i) Vorticity and (ii) Pressure perturbations over the cavity (M=1.5): a) Vortex formation at the leading edge and vortex pairing, b) Convection of vortex pair downstream, c) Discrete vortices along the shear layer, d) Vortex Impingement at the trailing edge and generation of new vortex at the leading edge }
\label{fig:36}
\end{figure}

\begin{figure}
	\begin{subfigure}{1\textwidth}
		\includegraphics[scale=0.6]{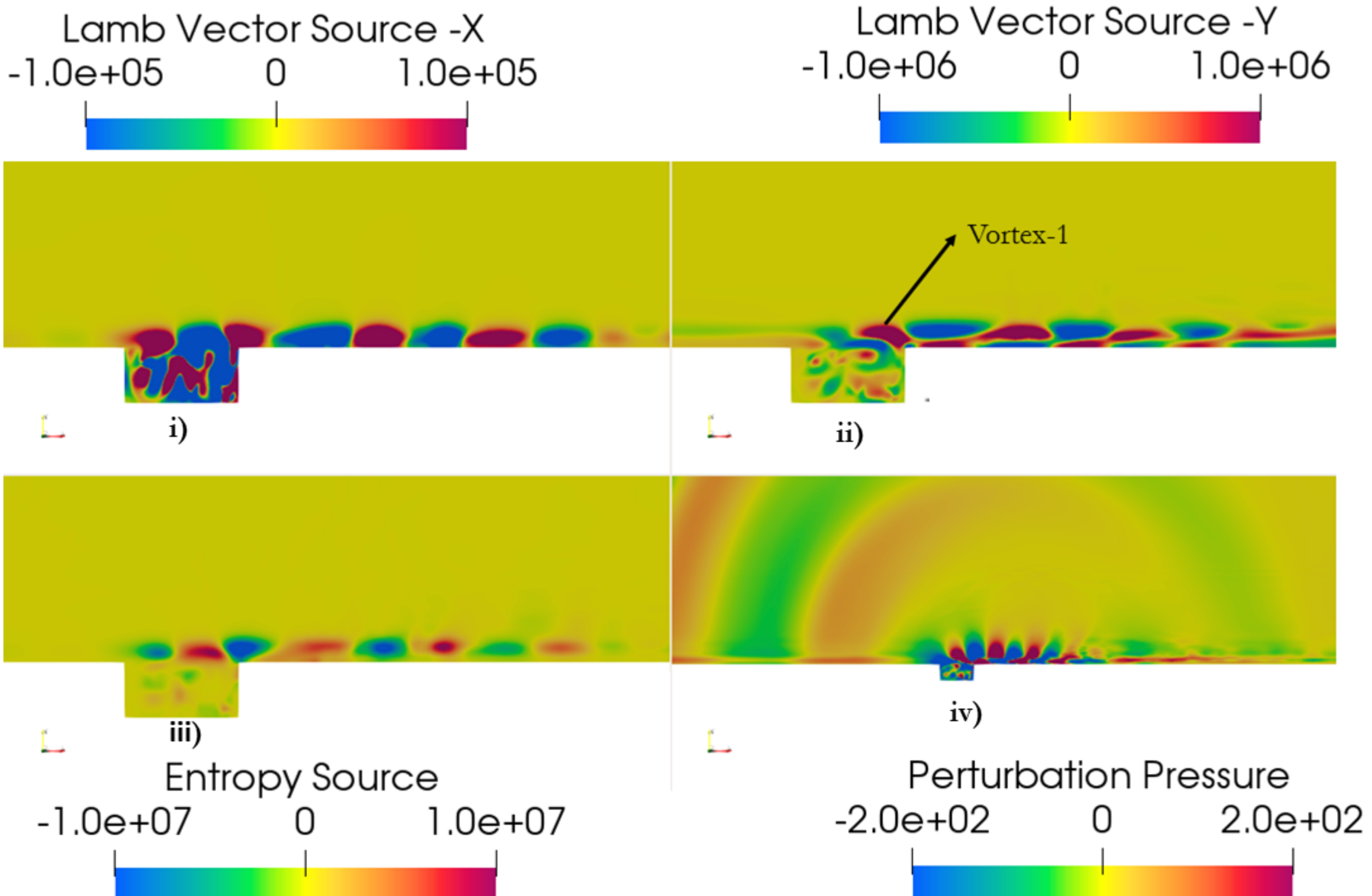}  
		\caption{vortex clipping}
		\label{fig:40(a)}
	\end{subfigure}
	
	\begin{subfigure}{1\textwidth}
		\includegraphics[scale=0.6]{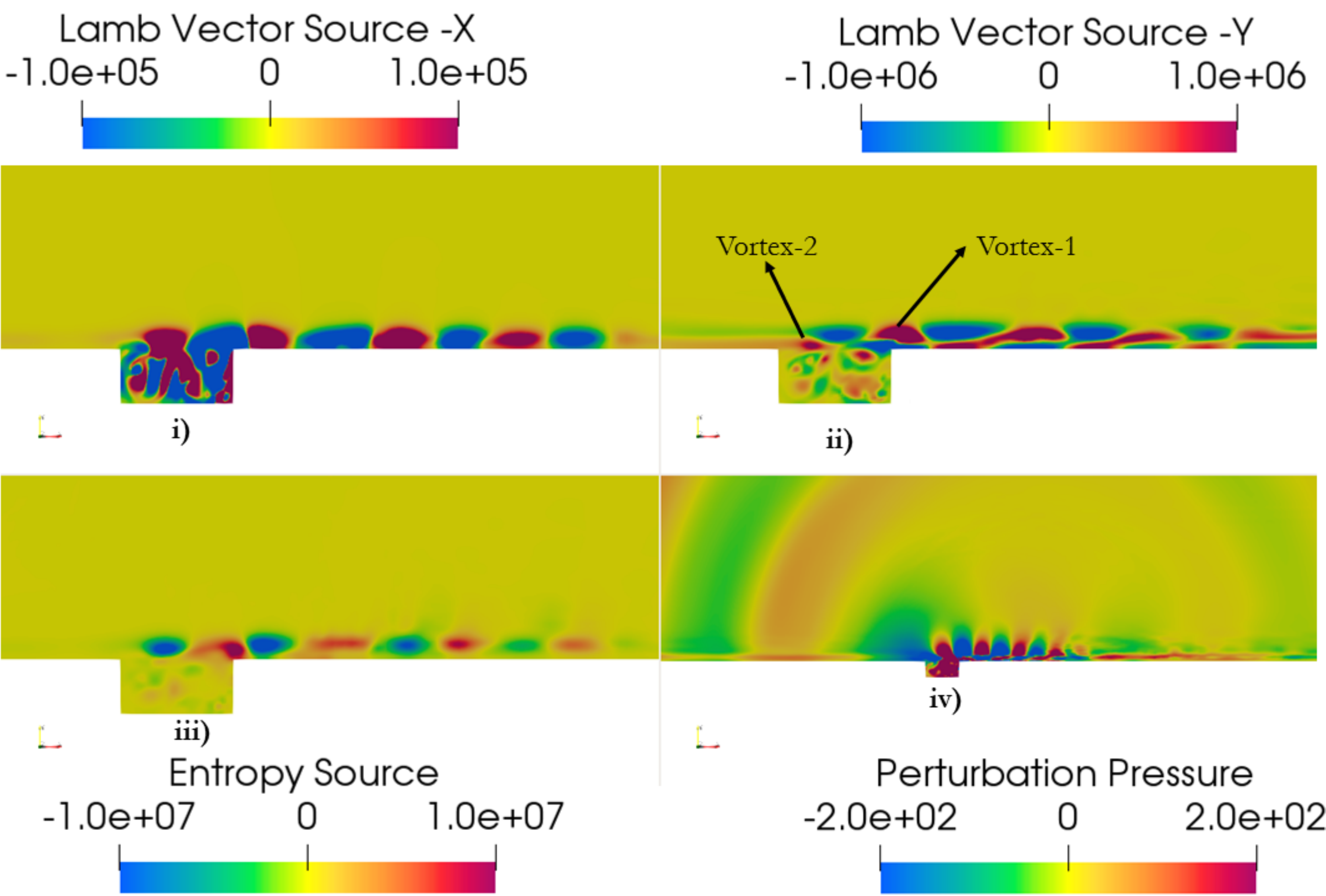}  
		\caption{New vortex from the leading edge}
		\label{fig:40(b)}
	\end{subfigure}
\end{figure}
\begin{figure}
	\ContinuedFloat
	\begin{subfigure}{1\textwidth}
		\includegraphics[scale=0.6]{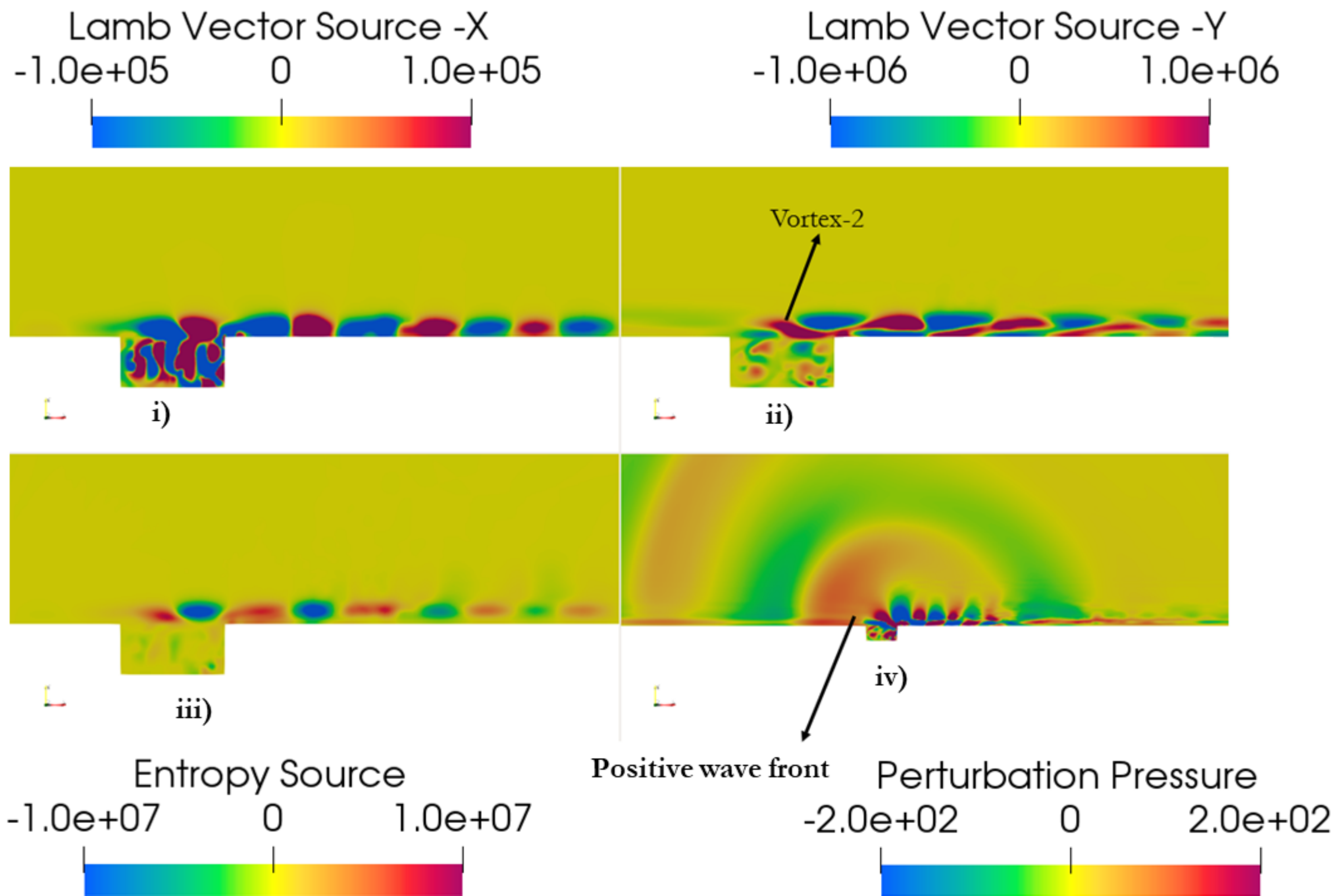}  
		\caption{Vortex bending}
		\label{fig:40(c)}
	\end{subfigure}
	
	\begin{subfigure}{1\textwidth}
		\includegraphics[scale=0.6]{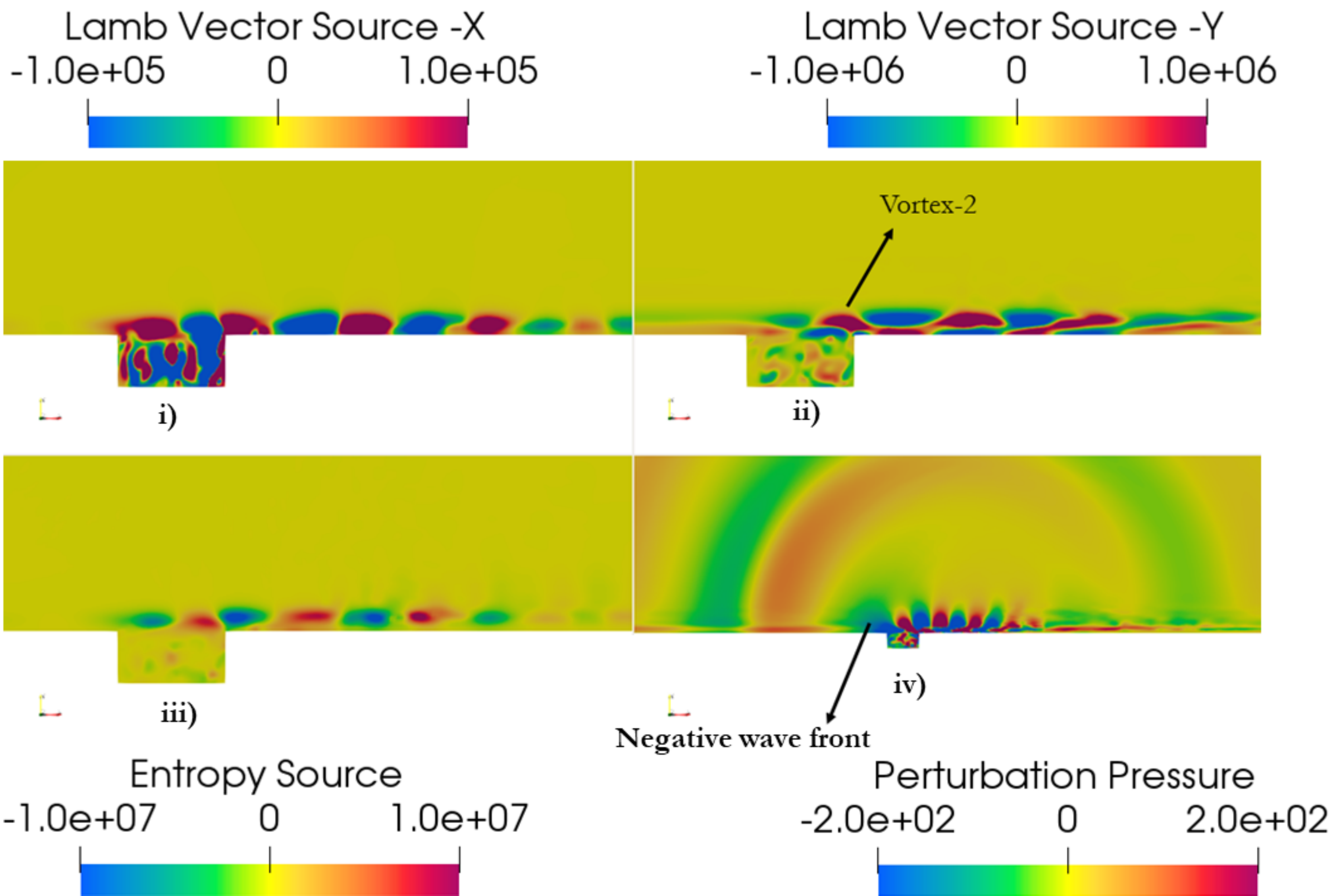}  
		\caption{New vortex clipping}
		\label{fig:40(d)}
	\end{subfigure}
	\caption{Evolution of sources ( (i) X-component of Lamb Vector (ii) Y-component of Lamb Vector (iii) Entropy Source)) and (iv) Perturbation pressure for subsonic cavity flow: a) Vortex Clipping, b) New vortex from the leading edge, c) Vortex bending, d) New vortex clipping }
	\label{fig:40}
\end{figure}
\begin{figure}
	\begin{subfigure}{1\textwidth}
		\includegraphics[scale=0.6]{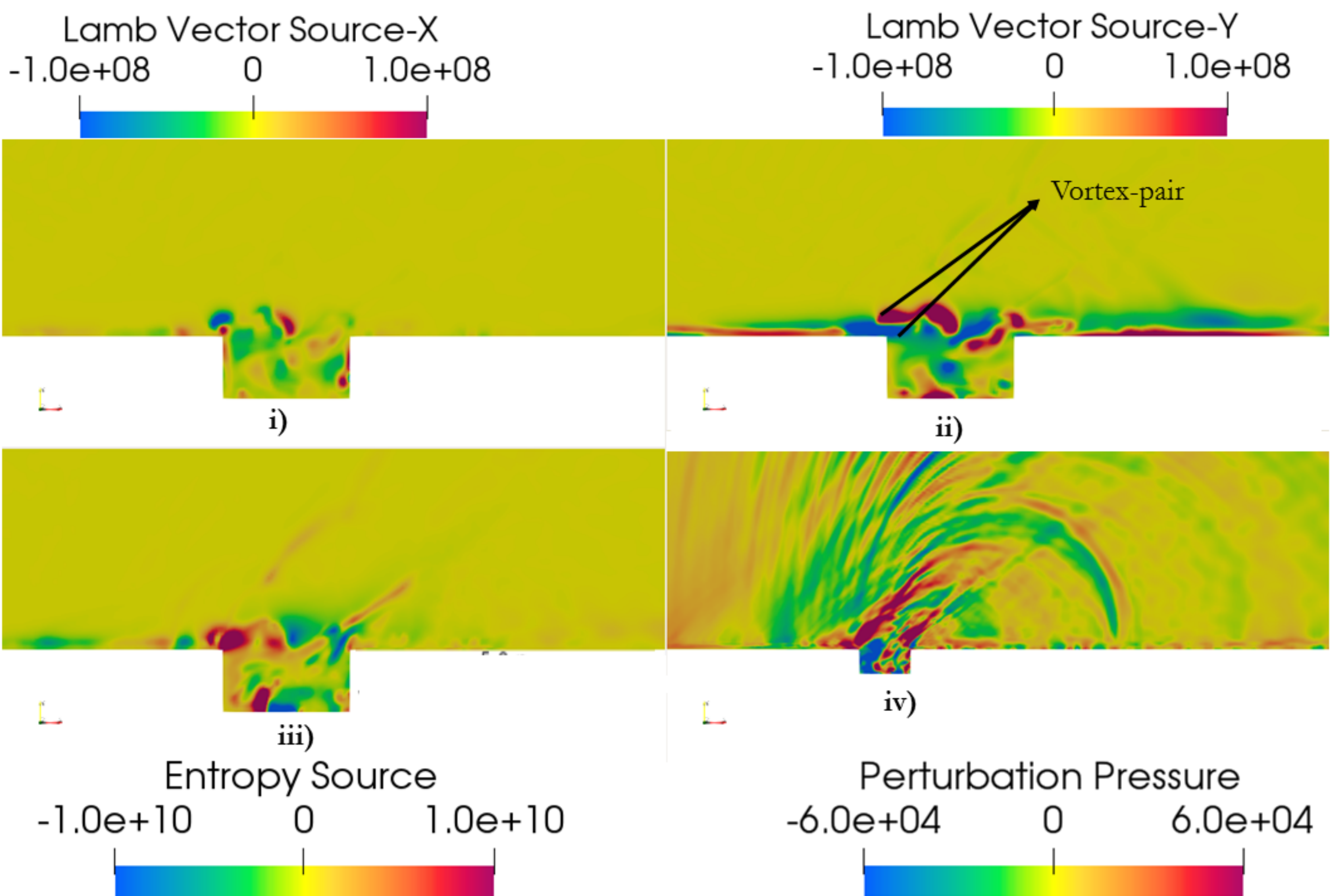}  
		\caption{vortex pair at the leading edge}
		\label{fig:41(a)}
	\end{subfigure}
	\begin{subfigure}{1\textwidth}
		\includegraphics[scale=0.6]{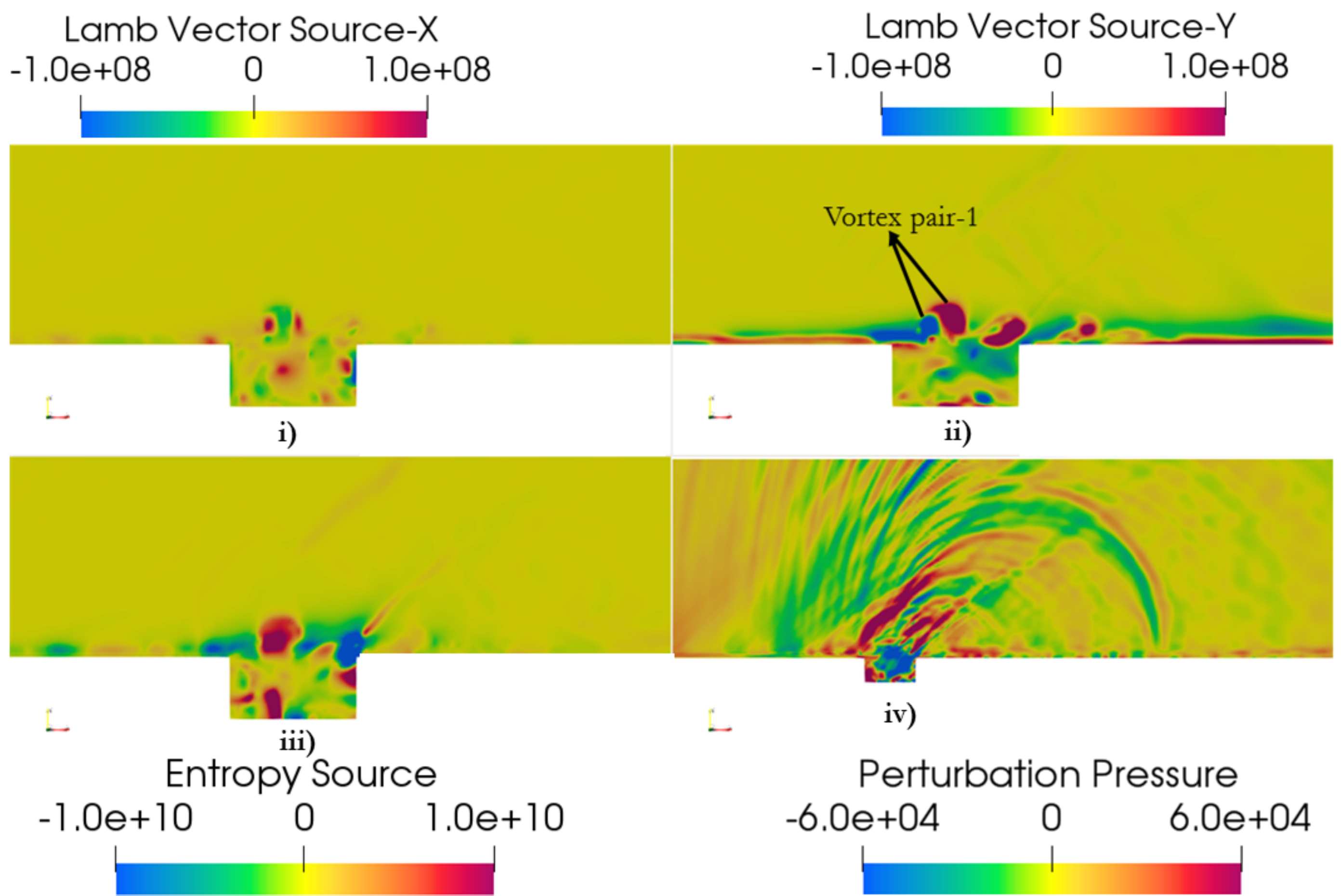}  
		\caption{vortex clipping at the trailing edge}
		\label{fig:41(b)}
	\end{subfigure}
\end{figure}
\begin{figure}
	\ContinuedFloat
	\begin{subfigure}{1\textwidth}
		\includegraphics[scale=0.6]{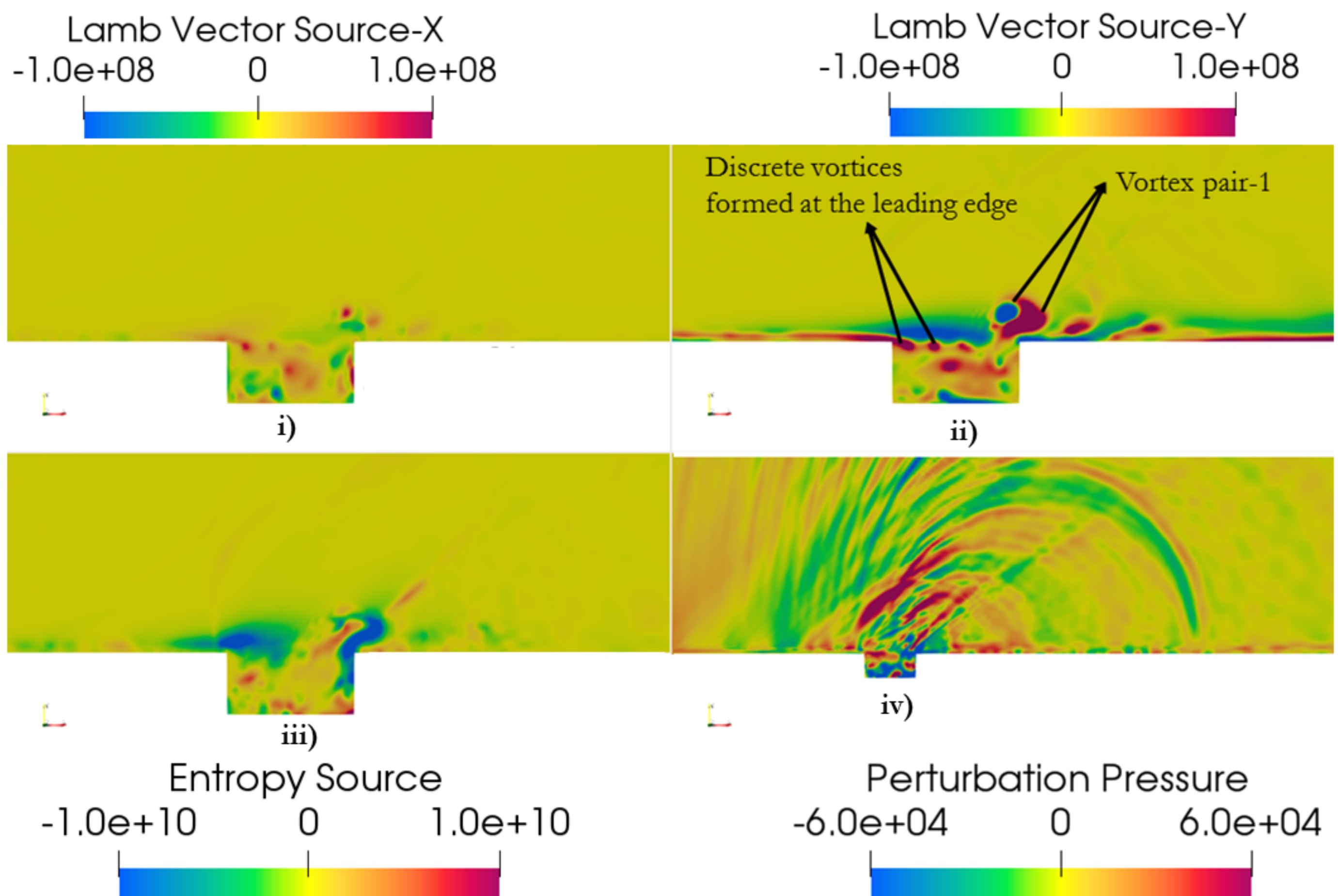}  
		\caption{Discrete vortices at the leading edge}
		\label{fig:41(c)}
	\end{subfigure}
	\begin{subfigure}{1\textwidth}
		\includegraphics[scale=0.6]{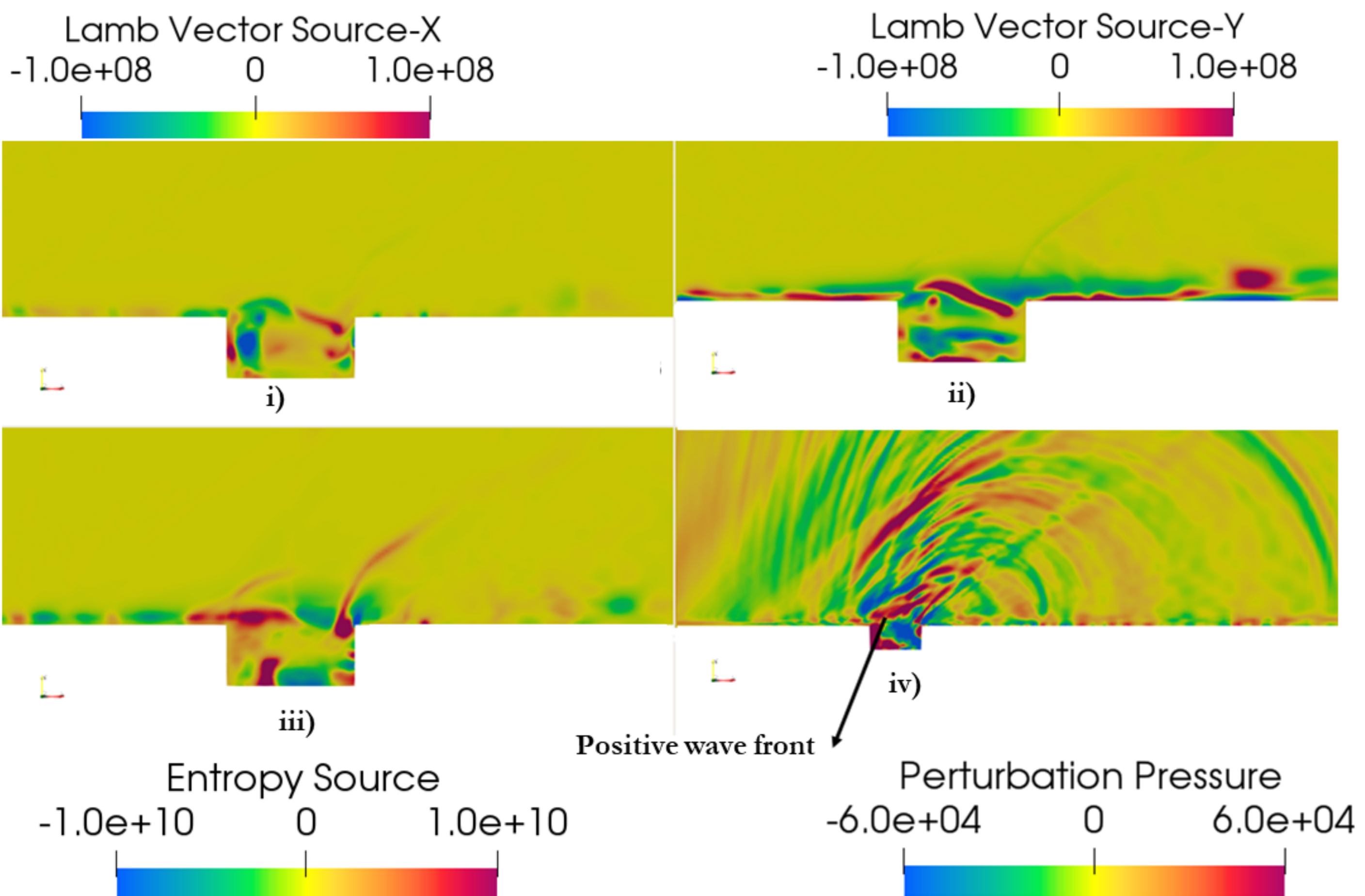}  
		\caption{Merging of vortices}
		\label{fig:41(d)}
	\end{subfigure}
\end{figure}
\begin{figure}
	\ContinuedFloat
	\begin{subfigure}{1\textwidth}
		\includegraphics[scale=0.6]{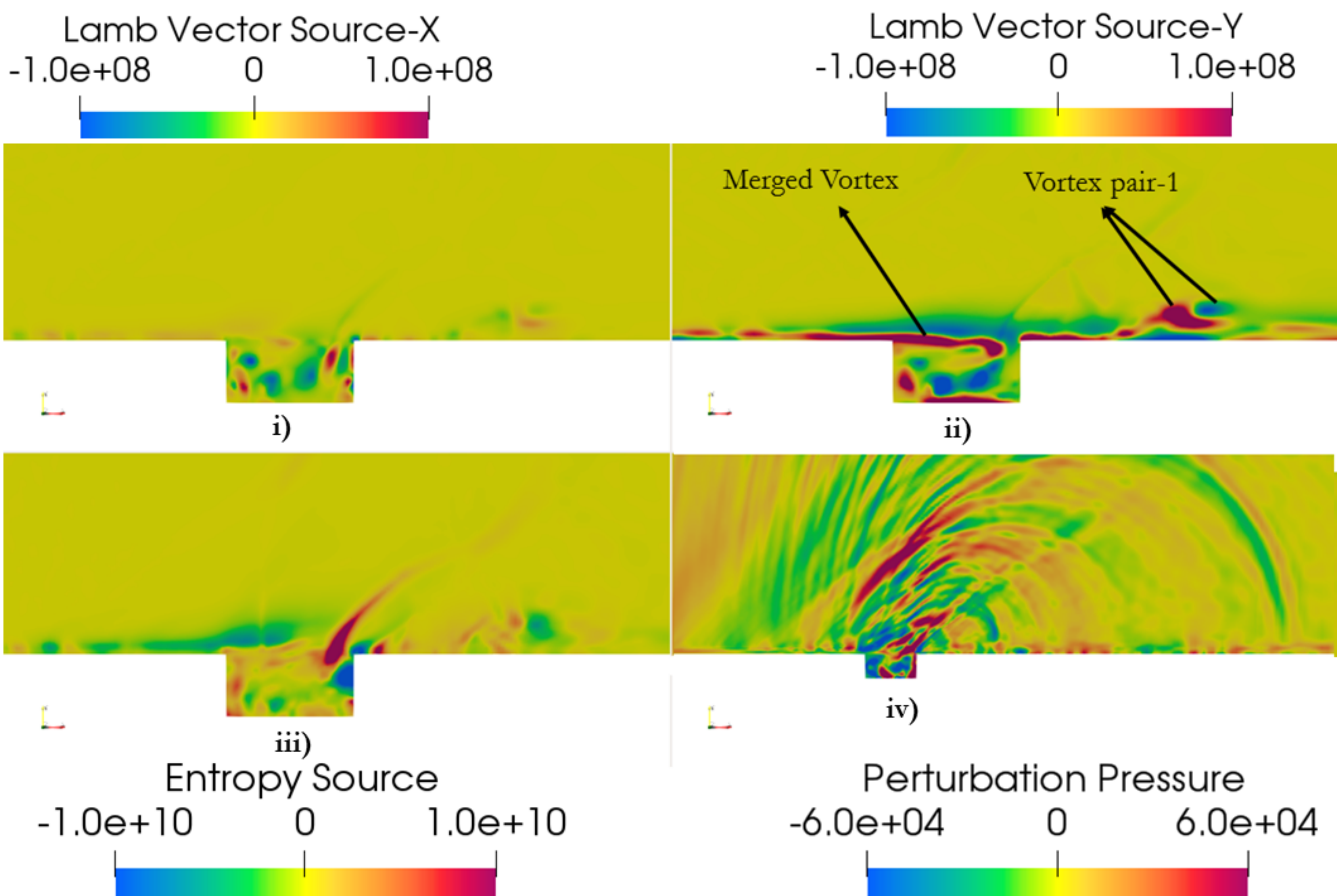}  
		\caption{vortex bending}
		\label{fig:41(e)}
	\end{subfigure}
	\caption{Evolution of sources ( (i) X-component of Lamb Vector (ii) Y-component of Lamb Vector (iii) Entropy Source)) and (iv) Perturbation pressure for supersonic cavity flow: a) Vortex pair at the leading edge, b) Vortex clipping at the trailing edge, c) Discrete vortices at the leading edge, d) Merging of vortices, e) Vortex bending }
	\label{fig:41}
\end{figure}  
\end{document}